\documentclass[12pt,letterpaper,usenames,dvipsnames]{article}
\usepackage{jheppub}

\usepackage{dsfont}
\usepackage{amssymb}
\usepackage{latexsym}
\usepackage{braket}
\usepackage{graphicx}
\usepackage{subcaption}
\usepackage[utf8]{inputenc}
\usepackage{multirow}
\usepackage{graphicx,tikz}
\usepackage[framemethod=TikZ]{mdframed} % fancy boxes
\usepackage{tikz-cd} %for complicated commutative diagrams
\usetikzlibrary{arrows,snakes,shapes.arrows,decorations.markings}
     \tikzset{>=triangle 90}
     \tikzstyle{bbc}=[draw,circle,fill=black,scale=.75]
     \tikzstyle{rc}=[circle,fill=red,scale=.6]
     \tikzstyle{wc}=[draw,circle,scale=.75]

\usepackage{amsmath}
\usepackage{hyperref}
\definecolor{darkred}{rgb}{0.8,0.1,0.1}
\hypersetup{colorlinks=true, linkcolor=darkred, citecolor=blue, linktoc=page}

\renewcommand{\thanks}[1]{\footnote{#1}}

\newcommand{\bea}{\begin{eqnarray}}
\newcommand{\eea}{\end{eqnarray}}
\newcommand{\ee}{\end{equation}}
\newcommand{\be}{\begin{equation}}

\usepackage{adjustbox}
\usepackage{multirow}
\usepackage{hhline}
\usepackage{arydshln}

\def\cH{{\cal H}}
\def\cI{{\cal I}}

\def\cM{{\cal M}}
\def\cN{{\cal N}}

\def\cR{{\cal R}}

\def\cS{{\cal S}}

\def\cW{{\cal W}}

\def\half{ {1\over 2}}

\def\ZZ{{\mathbb Z}}
\def\RR{{\mathbb R}}

\def\CC{{\mathbb C}}

\def\no{\nonumber}
\def\text{\mathrm}

\newcommand{\dittotikz}{%
    \tikz{
        \draw [line width=0.12ex] (-0.2ex,0) -- +(0,0.8ex)
            (0.2ex,0) -- +(0,0.8ex);
        \draw [line width=0.08ex] (-0.6ex,0.4ex) -- +(-1.5em,0)
            (0.6ex,0.4ex) -- +(1.5em,0);
    }%
}

\usetikzlibrary{decorations.pathreplacing,calligraphy}

\makeatletter
\def\@fpheader{\ }
\makeatother

\title{Exceptional moduli spaces for exceptional $\mathcal{N}=3$ theories}
\author[1]{Justin Kaidi}
\author[1,2,3]{Mario Martone}
\author[1,2]{Gabi Zafrir}

\affiliation[1]{Simons Center for Geometry and Physics, Stony Brook University, Stony Brook, NY 11794-3840, USA}
\affiliation[2]{C.~N.~Yang Institute for Theoretical Physics,  Stony Brook University,Stony Brook, NY 11794-3840, USA}
\affiliation[3]{Department of Mathematics, King's College London, The Strand, London WC2R 2LS, U.K.}

\emailAdd{jkaidi@scgp.stonybrook.edu, mario.martone@kcl.ac.uk, 
gzafrir@scgp.stonybrook.edu}

\abstract{It is expected on general grounds that the moduli space of 4d $\cN=3$ theories is of the form $\CC^{3r}/\Gamma$, with $r$ the rank and $\Gamma$ a crystallographic complex reflection group (CCRG). As in the case of Lie algebras, the space of CCRGs consists of several infinite families, together with some exceptionals. To date, no 4d $\cN=3$ theory with moduli space labelled by an exceptional CCRG (excluding Weyl groups) has been identified. In this work we show that the 4d $\cN=3$ theories proposed in \cite{Garcia-Etxebarria:2016erx}, constructed via non-geometric quotients of type-$\mathfrak{e}$ 6d (2,0) theories, realize nearly all such exceptional moduli spaces. In addition, we introduce an extension of this construction to allow for twists and quotients by outer automorphism symmetries. This gives new examples of 4d $\cN=3$ theories going beyond simple S-folds. }

\begin{document}
\maketitle

\section{Introduction}

It was long believed that the amount of   supersymmetry allowed for four-dimensional field theories was required to be $\cN=0,1,2,$ or $4$, with a conspicuous omission of $\cN=3$. The reason was simple: by superconformal representation theory, it can be shown that genuinely $\cN=3$ theories are necessarily isolated SCFTs \cite{Aharony:2015oyb,Cordova:2016emh,Chang:2018xmx}, and that a CPT-preserving Lagrangian invariant under twelve supercharges is always invariant under the maximal set of sixteen. 

It was only relatively recently that purely $\cN=3$ theories %\footnote{Henceforth, ``$\cN=3$ theories" will always refer to theories in four spacetime dimensions.}
were shown to exist \cite{Garcia-Etxebarria:2015wns,Lemos:2016xke,Nishinaka:2016hbw,Imamura:2016abe,Bourton:2018jwb,Amariti:2020lua,Zafrir:2020epd,Agarwal:2021oyl}, with an infinite family of them being realized in string theory as the worldvolume theories of stacks of $\mathrm{D}3$-branes probing so-called \textit{S-folds} \cite{Aharony:2016kai, Apruzzi:2020pmv, Giacomelli:2020jel, Heckman:2020svr,Agarwal:2016rvx, Giacomelli:2020gee, Bourget:2020mez,Imamura:2016udl,Arai:2018utu,Arai:2019xmp,Behan:2022uqr}---generalizations of the familiar orientifold plane. From the point of view of the low-energy effective theory, $\cN=3$ theories are very much like $\cN=4$ theories. In particular, there is no intrinsic difference between Coulomb and Higgs branches, and the moduli spaces are all locally flat, see e.g. \cite{Cordova:2016xhm}. Moreover, the $\cN=3$ moduli space has a \emph{triple special K\"ahler structure} \cite{Argyres:2019yyb}, which gives strong constraints on the set of allowed moduli spaces of vacua of the theories. Adding two extra assumptions, namely:

\begin{itemize}
\item[1.] The coordinate ring of an $\cN\geq 2$ Coulomb branch slice is freely-generated,

\item[2.] The moduli space is globally an orbifold,
\begin{equation}
\no
    \cM=\CC^{3r}/\Gamma~,
\end{equation}
with $r$ the rank of the $\cN=3$ theory,
\end{itemize}

\noindent a full classification becomes possible.\footnote{An explicit classification at rank-2, even lifting condition 1, has been carried out in \cite{Argyres:2019ngz}.} Indeed, consistent moduli spaces of $\cN=3$ theories satisfying the conditions above have been shown to be in one-to-one correspondence with crystallographic complex reflection groups (CCRGs) \cite{Caorsi:2018zsq,Argyres:2018wxu,Bonetti:2018fqz,Argyres:2019ngz,Tachikawa:2019dvq}. A thorough review of CCRGs can be found in Appendix \ref{app:CCRGs}. 

\begin{table}[tp]
\begin{center}
\begin{tabular}{c|c|c|c}
$ \Gamma$ & M-theory construction & $12 c$ & $\{ \Delta_i\}$
\\\hline
$\ZZ_6$& $\mathfrak{g}_2/\widetilde{\ZZ}_4, \mathfrak{g}_2/\ZZ_6 $ & 33 & $\{6\}$
\\
$G_4$ & $\mathfrak{d}_4/\widetilde{\ZZ}_3$ & 54 & $\{4,6\}$
\\
$G_5$ & $\mathfrak{e}_6 /\ZZ_6 $& 102 & $\{6,12\}$
\\
$G_8$ & $\mathfrak{e}_6 /\ZZ_4,\mathfrak{e}_7 /\ZZ_4, \mathfrak{f}_4 /\ZZ_4$  & 114 & $\{8,12\}$
\\
 $G_{12}$& $\mathfrak{f}_4/\widetilde{\ZZ}_4$  & 78 & $\{6,8\}$
\\
$G_{25}$& $\mathfrak{e}_6/\ZZ_3$ &  153 & $\{6,9,12\}$
\\
$G_{26}$& $\mathfrak{e}_7/\ZZ_6$ & 207 & $\{6,12,18\}$
\\
 $G_{31}$& $\mathfrak{e}_8/\ZZ_4$ & 372 & $\{8,12,20,24\}$
\\
 $G_{32}$ & $\mathfrak{e}_8/\ZZ_6$ & 492 & $\{12,18,24,30\}$
\end{tabular}
\end{center}
\caption{The exceptional 4d $\cN=3$ theories which we discuss in this paper. Each has moduli space $\CC^{3N}/\Gamma$, with $\Gamma$ an ECCRG listed above. The M-theory construction of each is denoted by $\mathfrak{g}/\ZZ_k$, which for simply-laced algebras $\mathfrak{g}$ represents the type-$\mathfrak{g}$ (2,0) theory compactified on $T^2$ with a $\ZZ_k$ S-fold. Non-simply-laced $\mathfrak{g}$ represents an appropriate outer automorphism twist  of a simply-laced (2,0) theory, as will be explained in the text.  In some cases the quotient $\ZZ_k$ also involves an outer automorphism, in which case we denote it by $\widetilde{\ZZ}_k$. }
\label{tab:mainresults}
\end{table}%

The previously-mentioned construction of $\cN=3$ theories via $\mathrm{D}3$-branes probing S-folds realizes only a small subset of the moduli spaces which are predicted from the classification of CCRGs. In particular, no theory with moduli space associated to an \textit{exceptional} CCRG (ECCRG) \cite{shephard1954finite} can be realized using the techniques of \cite{Garcia-Etxebarria:2015wns,Aharony:2015oyb}. This leads to the question of which, if any, string theoretic constructions can realize the remaining moduli spaces. 

In \cite{Garcia-Etxebarria:2016erx}, a new construction for  $\cN=3$ theories was introduced. As will be reviewed in Section \ref{sec:Mtheory}, this construction proceeds by first using M-theory on non-geometric backgrounds to engineer type-$\mathfrak{e}$ 6d (2,0) theories, and then doing a quotient along with a torus compactification to obtain a 4d theory with $\cN=3$ SUSY.  These  $\cN=3$ theories are expected to go beyond the class of theories accessible via  $\mathrm{D}3$-branes on S-folds,  though to date very little is known about them besides their string theory construction. In particular,  their moduli spaces remain completely unexplored. It is natural to hope that these theories may realize some or all of the ECCRGs, and in this work we will show that this hope is not misplaced.

Given the construction of  \cite{Garcia-Etxebarria:2016erx}, it is also natural to ask if one can extend it to obtain even more $\cN=3$ theories. For example, the type-$\mathfrak{g}$ (2,0) theory has discrete symmetries related to automorphisms of the Dynkin diagram of $\mathfrak{g}$, and we may ask if twists or quotients by those outer automorphism symmetries are allowed. In this paper we will argue that such twists and quotients are indeed consistent, and that they produce new 4d $\cN=3$ theories realizing ECCRGs.

Our main results are summarized in Table \ref{tab:mainresults}, with more detailed Hasse diagrams given in Figures \ref{fig:rank2e6}, \ref{fig:E67Hasse}, \ref{fig:E8Hasse}, and \ref{fig:F4Hasse}. By doing appropriate S-foldings and outer automorphism twists of exceptional (2,0) theories, we are able to obtain $\cN=3$ theories labelled by ECCRGs $G_4,\, G_5,\, G_8,\, G_{12},\, G_{25},\, G_{26},\, G_{31},$ and $G_{32}$.\footnote{Here we have listed only the cases of ECCRGs which are not Weyl groups. There are also the ECCRGs $G_{28}$, $G_{35}$, $G_{36}$, and $G_{37}$, which are the Weyl groups of $\mathfrak{f}_4, \mathfrak{e}_6, \mathfrak{e}_7,$ and $\mathfrak{e}_8$ respectively, and which can be realized in our construction by just doing a standard $T^2$ compactification. Since this is straightforward and gives rise to $\cN=4$ theories, we do not discuss it further.} This gives a construction for nearly all ECCRGs, with the only outliers being $G_{24}$, $G_{29}$, $G_{33}$, and $G_{34}$. It would be interesting to look for a further extension of our construction to realize those cases as well. 

We should also mention that while the techniques explained here realize nearly all cases satisfying conditions $1$ and $2$ above, lifting condition $1$ would allow for more geometries. It would be interesting to identify a UV realization of the $\cN=3$ theories realizing those geometries as well.

\paragraph{Outline:} This paper is organized as follows. In Section \ref{sec:Mtheory} we give a thorough review of the construction of \cite{Garcia-Etxebarria:2016erx}, which allows one to obtain $\cN=3$ theories from type-$\mathfrak{e}$ (2,0) theories. We will also provide generalizations of this construction allowing for twists and quotients by outer automorphism symmetries.  In Section \ref{sec:orientifoldofa} we give a review of the 4d $\cN=3$ theories obtained from compactification and orientifolding/S-folding of the type-$\mathfrak{a}_N$ (2,0) theories, with a focus on their moduli spaces. This serves as a warm-up for the more complicated examples, which begin in Section \ref{sec:exceptionalsec}. There we consider 4d $\cN=3$ theories obtained by S-folding type-$\mathfrak{e}$ (2,0) theories, and show that their moduli spaces are labelled by ECCRGs. %Again there is an ambiguity of whether they can be interpreted as discrete gaugings of known $\cN=3$ theories, but a more detailed analysis of the behavior of the (2,0) theory strongly suggests that this is not possible. 
Section \ref{sec:nonsimplac} extends the moduli space discussion to allow for outer automorphism twists, where ECCRGs again make an appearance.

For the readers' convenience we include Appendices \ref{app:CCRGs} and \ref{app:Hasse}, which give relevant background information on complex reflection groups and Hasse diagrams. We also include Appendix \ref{sec:dtype(2,0) theory}, in which we use the techniques of the main text to  study the moduli spaces of $\cN=3$ theories obtained by S-folding the type-$\mathfrak{d}_N$ (2,0) theory. We find that the moduli spaces are of the form $\CC^{3N}/\Gamma$ with $\Gamma=G(4,1,n), G(4,2,n), G(6,1,n),$ and $G(6,2,n)$, and hence that ECCRGs do not play a role. Furthermore, most of these theories are expected to be discrete gaugings of previously known theories, though the results of Section \ref{sec:e8sec} suggest otherwise for two of them.

\section{M-theory construction: review and extensions}
\label{sec:Mtheory}

\subsection{Basic setup}
We begin by reviewing the M-theory construction of $4d$ $\mathcal{N}=3$ SCFTs introduced in \cite{Garcia-Etxebarria:2016erx}. The starting point is M-theory on the space $\RR^{1,7}\times T^3$. It is known that upon compactification on $T^3$, M-theory has an $SL(3,\mathbb{Z})\times SL(2,\mathbb{Z})_{\rho}$ symmetry, where $SL(3,\mathbb{Z})$ is the geometric symmetry rotating the cycles of $T^3$, while $SL(2,\mathbb{Z})_{\rho}$ is a non-geometric symmetry exchanging M$2$-branes with M$5$-branes wrapped on $T^3$ \cite{Aharony:1996wp}. The subscript on the latter refers to the fact that it acts on the modular parameter $\rho:= \int_{T^3} C_3 + i \sqrt{\mathrm{det}\,G_{T^3}}$, with $C_3$ the M-theory three-form field and $G_{T^3}$ the metric on $T^3$. One way to understand this symmetry is to first reduce from M-theory on $T^3$ to  Type IIA on $T^2$, with the modular parameter for the $T^2$ being $\rho_{\mathrm{IIA}} := \int_{T^2} B_2 + i \sqrt{\mathrm{det}\, G_{T^2}}$. There is a well-known $SL(2,\ZZ)$ action on $\rho_{\mathrm{IIA}}$, with the  $T$ transformation coming from integral shifts of $B_2$. The $S$ transformation is a bit more complicated, and comes from T-duality along the two circles of $T^2$, combined with an exchange of the two circles. This is easiest to understand in the case of no background $B_2$ field (with torus radii $R_a, R_b$), in which case we have
\bea
\label{es:rhoIIA}
\rho_{\mathrm{IIA}} = i R_a R_b~. 
\eea
Indeed, each T-duality acts as 
\bea
\label{eq:Tdual}
\mathrm{T}\text{-}\mathrm{duality}: \qquad R_i \rightarrow {1\over R_i}~,
\eea
and hence the chain of T-dualities described above corresponds to the usual $S$ operation $\rho_{\mathrm{IIA}} \rightarrow - {1 \over \rho_{\mathrm{IIA}}}$. 

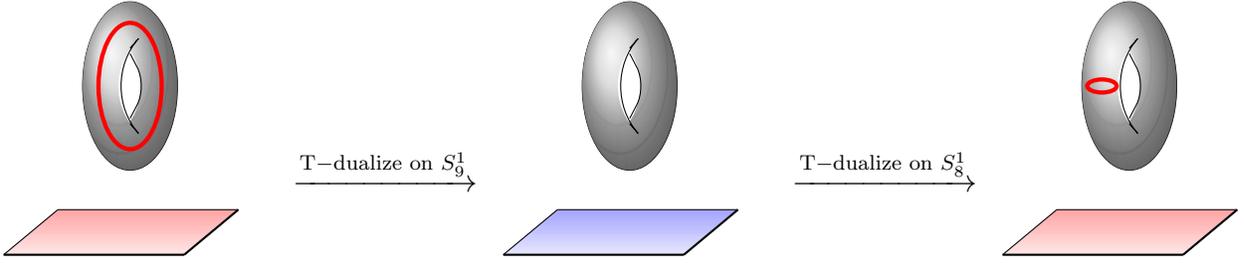
\begin{figure}[!tbp]
\begin{center}
\[
\hspace{-0.2 in}
\begin{tikzpicture}[scale=1.2,baseline=0]
%plane
 \shade[top color=red!40, bottom color=red!10]  (0,-0.7) -- (2,-0.7) -- (2.6,-0.2) -- (0.6,-0.2)-- (0,-0.7);
 \draw[thick] (0,-0.7) -- (2,-0.7);
\draw[] (0,-0.7) -- (0.6,-0.2);
\draw[]  (0.6,-0.2)--(2.6,-0.2);
\draw[thick]  (2.6,-0.2)-- (2,-0.7);
\end{tikzpicture}
\hspace{-1.2 in}
\begin{tikzpicture}[scale=0.7,baseline=-40]
\begin{scope}[rotate=90]
 %Torus
\draw[] (0,0) ellipse (1.6 and .9);
\shade[ball color = lightgray!80!white, opacity=0.55] (0,0) ellipse (1.6 and .9);
%Hole
\begin{scope}[scale=.8]
\clip (0,1.3) circle (1.55);
\fill[white] (0,-1.27) circle (1.55);
\end{scope}
\begin{scope}[scale=0.8]
\path[rounded corners=24pt] (-.9,0)--(0,.6)--(.9,0) (-.9,0)--(0,-.56)--(.9,0);
%\draw[][rounded corners=28pt] (-1,.1)--(0,-.6)--(1,.1);
\draw[][rounded corners=24pt] (-0.9,0)--(0,.8)--(0.9,0);
\draw[][rounded corners=8pt] (-0.77,0)--(0,-0.4)--(0.77,0);
%brane
\draw[ultra thick,red] (0,0) ellipse (1.5 and .75);
\end{scope}
  \end{scope}
\end{tikzpicture}
 \hspace{0.5 in} \,\,\xrightarrow{\text{T-dualize\,\,on\,\,} S_9^1}\,\,\,
 \begin{tikzpicture}[scale=1.2,baseline=0]
%plane
 \shade[top color=blue!40, bottom color=blue!10]  (0,-0.7) -- (2,-0.7) -- (2.6,-0.2) -- (0.6,-0.2)-- (0,-0.7);
 \draw[thick] (0,-0.7) -- (2,-0.7);
\draw[] (0,-0.7) -- (0.6,-0.2);
\draw[]  (0.6,-0.2)--(2.6,-0.2);
\draw[thick]  (2.6,-0.2)-- (2,-0.7);
\end{tikzpicture}
\hspace{-1.2 in}
\begin{tikzpicture}[scale=0.7,baseline=-40]
\begin{scope}[rotate=90]
%Torus
\draw[] (0,0) ellipse (1.6 and .9);
\shade[ball color = lightgray!80!white, opacity=0.55] (0,0) ellipse (1.6 and .9);
%Hole
\begin{scope}[scale=.8]
\clip (0,1.3) circle (1.55);
\fill[white] (0,-1.27) circle (1.55);
\end{scope}
\begin{scope}[scale=0.8]
\path[rounded corners=24pt] (-.9,0)--(0,.6)--(.9,0) (-.9,0)--(0,-.56)--(.9,0);
\draw[][rounded corners=24pt] (-0.9,0)--(0,.8)--(0.9,0);
\draw[][rounded corners=8pt] (-0.77,0)--(0,-0.4)--(0.77,0);
\end{scope}
  \end{scope}
\end{tikzpicture}
 \hspace{0.5 in} \,\,\xrightarrow{\text{T-dualize\,\,on\,\,} S_8^1}\,\,\,
\begin{tikzpicture}[scale=1.2,baseline=0]
%plane
 \shade[top color=red!40, bottom color=red!10]  (0,-0.7) -- (2,-0.7) -- (2.6,-0.2) -- (0.6,-0.2)-- (0,-0.7);
 \draw[thick] (0,-0.7) -- (2,-0.7);
\draw[] (0,-0.7) -- (0.6,-0.2);
\draw[]  (0.6,-0.2)--(2.6,-0.2);
\draw[thick]  (2.6,-0.2)-- (2,-0.7);
\end{tikzpicture}
\hspace{-1.2 in}
\begin{tikzpicture}[scale=0.7,baseline=-40]
\begin{scope}[rotate=90]
%Torus
\draw[] (0,0) ellipse (1.6 and .9);
\shade[ball color = lightgray!80!white, opacity=0.55] (0,0) ellipse (1.6 and .9);
%Hole
\begin{scope}[scale=.8]
\clip (0,1.3) circle (1.55);
\fill[white] (0,-1.27) circle (1.55);
\end{scope}
\begin{scope}[scale=0.8]
\path[rounded corners=24pt] (-.9,0)--(0,.6)--(.9,0) (-.9,0)--(0,-.56)--(.9,0);
\draw[][rounded corners=24pt] (-0.9,0)--(0,.8)--(0.9,0);
\draw[][rounded corners=8pt] (-0.77,0)--(0,-0.4)--(0.77,0);
%brane
\draw[ultra thick,red] (0,0.65) ellipse ( .155 and 0.35);
\end{scope}
  \end{scope}
\end{tikzpicture}
 \]
 \caption{Beginning with a $\mathrm{D}4$-brane (red) in Type IIA wrapped on $\RR^{1,3} \times S^1_9$, we T-dualize on $S^1_9$ to obtain a  $\mathrm{D}3$-brane (blue) in Type IIB on $\RR^{1,3}$. Performing a second T-duality on $S^1_8$ gives a $\mathrm{D}4$-brane on $\RR^{1,3} \times S^1_8$, which upon interchange $S^1_9  \leftrightarrow S^1_8$ is identical to the original configuration.}
\label{fig:D4D3}
\end{center}
\end{figure}

We now introduce M5-branes into this background. The world-volume is chosen to fill $\RR^{1,3}\times T^2 \subset \RR^{1,7}\times T^3$, as shown below:
\vspace{0.1 in}

\begin{table}[htp]
\begin{center}
\begin{tabular}{cccccccccccc}
& $x^0$ & $x^1$ & $x^2$ & $x^3$ & $x^4$ & $x^5$ & $x^6$ & $x^7$ & $x^8$ & $x^9$& $x^{10}$
\\
\hline
$\mathrm{M}5$ & $\times$ & $\times$ & $\times$ & $\times$ & & & & & & $\times$& $\times$
\\
& \multicolumn{8}{c}{$\underbrace{\hspace{2.5 in}}$}  & \multicolumn{3}{c}{$\underbrace{\hspace{1 in}}$} 
\\
& \multicolumn{8}{c}{$\RR^{1,7}$}  & \multicolumn{3}{c}{$T^3$} 
\end{tabular}
\end{center}
\end{table}%
\vspace{-0.2 in}
\noindent
 The presence of the M5-branes breaks $SO(1,7) \times SL(3,\mathbb{Z})\times SL(2,\mathbb{Z})_{\rho}$ to $SO(1,3) \times SO(4)_R \times SL(2,\mathbb{Z})_{\tau}\times SL(2,\mathbb{Z})_{\rho}$, where $SL(2,\mathbb{Z})_{\tau}$ is the subgroup of $SL(3, \ZZ)$ acting on $T^2=S^1_9 \times S^1_{10}$. The preservation of $SO(1,3)\times SO(4)_R  \times SL(2,\mathbb{Z})_{\tau}$ is obvious, but the preservation of $SL(2,\mathbb{Z})_{\rho}$, and in particular of the $S$ transformation of $SL(2,\mathbb{Z})_{\rho}$, is non-trivial. To see that it is preserved, recall that $S$ can be realized by reducing to Type IIA, doing a double T-duality, exchanging the two circles, and then lifting back to M-theory. Choosing the M-theory circle to be $S^1_{10}$, we can then reduce to D4-branes wrapping $S^1_9$ in Type IIA. Performing T-duality along $S^1_9$ leads to unwrapped D3-branes in Type IIB, and performing a second T-duality along $S^1_8$ gives D4-branes wrapping $S^1_8$ in Type IIA. Exchanging the two circles $S^1_9 \leftrightarrow S^1_8$ and lifting back to M-theory gives the original configuration---see Figure \ref{fig:D4D3}.

Given these symmetries, we can consider various (potentially non-geometric) orbifolds involving them. We will be interested in a quotient by 
\bea
\label{eq:Sfolddef}
\mathsf{O}_k := \mathsf{R}_k \mathsf{T}_k \mathsf{B}_k 
\eea
where $\mathsf{R}_k$ is the generator of a $\mathbb{Z}_k$ subgroup of the $SO(4)_R$ symmetry group acting on the four extended coordinates orthogonal to the M5-branes, $\mathsf{T}_k$ is the generator of a $\mathbb{Z}_k$ subgroup of $SL(2,\mathbb{Z})_{\tau}$, and $\mathsf{B}_k$ is the generator of a $\mathbb{Z}_k$ subgroup of $SL(2,\mathbb{Z})_{\rho}$. In other words, we consider a quotient by the diagonal combination of these three $\ZZ_k$ symmetries. We refer to such a quotient as an \textit{S-fold}.\footnote{Technically, this operation is the T-dual of an S-fold as defined in \cite{Garcia-Etxebarria:2015wns,Aharony:2016kai}. We hope that this abuse of terminology will not cause too much confusion. }
 Note that only when $k=2,3,4,$ or $6$ are the $\ZZ_k$ legitimate symmetries---for other values of $k$ there are no fixed points for the action on $\tau$ and $\rho$.

We now ask for the amount of SUSY preserved under this quotient. To understand this, it is convenient to first consider the limit in which the $T^2$ wrapped by the $\mathrm{M}5$-brane is shrunk to zero size, resulting in $4d$ $\mathcal{N}=4$ super Yang-Mills (SYM) theory with enhanced symmetry $SO(1,3)\times SO(6)_R$ and S-duality group $SL(2,\mathbb{Z})_\rho$. Note that we may associate a $U(1)$ bundle to $SL(2, \ZZ)_\rho$, where a transition function implementing $\rho \rightarrow {a \rho + b \over c \rho + d}$ in the original $SL(2,\ZZ)$ bundle is replaced with a transition function given by $e^{i\, \mathrm{arg}(c \rho+d)}$ \cite{Kapustin:2006pk}. The charges of the $\cN=4$ supercharges under $SO(1,3)\times SO(6)_R \times U(1)_\rho$ are 
\bea
({\bf 2}_{SO(3,1)},{\bf 4}_{SO(6)_R})_{-\frac{1}{2}}\oplus (\overline{\bf 2}_{SO(3,1)},\overline{\bf 4}_{SO(6)_R})_{\frac{1}{2}}~,
\eea
where the subscript denotes the charges under $U(1)_\rho$.

We now return to the original configuration of an $\mathrm{M}5$-brane wrapping $T^2$. When the torus is non-vanishing we expect only an $SO(4)_R \times U(1)_{\tau} \subset SO(6)_R$ to be manifest, with the $U(1)_{\tau}$ bundle being the one associated with the $SL(2,\mathbb{Z})_{\tau}$ symmetry of the torus. Decomposing the supercharges into representations of $SO(1,3) \times SO(4)_R \times U(1)_{\tau} \times U(1)_\rho$ gives

\bea \label{Qrep}
 ({\bf 2};{\bf 2},0)^{\frac{1}{2}}_{-\frac{1}{2}}\oplus ({\bf 2};{\bf 1},1)^{-\frac{1}{2}}_{-\frac{1}{2}}\oplus ({\bf 2};{\bf 1},-1)^{-\frac{1}{2}}_{-\frac{1}{2}}\oplus (\overline{\bf 2};{\bf 2},0)^{-\frac{1}{2}}_{\frac{1}{2}} \oplus (\overline{\bf 2};{\bf 1},1)^{\frac{1}{2}}_{\frac{1}{2}} \oplus (\overline{\bf 2};{\bf 1},-1)^{\frac{1}{2}}_{\frac{1}{2}}~.
\eea 
For reasons that will become apparent in a moment, we have used $SO(4)_R \cong SU(2)_1 \times SU(2)_2$ and labelled representations by $(SO(1,3);SU(2)_1,U(1)_2)^{U(1)_{\tau}}_{U(1)_{\rho}}$ where $U(1)_2$ is the Cartan for $SU(2)_2$. 

We now consider the action of the quotient by $\mathsf{O}_k$ on the supercharges. The operations $\mathsf{T}_k \in U(1)_{\tau}$ and $\mathsf{B}_k \in U(1)_{\rho}$ give rise to $\pm \frac{\pi i}{k}$ rotations of the supercharges. For the operation $\mathsf{R}_k \in SO(4)_R$, it is convenient to think of it as acting on $\mathbb{C}^2$ spanned by the coordinates $z_1$ and $z_2$. In that case we are interested in the quotient acting as $z_1 \rightarrow e^{\frac{2\pi i}{k}} z_1$, $z_2 \rightarrow e^{\frac{2\pi i}{k}} z_2$. The $SO(4)_R\cong SU(2)_1\times SU(2)_2$ symmetry acts on $\mathbb{C}^2$ such that $(z_1, z_2)$ forms a doublet of $SU(2)_2$ and $(z_1, \bar{z}_2)$ forms a doublet of $SU(2)_1$. Thus the quotient we are interested in breaks $SU(2)_2$ to $U(1)_2$. We can now combine the effect of all three transformations on \eqref{Qrep}, giving:
\bea
 ({\bf 2};{\bf 2},0)^{\frac{1}{2}}_{-\frac{1}{2}} &\rightarrow&  ({\bf 2};{\bf 2},0)^{\frac{1}{2}}_{-\frac{1}{2}}~, \hspace{1 in}({\bf 2};{\bf 1},1)^{-\frac{1}{2}}_{-\frac{1}{2}} \rightarrow ({\bf 2};{\bf 1},1)^{-\frac{1}{2}}_{-\frac{1}{2}}~,
 \no\\
 ({\bf 2};{\bf 1},-1)^{-\frac{1}{2}}_{-\frac{1}{2}} &\rightarrow& e^{- {4 \pi i \over k}}({\bf 2};{\bf 1},-1)^{-\frac{1}{2}}_{-\frac{1}{2}}~, \hspace{0.5 in}(\overline{\bf 2};{\bf 2},0)^{-\frac{1}{2}}_{\frac{1}{2}}\rightarrow (\overline{\bf 2};{\bf 2},0)^{-\frac{1}{2}}_{\frac{1}{2}}~,
 \no\\
 (\overline{\bf 2};{\bf 1},1)^{\frac{1}{2}}_{\frac{1}{2}}&\rightarrow& e^{ {4 \pi i \over k}}(\overline{\bf 2};{\bf 1},1)^{\frac{1}{2}}_{\frac{1}{2}}~, \hspace{0.8in}(\overline{\bf 2};{\bf 1},-1)^{\frac{1}{2}}_{\frac{1}{2}} \rightarrow (\overline{\bf 2};{\bf 1},-1)^{\frac{1}{2}}_{\frac{1}{2}}~.
\eea
We see that for  $k=3,4,6$ the proposed quotient preserves twelve of the sixteen supercharges, giving an $\mathcal{N}=3$ theory in $4d$. For $k=2$, all sixteen supercharges are preserved, and we get an $\mathcal{N}=4$ theory.

\subsection{Exceptional 4d $\cN=3$ theories}

The construction described above is equivalent to the standard Type IIB S-fold construction of \cite{Garcia-Etxebarria:2015wns}. We next describe how to generalize it by replacing the M5-branes describing the $\mathfrak{a}_N$-type $(2,0)$ theory with $\mathfrak{d}_N$- and $\mathfrak{e}$-type $(2,0)$ theories. 

\subsection*{$\mathfrak{d}_N$-type}

The generalization to $\mathfrak{d}_N$-type theories is rather straightforward and can be done by the addition of an $\mathrm{OM}5$-plane parallel to the $\mathrm{M}5$-branes. This gives rise to an additional $\mathbb{Z}_2$ quotient of the transverse space. This quotient commutes with the action of $SO(4)_R$ and $U(1)_{\tau}$, and thus we expect the quotients by $\mathsf{R}_k$ and $\mathsf{T}_k$ to work exactly as before. The only subtle issue here is whether $SL(2,\mathbb{Z})_{\rho}$ remains unbroken upon the addition of the $\mathrm{OM}5$-plane. 

One way to see that this is so is to observe that the low-energy theory on the M5-branes reduced on $T^2$ will be $\mathcal{N}=4$ SYM with $SO(2N)$ gauge group, which still possesses an $SL(2,\mathbb{Z})$ duality symmetry, which as we saw before descends from $SL(2, \ZZ)_\rho$. A more direct confirmation can be obtained by using the string theory definition of the $S$ transformation: concretely, we again take the M-theory circle to be $S^1_{10}$, and upon reducing to Type IIA the $\mathrm{OM}5$-plane decomposes into two O$4^-$-planes wrapping $S^1_9$ and located at opposite points of $S^1_8$ \cite{Hanany:2000fq}. Performing T-duality on $S^1_8$ leads to an O$5^-$-plane in Type IIB, and performing a second T-duality on $S^1_9$ gives back two O$4^-$-planes, with the wrapped and unwrapped circles exchanged. Finally, exchanging $S^1_8 \leftrightarrow S^1_9$ gives back the original configuration.

Hence we conclude that the presence of the $\mathrm{OM}5$-plane is consistent with the existence of the $\mathsf{O}_k$ symmetry, and that the quotient by this operation can be carried out as before to get potentially new 4d $\cN=3$ theories.  

\subsection*{$\mathfrak{e}$-type}

We now consider the more involved generalization to $\mathfrak{e}$-type theories. In general, the $(2,0)$ theories of type $\mathfrak{ade}$ can be engineered in string theory by reducing Type IIB on $\mathbb{C}^2/\Gamma$, with $\Gamma$ a discrete subgroup of $SU(2)$. These discrete subgroups are known to admit an ADE classification. As we ultimately hope to map the configuration to an M-theory configuration on $T^3$, it is convenient to compactify two directions such that $\mathbb{C}^2$ is replaced with $T^2$ fibered over $\mathbb{C}$, with the fibration done in such a way that upon decompactfication $\mathbb{C}^2/\Gamma$ is recovered. More concretely, as the $T^2$ fiber is taken around the fixed point in $\CC$, it should come back to itself up to an action of $SL(2,\mathbb{Z})_\tau$. The possible such actions were classified by Kodaira, and are again known to obey an ADE classification. The monodromy action is then completely determined by the choice of $\Gamma$, i.e. the choice of $(2,0)$ theory. For the case of $\mathfrak{e}_6$, $\mathfrak{e}_7$, and $\mathfrak{e}_8$ that will be of interest to us here, the monodromies are precisely the generators of the $\mathbb{Z}_3$, $\mathbb{Z}_4$, and $\mathbb{Z}_6$ subgroups of $SL(2,\mathbb{Z})_\tau$.

We next T-dualize to Type IIA. This gives a similar construction, but with the fibration now involving the T-duality symmetry of Type IIA. In other words, the $SL(2,\mathbb{Z})$ action is no longer the geometric one acting on the torus, but instead becomes an element of the double T-duality symmetry of Type IIA on a torus. %\footnote{We can phrase this in another way by considering Type IIA reduced on $T^2$. In that case, when we go around the singularity in $\mathbb{C}$ we end up with Type IIA reduced on the torus acted upon by the appropriate element of that symmetry.}. 
We can see this by considering the imaginary part of $\tau$,  the modular parameter of the original $T^2$. In the Type IIB description, this quantity was given by $i \frac{R_1}{R_2}$, with $R_1, R_2$ the radii of the two circles. However, in Type IIA this becomes $i R_1 R_2$ by (\ref{eq:Tdual}). This is none other than the imaginary part of the modular parameter $\rho_{\mathrm{IIA}}$ introduced previously, which was acted on by the non-geometric $SL(2,\mathbb{Z})_{\rho}$. 

We now lift to M-theory on a $T^3$ fibered over $\mathbb{C}$, with $\rho$ being associated with $T^3$. The fibration is done in such a way that upon reduction on $S^1_{10}$, going around the singularity gives the same $SL(2,\mathbb{Z})_{\rho}$ monodromy as before.  When the monodromy element is of type $\mathfrak{a}_N$ or $\mathfrak{d}_N$, we expect the result to be equivalent to a configuration of M5-branes in flat space or in the presence of an $\mathrm{OM}5$-plane, as discussed before. However, when the monodromy element is of type $\mathfrak{e}_6$, $\mathfrak{e}_7$, or $\mathfrak{e}_8$ we expect to get a more subtle configuration related to the $\mathfrak{e}$-type $(2,0)$ theory. Unlike in the $\mathfrak{a}_N$ and $\mathfrak{d}_N$ cases, in the $\mathfrak{e}$-type case the monodromy corresponds to one of the $\ZZ_3, \ZZ_4,$ or $\ZZ_6$ discrete subgroups of $SL(2,\mathbb{Z})_{\rho}$, which are only symmetries for special values of $\rho$. This means that we must fix $\rho$ to an appropriate value, which in particular means that the volume of $T^3$ is fixed. Hence in the $\mathfrak{e}$-type case we cannot decompactify the geometry to obtain a brane system in flat space as before.  

\subsection*{$\mathfrak{e}$-type S-folds}

We now combine the non-geometric fibration used to get the $\mathfrak{e}$-type $(2,0)$ theories with the non-geometric orbifolding by $\mathsf{O}_k$. To do so we again follow the discussion in \cite{Garcia-Etxebarria:2016erx} and consider M-theory on $\RR^{1,5}\times T^5$. It is convenient to separate the directions as $\RR^{1,3}\times \mathbb{C} \times T^2_a \times S^1 \times T^2_b$. We first employ the non-geometric fibration, in which we fiber the three-torus $S^1 \times T^2_a$ over $\mathbb{C}$. By the previous subsection, the reduction of M-theory on this space will give rise to the $\mathfrak{e}$-type $(2,0)$ theory on $\RR^{1,3}\times T^2_b$.

We next perform the $\mathsf{O}_k$ quotient on $\mathbb{C} \times T^2_a \times S^1 \times T^2_b$. Here $S^1$ plays the role of the circle direction $S^1_8$ that was previously orthogonal to the M5-branes and $\mathbb{C} \times T^2_a$ plays the role of the transverse space $\mathbb{C}^2$ spanned by $x^4, \dots, x^7$ (though we stress that in the current case there are no M5-branes proper, only a $(2,0)$ theory on $\RR^{1,3} \times T^2_b$). 
%Recall that we had previously replaced $\mathbb{C}^2/\Gamma$ by the fibration of $T^2$ over $\mathbb{C}$, and hence the replacement of $\mathbb{C}^2$ by this space. Similarly, $T^2_a \times S^1$ plays the role of the $T^3$ we had previously. 
With these identifications the $\mathsf{R}_k$ quotient acts as a rotation of $\mathbb{C} \times T^2_a$, whereas $\mathsf{T}_k$ and $\mathsf{B}_k$ are elements of the two $SL(2,\mathbb{Z})$ subgroups that are expected to arise when M-theory is reduced on $T^2_b \times S^1$---one is the standard geometric $SL(2,\mathbb{Z})_{\tau} \subset SL(3, \ZZ)$ associated with $T^2_b$, while the other is the $SL(2,\mathbb{Z})_{\rho}$ of M-theory compactified on the full $T^3$. Reducing M-theory on $\mathbb{C} \times T^2_a \times S^1 \times T^2_b$, we expect to find a $4d$ theory with $\mathcal{N}=3$ SUSY. This follows since without the quotient we expect to get $\mathfrak{e}$-type $\mathcal{N}=4$ SYM theory, and as we have noted above, the quotient by the combined symmetry $\mathsf{O}_k$ reduces this to $\mathcal{N}=3$. 

Several comments regarding the above construction are in order:

\begin{enumerate}
  
  \item First, one might worry whether the quotient and fibration done above are really compatible. Specifically, we might be worried that some of the symmetries we are quotienting by do not actually exist in the presence of the fibration. For example, in the definition of the fibration we have compactified some of the directions of $\mathbb{C}^2$ to be (locally) $\mathbb{C}\times T^2_a$. This will break the $SO(4)_R$ symmetry to a $U(1)$ subgroup acting only on $\mathbb{C}$, and so naively we lose the action on $T^2_a$. However, for our purposes we only need the quotient for groups that act as isometries of the torus, and as such we can still take the quotient on $T^2_a$ if we tune its complex structure moduli such that it has the necessary isometry. One might also worry that the fibration might break the $SL(2,\mathbb{Z})_{\tau} \times SL(2,\mathbb{Z})_{\rho}$ symmetry. In fact, we \textit{are} breaking the $SL(3,\mathbb{Z})$ symmetry of $T^2_a \times S^1$ since the $S^1$ participates in the fibration, while $T^2_a$ does not. This however is expected, as that symmetry was also previously broken by the presence of the M5-branes. It is only the $SL(2,\mathbb{Z})_{\tau} \subset SL(3,\mathbb{Z})$ which is important for our purposes, and this remains untouched.

\item The construction described so far requires that some geometric parameters of the torus are tuned to specific values. Specifically, for the fibration of $S^1 \times T^2_b$ over $\mathbb{C}$ to be sensible we may need to tune $\rho$ associated with $T^3 = S^1 \times T^2_b$ to a special value compatible with the fibration. Similarly, for the quotient we need to tune the parameter associated with $T^3 = S^1 \times T^2_a$ to a special value compatible with the quotient. Additionally, the geometric complex structure moduli of both $T^2_a$ and $T^2_b$ must be tuned to values compatible with the quotient. One might worry that there is no choice of parameters which can satisfy all of these constraints. This however is not the case: there will be a solution for every choice of $\mathfrak{e}$-type theory and $k=2,3,4,6$, as follows from the fact that we have four complex equations in nine variables (the radii of the five circles, the $G_3$ flux on $S^1 \times T^2_a$ and $S^1 \times T^2_b$, and the angle variables for the complex moduli of $T^2_a$ and $T^2_b$). 

\item Since the orbifold that we are considering is non-geometric, there is a subtle question of whether it gives a truly valid M-theory background. This is particularly the case since there is a singularity at the origin of $\mathbb{C}$, and one might worry that the proper definition of the orbifold would require specifying additional boundary conditions at the singularity. Relatedly, one might worry about the presence of twisted states coming from the orbifold.\footnote{We thank Yuji Tachikawa for discussions on these points.} We do not currently have a satisfactory answer to these concerns, but the fact that $\Gamma=A_N$ reproduces the standard S-fold construction---which can be described alternatively as a fully geometric quotient $(\mathbb{C}^2 \times T^2)/\mathbb{Z}_k$ in F-theory---gives us some confidence that this construction is sensible. For the purposes of this paper we shall assume that the construction is sensible and use it to derive the moduli spaces of the resulting theories. As we shall see, the moduli spaces which we obtain pass numerous non-trivial consistency checks and as such appear to correspond to legitimate $\mathcal{N}=3$ SCFTs. We take this as a further indication that the M-theory construction described here is indeed sensible.

\item Finally, the (2,0) theory is in general a \textit{relative} theory \cite{Aharony:1998qu,Witten:1998wy,Witten:2009at,Freed:2012bs}, which means that a naive torus compactification of the theory will also be relative. In order to obtain a well-defined $\cN=3$ theory, it is necessary to choose a \textit{polarization}---i.e. a maximal isotropic sublattice of $Z_\mathfrak{g} \times Z_\mathfrak{g}$, with $Z_\mathfrak{g}$ the center of the simply connected group of type $\mathfrak{g}$ \cite{Tachikawa:2013hya}. In general there can be multiple ways of choosing such a polarization, giving rise to multiple legitimate 4d theories for a given relative (2,0) theory; part of this degeneracy (though potentially not all \cite{GarciaEtxebarria:2019caf}) resides in the choice of line operators and global structure of the gauge group. In the current work we will work only at the level of the algebra, leaving interesting details about the group structure for future study.

\end{enumerate}

\subsection{Generalizations}
\label{sec:generalization}
We now introduce further generalizations of the construction described above. These generalizations involve quotienting by discrete outer automorphism symmetries of the $(2,0)$ theory. The $(2,0)$ theory of type $\mathfrak{ade}$ has discrete symmetries that act as outer automorphisms of the associated $\mathfrak{ade}$ algebra. In the realization in terms of $\mathbb{C}^2/\Gamma$ in Type IIB, we can perform a blowup at the singular point, whereby the discrete symmetry should be realized as a geometric symmetry permuting the blowup cycles in accordance with the action on the Dynkin diagram \cite{Bershadsky:1996nh}. As such, we expect the outer automorphism symmetry to exist as a (geometric) symmetry when we reduce Type IIB string theory on $\mathbb{C}^2/\Gamma$, and hence also as a (potentially non-geometric) symmetry in the reduction of M-theory on $T^3$ fibered over $\mathbb{C}$. We can use this additional symmetry to generate new $\cN=3$ theories by either twisting or quotienting. We now describe both of these possibilities.

\subsubsection*{Twisting}

We first consider twisting by the outer automorphism symmetries. Begin by considering M-theory on $T^3$ fibered over $\mathbb{C}$ such that we get a $(2,0)$ theory of type $\mathfrak{g} \in \mathfrak{ade}$ in the remaining six directions. As we have just mentioned, in general this theory has a discrete symmetry related to the outer automorphism symmetry of $\mathfrak{g}$. We can now consider compactifying the theory on a circle with holonomy for this symmetry. This will give 5d SYM theory with a \textit{non-simply-laced} gauge group. Compactifying on an additional circle leads to $\mathcal{N}=4$ SYM theory, again with non-simply-laced gauge group. This setup as been studied for instance in \cite{Vafa:1997mh}. The full setup can be thought of as M-theory  on $\RR^{1,3} \times \mathbb{C}\times T^3 \times S^1_a \times S^1_b$, with the $T^3$ fibered over $\mathbb{C}$ and an outer automophism holonomy around $S^1_a$. 

We now add the $\mathsf{O}_k$ quotient on top of this configuration. It is not immediately obvious that this is possible, since the nature of some of the symmetries involved in $\mathsf{O}_k$ changes in the presence of the outer automorphism twist. In particular, it was noted in \cite{Vafa:1997mh,Argyres:2006qr} that in the presence of an outer automorphism twist the action of the $S$ element of $SL(2,\mathbb{Z})_{\rho}$ becomes 
\bea
\label{eq:newSdef}
\widetilde{S}: \,\,\, \rho \rightarrow - {1\over n_\mathfrak{g} \rho}~, \hspace{0.5 in} n_\mathfrak{g} = \left\{ \begin{matrix} \,\,2 &\hspace{0.5in} \mathfrak{g}= \mathfrak{b}_N,\, \mathfrak{c}_N, \mathfrak{f}_4 \\ \,\,3 &  \mathfrak{g}=\mathfrak{g}_2  \end{matrix} \right. ~, 
\eea
 where $\mathfrak{g}$ now represents the gauge group of the non-simply-laced 4d SYM theory and $n_{\mathfrak{g}}$ is the order of the outer automorphism used to obtain it. 
 This can be understood from the perspective of T-duality in Type IIA: indeed, because of the holonomy around $S^1_a$, we now need to go $n_{\mathfrak{g}}$ times around the circle to come back to the same configuration, and hence the radius of $S^1_a$ is effectively multiplied by $n_{\mathfrak{g}}$. This means that instead of acting as in (\ref{eq:Tdual}), T-duality now sends  $R_a \rightarrow {1\over n_{\mathfrak{g}} R_a}$ while acting on $R_b$ in the previous manner, and hence in total $\rho_{\mathrm{IIA}} = i{R_a R_b}$ (for $B_{12}=0$) transforms as in (\ref{eq:newSdef}).  As will be discussed in more detail in Section \ref{sec:nonsimplac}, the transformation $\widetilde{S}$ together with the transformation $T: B_{12} \rightarrow B_{12}+1$ generates a so-called  \textit{ Hecke group} $\cH({n_{\mathfrak{g}}})$ \cite{Argyres:2006qr}, and thus we conclude that in the presence of outer automorphism twists we must replace $SL(2,\mathbb{Z})_{\rho}$ with the appropriate Hecke group $\cH({n_{\mathfrak{g}}})_\rho$.    

%The action of the outer automorphsim in general also changes the fibration of $T^3$ over $\mathbb{C}$. For instance, if $\Gamma=A_{2n-1}$ on one side we get $\Gamma=D_{n+1}$ on the other. As such generically, this action is not a symmetry of the background. However, when $\Gamma=E_6$ or $\Gamma=D_4$ with the outer automorphism of order $3$, then the fibration does not change and this transformation becomes a symmetry. 

The above comments also imply that the $SL(2, \ZZ)_\tau$ symmetry associated with $T^2_b$ should be replaced by a Hecke group $\cH({n_{\mathfrak{g}}})_\tau$. This follows from the fact that the modular parameters $\tau$ and $\rho$ of $T^2_a$ and $T^2_b\times S^1$ are exchanged under T-duality. On the other hand, the symmetry $\mathsf{R}_k$ which acted as rotations of $\mathbb{C}\times T^2_a$ remains largely unchanged: upon twisted compactification we again obtain a $\ZZ_k$ symmetry acting as a subgroup of $U(1)$ rotations in $\mathbb{C}$, together with a discrete cyclic group that emerges when the complex structure modulus of $T^2_a$ is fixed to special values.

%The reason for the change is that $T^2_b$ involves a cycle on which the holonomy is turned on, which breaks part of the symmetry. In general we expect the S transformation to be broken by this choice, as is indeed the case for most choices of $\Gamma$. This follows from considering the resulting $4d$ theory, which for $\Gamma=A$ and $D$ with the outer automorphism of order two leads to 4d $\mathcal{N}=4$ SYM theories with gauge groups of type $B$ and $C$. The latter are known to be exchanged under S-duality so in this case such a duality is not a symmetry\footnote{For the special case of rank $2$ S-duality is a symmetry as $B_2=C_2$, but we shall not consider this case in detail here.}. However, for the cases of $\Gamma=E_6$ or $\Gamma=D_4$ with the outer automorphism of order $3$, where S-duality maps the gauge group back to itself, we expect to do have S-duality as a symmetry. Since the duality transformation of the resulting four dimensional theories is the Hecke group\cite{Argyres:2006qr}, we expect it to also be the symmetry group associated with $T^2_a$.   

To summarize, in the presence of outer automorphism twists the $SL(2,\ZZ)_\tau\times SL(2, \ZZ)_\rho$ symmetries are modified to $\cH({n_{\mathfrak{g}}})_\tau \times \cH({n_{\mathfrak{g}}})_\rho$. We can then consider performing a quotient by a discrete subgroup of the two Hecke groups, together with a rotation of $\mathbb{C}\times T^2_a$. To do so, we must first determine the discrete subgroups of Hecke groups that fix a specific value of the modular parameter, i.e. the analogs of the discrete subgroups $\ZZ_3, \ZZ_4, \ZZ_6 \subset SL(2, \ZZ)$ fixing $\tau = i, e^{\pm {i \pi \over 3}}$. This will be done in Section \ref{sec:nonsimplac}, where the 4d $\cN=3$ theories arising from outer automorphism twists are studied in detail.

\subsubsection*{Quotienting}

We next consider quotients by the outer automorphisms symmetries. 
%As before, we begin with M-theory on $\RR^{1,3}\times \mathbb{C} \times T^2_a \times S^1 \times T^2_b$ with $S^1 \times T^2_a$ fibered over $\mathbb{C}$. As has been discussed at great lengths, upon compactification on $\mathbb{C} \times T^2_a \times S^1 \times T^2_b$ we obtain $SL(2,\mathbb{Z})_{\tau}\times SL(2,\mathbb{Z})_{\rho}$ symmety, which could be used to perform a question.  We now consider a more general quotient where we also involve the discrete outer automorphism symmetry.
Consider first the case in which $\mathfrak{g}$ has a $\mathbb{Z}_2$ outer automorphism symmetry, which holds for all cases except $\mathfrak{e}_7$ and $\mathfrak{e}_8$. We can then consider performing a quotient by a $\mathbb{Z}_4$ group involving this symmetry. In particular, we first consider the $\mathbb{Z}_4$ symmetry generated by $\mathsf{O}_4$ as before, which acts geometrically on $\mathbb{C} \times T^2_a$ and $T^2_b$, as well as non-geometrically as a subgroup of the $SL(2,\mathbb{Z})_{\rho}$ symmetry of M-theory compactified on $T^3$. We can now consider a new $\mathbb{Z}_4$ symmetry whose generator is given by the combination of the generator of the previous $\mathbb{Z}_4$ symmetry together with the $\mathbb{Z}_2$ outer automorphism symmetry, which we denote by $\widetilde{\ZZ_4}$. Quotienting by $\widetilde{\ZZ_4}$ can sometimes lead to new  $\cN=3$ theories, as we will see below. 

Similarly, we can consider the case where $\mathfrak{g}=\mathfrak{d}_4$ and there is a $\mathbb{Z}_3$ outer automorphism symmetry. We may then consider a $\mathbb{Z}_3$ or $\ZZ_6$ quotient where the generator is a combination of $\mathsf{O}_3$ or $\mathsf{O}_6$ and the $\mathbb{Z}_3$ outer automorphism, which we denote by $\widetilde{\ZZ_3}$ and $\widetilde{\ZZ_6}$ respectively. This will again be seen to give new $\cN=3$ theories.

\section{Moduli spaces from orientifolds and S-folds}
\label{sec:orientifoldofa}

The remainder of this work is devoted to understanding the moduli spaces of the 4d $\cN=3$ theories obtained from the above M-theory constructions. We will see that in all cases, the moduli space can be written as $\CC^{3N}/\Gamma$, with $\Gamma$ a crystallographic complex reflection group \textit{that can be realized as a subgroup of the Weyl group of a simple Lie algebra, i.e. $\Gamma \subset \cW(\mathfrak{g})$}.

To see this, it will first be necessary to develop some systematic tools for studying moduli spaces. We begin by reviewing the case of  $\cN=3$ theories arising from type-$\mathfrak{a}_{N-1}$ (2,0) theories. These can alternatively be realized via standard S-folds on a stack of  $N$  $\mathrm{D}3$-branes in Type IIB. The moduli space of the resulting theory is given geometrically by the motion of the branes in the six transverse directions. Let us first consider the situation with no S-folding. Since the branes are indistinguishable, the total moduli space is 
\bea
\CC^{3N} / S_N \cong \CC^{3N}/ \cW(\mathfrak{a}_{N-1})~,
\eea
with $S_N$ the group of permutations on $N$ elements, which is equivalent to the Weyl group of $\mathfrak{a}_{N-1}$. A generic point on the moduli space can be given by a matrix
\bea
\label{eq:basicPhimat}
\Phi = \left( \begin{matrix} \phi_1 & & & \\   &\phi_2 & &\\   && \ddots &\\   & & & \phi_{N}\end{matrix} \right) ~, \hspace{0.5 in} \phi_i \in \CC^3~.
\eea
The Weyl group $\cW(\mathfrak{a}_{N-1})$ acts by permuting the $\phi_i$, and hence the singular loci on the moduli space are the locations at which, for some $i \neq j$, we have $\phi_i = \phi_j$. 
At these points one or more of the D3-branes coincide, giving rise to additional massless degrees of freedom.

\subsection{Orientifolding}
\label{sec:orientifold}

\begin{figure}[tbp]
\begin{center}
\begin{tikzpicture}[decoration={markings,
mark=at position .5 with {\arrow{>}}}]
\begin{scope}[scale=1.3]
 \shade[top color=red!40, bottom color=red!10]  (0,-0.2) -- (2,-0.2) -- (2.6,0.3) -- (0.6,0.3)-- (0,-0.2);
 \shade[top color=red!40, bottom color=red!10]  (0,-0.1) -- (2,-0.1) -- (2.6,0.4) -- (0.6,0.4)-- (0,-0.1);
  \shade[top color=blue!40, bottom color=blue!10]  (0,0) -- (2,0) -- (2.6,0.5) -- (0.6,0.5)-- (0,0);
   \shade[top color=blue!40, bottom color=blue!10]  (0,0.1) -- (2,0.1) -- (2.6,0.6) -- (0.6,0.6)-- (0,0.1);

\draw[thick] (0,0.1) -- (2,0.1);
\draw[thick] (0,0.1) -- (0.6,0.6);
\draw[thick]  (0.6,0.6)--(2.6,0.6);
\draw[thick]  (2.6,0.6)-- (2,0.1);

\draw[thick] (0,0) -- (2,0);
\draw[] (0,0) -- (0.6,0.5);
\draw[]  (0.6,0.5)--(2.6,0.5);
\draw[thick]  (2.6,0.5)-- (2,0);

\draw[thick] (0,-0.1) -- (2,-0.1);
\draw[] (0,-0.1) -- (0.6,0.4);
\draw[]  (0.6,0.4)--(2.6,0.4);
\draw[thick]  (2.6,0.4)-- (2,-0.1);

\draw[thick] (0,-0.2) -- (2,-0.2);
\draw[] (0,-0.2) -- (0.6,0.3);
\draw[]  (0.6,0.3)--(2.6,0.3);
\draw[thick]  (2.6,0.3)-- (2,-0.2);

\draw[ thick,dashed] (-0.8,0.05) -- (2.8,0.05);

\node[left] at (0,0.24) {$N \{$};
\node[left] at (0,-0.15) {$N \{$};
\end{scope}
%%%%%%%%%%%%%

\begin{scope}[scale=1.3,xshift=4cm]
 \shade[top color=red!40, bottom color=red!10]  (0,-0.7) -- (2,-0.7) -- (2.6,-0.2) -- (0.6,-0.2)-- (0,-0.7);
 \shade[top color=red!40, bottom color=red!10]  (0,-0.1) -- (2,-0.1) -- (2.6,0.4) -- (0.6,0.4)-- (0,-0.1);
  \shade[top color=blue!40, bottom color=blue!10]  (0,0) -- (2,0) -- (2.6,0.5) -- (0.6,0.5)-- (0,0);
    \shade[top color=blue!40, bottom color=blue!10]  (0,0.6) -- (2,0.6) -- (2.6,1.1) -- (0.6,1.1)-- (0,0.6);
  
\draw[thick] (0,0.6) -- (2,0.6);
\draw[thick] (0,0.6) -- (0.6,1.1);
\draw[thick]  (0.6,1.1)--(2.6,1.1);
\draw[thick]  (2.6,1.1)-- (2,0.6);

\draw[thick] (0,0) -- (2,0);
\draw[thick] (0,0) -- (0.6,0.5);
\draw[thick]  (0.6,0.5)--(2.6,0.5);
\draw[thick]  (2.6,0.5)-- (2,0);

\draw[thick] (0,-0.1) -- (2,-0.1);
\draw[] (0,-0.1) -- (0.6,0.4);
\draw[]  (0.6,0.4)--(2.6,0.4);
\draw[thick]  (2.6,0.4)-- (2,-0.1);

\draw[thick] (0,-0.7) -- (2,-0.7);
\draw[thick] (0,-0.7) -- (0.6,-0.2);
\draw[thick]  (0.6,-0.2)--(2.6,-0.2);
\draw[thick]  (2.6,-0.2)-- (2,-0.7);

\draw[ thick,dashed] (-0.8,0.05) -- (2.8,0.05);

\end{scope}

\begin{scope}[scale=1.3,xshift=8cm]
 \shade[top color=red!40, bottom color=red!10]  (0,-0.7) -- (2,-0.7) -- (2.6,-0.2) -- (0.6,-0.2)-- (0,-0.7);
 \shade[top color=red!40, bottom color=red!10]  (0,-0.6) -- (2,-0.6) -- (2.6,-0.1) -- (0.6,-0.1)-- (0,-0.6);
  \shade[top color=blue!40, bottom color=blue!10]  (0,0.5) -- (2,0.5) -- (2.6,1) -- (0.6,1)-- (0,0.5);
    \shade[top color=blue!40, bottom color=blue!10]  (0,0.6) -- (2,0.6) -- (2.6,1.1) -- (0.6,1.1)-- (0,0.6);
  
\draw[thick] (0,0.6) -- (2,0.6);
\draw[thick] (0,0.6) -- (0.6,1.1);
\draw[thick]  (0.6,1.1)--(2.6,1.1);
\draw[thick]  (2.6,1.1)-- (2,0.6);

\draw[thick] (0,0.5) -- (2,0.5);
\draw[] (0,0.5) -- (0.6,1);
\draw[]  (0.6,1)--(2.6,1);
\draw[thick]  (2.6,1)-- (2,0.5);

\draw[thick] (0,-0.6) -- (2,-0.6);
\draw[thick] (0,-0.6) -- (0.6,-0.1);
\draw[thick]  (0.6,-0.1)--(2.6,-0.1);
\draw[thick]  (2.6,-0.1)-- (2,-0.6);

\draw[thick] (0,-0.7) -- (2,-0.7);
\draw[] (0,-0.7) -- (0.6,-0.2);
\draw[]  (0.6,-0.2)--(2.6,-0.2);
\draw[thick]  (2.6,-0.2)-- (2,-0.7);

\draw[ thick,dashed] (-0.8,0.05) -- (2.8,0.05);

\end{scope}

\end{tikzpicture}
\caption{Left: $N$ $\mathrm{D}3$-branes probing an orientifold plane (dotted line), together with their mirrors. Center: a configuration in which $N-1$ branes probe the orientifold and one is pulled away; the positions of the internal stack span a $\CC^{3(N-1)}/ \cW( \mathfrak{c}_{N-1})$ moduli space. Right: a configuration in which all $N$ branes are pulled together away from the orientifold; their positions span a $\CC^{3N}/\cW( \mathfrak{a}_{N-1})$ moduli space.}
\label{fig:orientifold}
\end{center}
\end{figure}
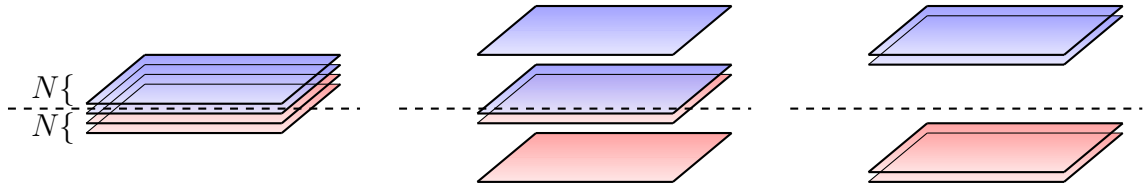

We now consider this setup in the presence of an orientifold $\mathrm{O}3^+$-plane. More precisely, we consider a configuration of $N$ $\mathrm{D}3$-branes and $N$ mirror $\mathrm{D}3$-branes, as shown in Figure \ref{fig:orientifold}. Let us understand the moduli space of this theory. Before orientifolding, a generic point in the moduli space is given by a matrix of the form  (\ref{eq:basicPhimat}), with $N \rightarrow 2N$. After orientifolding each brane must have a mirror partner, which means that sending $\phi_i \rightarrow - \phi_i$ for any $i$ should either leave $\Phi$ invariant, or give rise to a \textit{permutation} of its entries, i.e. to an action of $\cW(\mathfrak{a}_{2N-1})$ (for example, if $\phi_2 = -\phi_1$, then $\phi_1 \rightarrow - \phi_1$ is equivalent to a permutation $\phi_1 \leftrightarrow \phi_2$). In other words, the portion of the moduli space which survives is the portion on which the orbifold action can be realized as an element of the Weyl group of $\mathfrak{a}_{2N-1}$. This fixes half of the entries of $\Phi$ in terms of the other half, and reduces the naive moduli space without quotienting from $\CC^{6N}$ to $\CC^{3N}$.

To get the true moduli space, we must now do an additional quotient. In addition to quotienting $\CC^{3N}$ by permutations between the $N$ independent $\phi_i$ (since the branes are indistinguishable), we must also quotient by $N$ copies of $\ZZ_2$ implementing the identifications $\phi_i \rightarrow - \phi_i$ for each $i=1,\dots,N$. Since $\cW(\mathfrak{a}_{N-1})$ permutes between these $N$ copies of $\ZZ_2$, the correct quotient is by the semi-direct product 
\bea
\CC^{3N}/\cW(\mathfrak{a}_{N-1}) \rtimes \ZZ_2^N~.
\eea
In fact, the group $\cW(\mathfrak{a}_{N-1}) \rtimes \ZZ_2^N$ is none other than the Weyl group of $\cW(\mathfrak{c}_{N})$. This makes sense: the theory living on a stack of $N$ $\mathrm{D}3$-branes probing an $\mathrm{O}3^+$-plane is known to be $\mathfrak{c}_{N}$ SYM theory.\footnote{For branes probing an $\widetilde{\mathrm{O}3}^+$-plane the theory is also of type  $\mathfrak{c}_{N}$, while for branes probing an $\widetilde{\mathrm{O}3}^-$ it is of type $\mathfrak{b}_{N}$. Since $\cW(\mathfrak{b}_{N})=\cW(\mathfrak{c}_{N})$, the same comments hold in those cases as well. On the other hand, for an ${\mathrm{O}3}^-$-plane the theory is of type $\mathfrak{d}_N$, and the discussion requires modification. See Appendix \ref{sec:dtype(2,0) theory} for details. }

We now aim to understand the singular loci of this $\mathfrak{c}_{N}$ SYM theory. This corresponds to identifying all possible brane configurations for which extra massless degrees of freedom exist, \emph{i.e.}  configurations in which  at least two branes overlap. For $N=2$, such configurations are shown in Figure \ref{fig:orientifold}. For $N>2$ analogous configurations exist, and there are intermediate cases as well. 

To understand the low-energy effective theories describing the extra massless states on these singular loci, we can begin by understanding their moduli spaces. These will generically be of the form $\CC^{3r'}/\Gamma$, where $r'$ is the number of coincident branes (i.e. the rank of the effective low-energy theory) and $\Gamma$ is an appropriate subgroup of $\cW(\mathfrak{c}_{N})$. In the current case it is easy to read off $\Gamma$ directly from the brane configuration---it  is the subgroup which leaves it invariant---but in order to build up to more complicated examples, we will instead take a different approach. We begin by rewriting the matrix $\Phi$ as
\bea
\Phi = \left(\begin{matrix} \phi_1 \\ \vdots \\ \phi_{2N}\end{matrix}  \right) ~,
\eea
 which is a $2N$-dimensional vector with entries valued in $\CC^3$.  
Orientifolding means that we would like to restrict to $\Phi$ such that 
\bea
\label{eq:orbifoldconst}
\mathsf{O}_2 \Phi = w \,\Phi~,
\eea
where $\mathsf{O}_2 = - \mathds{1}$ and $w \in \cW(\mathfrak{a}_{2N-1})$ is a permutation. As said above, this constraint will generically fix half of the components of $\Phi$ in terms of the other half. In particular, we can now write 
\bea
\Phi = \binom{v}{Mv}~,
\eea
with $v = (\phi_1, \dots, \phi_N)$ the vector of free coordinates and $M$ an $N\times N$-matrix chosen such that (\ref{eq:orbifoldconst}) is satisfied.\footnote{We take $M$ to act on $N$-vectors with entries valued in $\CC^3$. This could instead be phrased as $M \otimes \mathds{1}_3$ acting on $3N$-vectors with entries valued in $\CC$.} Recall that previously we concluded that the moduli space of the orientifold was $\CC^{3N} /\cW(\mathfrak{c}_{N}) $. The vector $v$ serves as the coordinates on $\CC^{3N}$, and the quotient by $\cW(\mathfrak{c}_{N}) $ can be understood by asking for the subset of $\cW(\mathfrak{a}_{2N-1})$ which can be realized as an action on the $N$-(complex-)dimensional space spanned by $v$. In other words, we should quotient by all $w \in \cW(\mathfrak{a}_{2N-1})$ such that the relation
\bea
\label{eq:singlocmodeq}
w  \binom{v}{Mv} =  \binom{Av}{MAv}
\eea
can be satisfied for some $A \in GL(N, \CC)$. The set of all $w$ such that this holds can be shown to be the group $\cW(\mathfrak{c}_{N})$ identified before. 

In this language it is now straightforward to discuss the singular loci of the theory. Consider for example the case of $N=2$ with the singular locus corresponding to the rightmost configuration of  Figure \ref{fig:orientifold}. In this case we may take 
\bea
M= - \mathds{1}_{2 \times 2}~, \hspace{0.5 in} v = \left(\phi ,\, \phi\right) ~.
\eea
The moduli space on the singular locus $\Phi = \phi \left(1 ,1,-1,-1\right) $ is then just $\CC^{6}/\Gamma$, where $\Gamma$ is the set of elements in $\cW(\mathfrak{a}_{3})$ which leave $\Phi$ invariant. This is clearly just the set of permutations of the two elements of $v$, i.e. $\cW(\mathfrak{a}_{1})$. The moduli space $\CC^{6}/\cW(\mathfrak{a}_{1})$ is exactly as expected for a stack of two $\mathrm{M}5$-branes away from an orientifold plane. 

Next consider the singular locus corresponding to the middle panel of Figure \ref{fig:orientifold}. Now we have 
\bea
M= - \mathds{1}_{2 \times 2}~, \hspace{0.5 in} v =  \left(\phi ,\, 0 \right) ~.
\eea
The moduli space on the singular locus $\Phi = \phi \left(1 ,0,-1,0\right) $ is again obtained by asking which elements of $\cW(\mathfrak{a}_{3})$ leave it fixed. In this case, we clearly have invariance under shifts $v_2 \rightarrow - v_2$, which spans $\ZZ_2\cong\cW(\mathfrak{c}_{1})$. The moduli space of the theory on this locus is then $\CC^{3}/\cW(\mathfrak{c}_{1})$, as expected for a $\mathrm{D}3$-brane probing an ${\mathrm{O}3}^+$-plane.

\subsection{S-folds} 

The above may be generalized straightforwardly to S-folds. A $\ZZ_k$ S-fold for $k=3,4,6$ gives rise to a  configuration of $N$ $\mathrm{D}3$-branes and $(k-1)N$ mirror $\mathrm{D}3$-branes. The moduli space then involves a quotient by 
\bea
\cW\left(\mathfrak{a}_{ N-1}\right) \rtimes \ZZ_k^{N} \,\,\cong\,\, G\left(k,1,  N \right)~
\eea
with $G(m,p,n)$ an imprimitive complex reflection group---see Appendix \ref{app:CCRGs} for a review. The first factor on the left-hand side captures permutations of the coordinates which remain independent after the S-folding, while the second factor captures transformations $\phi_i \rightarrow \zeta_k \phi_i$ with $\zeta_k$ a primitive $k$-th root of unity.  
Note that for $k=2$, this reduces to the orientifold case studied before upon noting that  $\cW(\mathfrak{c}_N) \cong G(2,1,N)$.

Thus far we have phrased our discussion in terms of branes in Type IIB, but as described in \cite{Garcia-Etxebarria:2016erx} this is equivalent to compactification of the type-$\mathfrak{a}_{N-1}$ (2,0) theory on $T^2$ together with a quotient by $\mathsf{O}_k$ defined in (\ref{eq:Sfolddef}). What we have seen is that this gives rise to an $\cN=3$ theory with moduli space%\footnote{Note that the dimension of the moduli space increases from $5N$ to $6N$ when we pass from 6d to 4d.}
\bea
\CC^{3 N }/ G\left(k,1, N\right)~. 
\eea
We would now like to generalize the above discussion to $\cN=3$ theories descending from exceptional (2,0) theories. In doing so, we will see that many  theories have moduli spaces labelled by ECCRGs.

\section{Exceptional $\cN=3$ theories}
\label{sec:exceptionalsec}
We now generalize the construction in the previous section to show that $\cN=3$ theories coming from type-$\mathfrak{e}$ (2,0) theories have  moduli spaces of the form $\CC^{3N}/\Gamma$ with $\Gamma$ an ECCRG. Unlike in the case of type-$\mathfrak{a}_{N}$ (2,0) theories discussed above, in which there was a simple alternative  picture in terms of branes in Type IIB, in the current case the non-geometric nature of the construction complicates the discussion. Fortunately, even without a brane picture we can still access the moduli space as follows. We begin by reducing to four dimensions without the S-fold, which gives an $\cN=4$ theory whose moduli space is known to be of the form $\CC^{3N}/\cW(\mathfrak{e})$. Adding in the S-fold changes the moduli space, but this occurs in a controlled way, and is completely captured by the R-symmetry transformation generated by $\mathsf{R}_k$. We can read off this action from the M-theory picture presented in Section \ref{sec:Mtheory} by considering the action of $\mathsf{R}_k$ on $\CC\subset \CC^{3}$; the $\cN=3$ supersymmetry then dictates that this action should extend to the whole $\CC^{3}$ in the 4d theory. The result is a moduli space of the form $\CC^{3N}/\Gamma$, where $\Gamma$ is a subgroup of the Weyl group $\cW(\mathfrak{e})$. With this general outline in mind, we now begin by discussing the set of ECCRGs that can be realized as subgroups of Weyl groups.

\subsection{ECCRGs as subgroups of Weyl groups}
\label{sec:exceptionalsubgroups}

The ECCRGs which give subgroups of Weyl groups were already classified in \cite{brewer1997complex}, and we will now review these results in our own language.
In the discussion here we will focus on the \textit{invariant polynomials} of the relevant Weyl groups; the notion of invariant polynomials is reviewed in Appendix \ref{app:CCRGs}.  The final results of the discussion are summarized in Table \ref{tab:subgroups}. Readers uninterested in the derivation can skip the remainder of this subsection.

The basic idea that we make use of is that for simply-laced groups, the $\mathsf{O}_k$ operation defined in (\ref{eq:Sfolddef}) acts on the order-$n$ invariant polynomials $u_n$ of $\Gamma$ as $u_n \rightarrow e^{2\pi n i/k} u_n$. This action comes entirely from the $\mathsf{R}_k$ part of the transformation, i.e. the portion acting as an element of $SO(4)_R$.  The invariant polynomials which remain after S-folding are the subset which are inert under this transformation, and the relevant subgroups are the ones labelled by the surviving invariant polynomials.

\begin{table}[!tp]
\begin{center}
\begin{tabular}{c|c}
$\mathfrak{g}$ & subgroups
\\\hline
$\mathfrak{g}_2$ & $\ZZ_6$
\\
$\mathfrak{d}_4$ & $G_4$
\\
$\mathfrak{f}_4$ & $G_8, G_{12}$
\\
$\mathfrak{e}_6$ & $G_5, G_8, G_{25}, G_{28}$
\\
$\mathfrak{e}_7$ & $G_8, G_{26}$
\\
$\mathfrak{e}_8$ & $G_{31}, G_{32}$
\end{tabular}
\end{center}
\caption{The subgroups of Weyl groups of $\mathfrak{g}$ which appear in our analysis. The non-simply-laced cases will be discussed in Section \ref{sec:ECCRGfromnonsimp}.}
\label{tab:subgroups}
\end{table}%

\paragraph{$\mathfrak{d}_4$:} We begin with $\mathfrak{d}_4$, which is not exceptional \textit{per se}, but which nevertheless will allow us to access ECCRGs. Indeed, it can be shown that the Weyl group $\cW(\mathfrak{d}_4)$ of $\mathfrak{d}_4$ contains the ECCRG $G_4$. To see this, note that the dimension of the invariant polynomials of $\cW(\mathfrak{d}_4)$ are $2,4,4,$ and $6$. The two invariants of degree 4 transform in the irreducible two-dimensional representation of $S_3$ under triality, while the invariants of degree 2 and 6 are invariant under this action, c.f. Appendix \ref{app:CCRGs}. Now consider the simultaneous action of a $\ZZ_3$ S-folding and a $\ZZ_3$ triality quotient. Under this combined action the invariant of degree 2 is projected out, while that of degree 6 is retained. One linear combination of the degree 4 invariants is also invariant under this transformation. This leads to a subgroup with invariants of degree 4 and 6, which are the correct invariants for $G_4$.

\paragraph{$\mathfrak{e}_6$:} The Weyl group $\cW(\mathfrak{e}_6)$ contains the ECCRGs $G_5$, $G_8$, $G_{25}$, and $G_{28} = \cW(\mathfrak{f}_4)$. To see this, note that the dimensions of the invariant polynomials of $\cW(\mathfrak{e}_6)$ are $2,5,6,8,9,$ and $12$. A $\ZZ_2$ S-fold keeps only the invariants of degree 2,6,8, and 12, which are the invariants for $\cW(\mathfrak{f}_4)$. On the other hand, a $\ZZ_3$ S-fold keeps only the invariants of degree $6,9$, and $12$, giving the correct invariants for $G_{25}$. A $\ZZ_4$ S-fold preserves the invariants of degree $8$ and $12$, which are the appropriate invariants for $G_8$, while a $\ZZ_6$ S-fold preserves the invariants of degree 6 and 12, which are the appropriate invariants for $G_5$. In summary, we have 
\bea
\ZZ_2: \quad \cW(\mathfrak{f}_4)~, \hspace{0.5 in} \ZZ_3 : \quad G_{25}~,  \hspace{0.5 in} \ZZ_4 : \quad G_{8}~,  \hspace{0.5 in} \ZZ_6 : \quad G_{5} ~.
\eea
Incidentally, note that since $\ZZ_2$ takes $\cW(\mathfrak{e}_6)$ to $\cW(\mathfrak{f}_4)$, we might expect that the $\ZZ_4$ and $\ZZ_6$ S-folds should also be realizable as S-folds of the type-$\mathfrak{e}_6$ (2,0) theory with appropriate outer automorphism twist. In Section \ref{sec:ECCRGsubgroupsnonsimp} we will see that this expectation is correct for $G_8$, but not for $G_5$ due to subtleties in the definition of S-folding for non-simply-laced groups. 

\paragraph{$\mathfrak{e}_7$:} The Weyl group $\cW(\mathfrak{e}_7)$ contains the ECCRGs $G_8$ and $G_{26}$.
 To see this, note that the dimensions of the
invariant polynomials of $\cW(\mathfrak{e}_7)$ are $2, 6, 8, 10, 12, 14,$ and 18. Clearly $\ZZ_2$ S-folding does not project out any invariants. On the other hand, $\ZZ_3$ S-folding leaves invariants of degree $6,12,$ and $18$, as appropriate for $G_{26}$. Because the $\ZZ_6$ S-folding is just a combination of $\ZZ_2$ and $\ZZ_3$, we conclude that the same is true of $\ZZ_6$. Finally, upon $\ZZ_4$ S-folding we are left with invariants of degree $8$ and $12$, as appropriate for $G_8$. In summary, we have 
\bea
\ZZ_4: \quad G_8~, \hspace{0.6in} \ZZ_3\,\,\&\,\, \ZZ_6: \quad G_{26}~. 
\eea
%Note that the $G_8$ can be thought of as being embedded inside the $\mathfrak{f}_4$ subgroup of $\mathfrak{e}_7$. 

\paragraph{$\mathfrak{e}_8$:} The Weyl group $\cW(\mathfrak{e}_8)$ contains the ECCRGs  $G_{31}$ and $G_{32}$. To see this, note that the dimensions of the
 invariant polynomials are $2,8,12,14,18,20,24,$ and $30$. The $\ZZ_2$ S-folding again acts trivially on the invariant polynomials, while $\ZZ_3$ and $\ZZ_6$ preserve invariants of degree $12, 18, 24,$ and $30$, as appropriate for $G_{32}$. On the other hand, the $\ZZ_4$ S-folding leaves invariants of degree $8,12,20,$ and $24$, which are the invariants of $G_{31}$. We thus have 
 \bea
\ZZ_4: \quad G_{31}~, \hspace{0.6in} \ZZ_3\,\,\&\,\, \ZZ_6: \quad G_{32}~. 
\eea
This completes the identification of subgroups of Weyl groups of simply-laced Lie algebras obtainable by S-folding. 
 
\subsection{$\cN=3$ theories and ECCRGs}
Having identified the relevant ECCRGs, we can now consider the corresponding $\cN=3$ theories. We proceed in the same order as above.

\subsubsection{Type-$\mathfrak{d}_4$ }
\label{sec:G4}
To obtain an $\cN=3$ theory with moduli space $\CC^6/G_4$, we begin with the type-$\mathfrak{d}_4$ (2,0) theory. As discussed in the previous subsection, in order to get $G_4$ from $\cW(\mathfrak{d}_4)$, we must do a $\ZZ_3$ S-folding, together with a $\ZZ_3$ triality quotient; the latter was mentioned in Section \ref{sec:generalization} and will be referred to as a $\widetilde{\ZZ_3}$ S-fold. That this gives $G_4$ can be seen explicitly by following the analysis in Section \ref{sec:orientifold}. Our starting point is a theory with moduli space $\CC^{12} / \cW(\mathfrak{d}_4)$ with coordinates given by a vector $\Phi = (\phi_1, \dots, \phi_4)$. The $\widetilde{\ZZ_3}$ S-folding can be implemented on the moduli space by the following matrix 
\bea
\mathds{O} = e^{2\pi i /3} \times \half \left(\begin{matrix} 1 & \,\,\,1 & \,\,\,1 & -1 \\ 1 & \,\,\,1 & -1 & \,\,\,1 \\ 1 & -1 & \,\,\,1 & \,\,\,1 \\ 1 & -1 & -1 & -1 \end{matrix} \right) ~,
\eea
where the first factor corresponds to the usual $\ZZ_3$ action $\mathsf{O}_3$, while the second corresponds to the triality transformation, c.f. Appendix \ref{app:CCRGs}. By gauging this, we mean that we restrict to a slice of the moduli space satisfying 
\bea
\mathds{O} \Phi = w\, \Phi~, \hspace{0.5in} w \in  \cW(\mathfrak{d}_4)~.
\eea
We now choose any element $w$ satisfying the above constraint on a codimension-2 locus of the moduli space. For example, we may take the explicit element 
\bea
w = \left(\begin{matrix} -1 & 0 & 0 & 0 \\ 0 & 0 & 0 & -1 \\ 0 & \,\,1 &0 & 0 \\ 0 & 0 & \,\,1 & 0 \end{matrix} \right) ~
\eea
in the defining four-dimensional representation of $\cW(\mathfrak{d}_4)$ (given in Appendix \ref{app:CCRGs}), which leaves fixed 
\bea
\label{eq:d4Phi}
\Phi = \binom{v}{Mv} ~, \hspace{0.5 in} M = \left( \begin{matrix} e^{\pi i /3}-1 & - e^{\pi i /3} \\ - e^{\pi i /3} & 1- e^{\pi i /3}\end{matrix} \right) ~,
\eea
with $v = (v_1, v_2)$ the coordinates on the remaining moduli space.
To identify the global structure of this reduced moduli space, we ask which elements of $ \cW(\mathfrak{d}_4)$ can be recast as an action on the two-dimensional subspace spanned by $v$, as in (\ref{eq:singlocmodeq}). By explicit implementation in GAP \cite{GAP4} or Magma \cite{MR1484478}, it can be shown that a total of 24 elements of $ \cW(\mathfrak{d}_4)$ can be recast in this manner. Indeed, the corresponding two-dimensional representation is generated by 
\bea
s = \left(\begin{matrix}0 & -1 \\ - e^{\pi i /3} & - e^{2\pi i /3} \end{matrix} \right) ~, \hspace{0.4 in}t = \left(\begin{matrix}1-e^{\pi i /3} & e^{\pi i/3} \\ 1 & 0 \end{matrix} \right) ~. 
\eea
 These matrices satisfy 
 \bea
 s^3 = t^3 = 1~, \hspace{0.5 in} sts = tst
 \eea
 which are exactly the defining relations for $G_4$. 
 
We thus conclude that the torus compactification of the type-$\mathfrak{d}_4$ (2,0) theory with $\widetilde{\ZZ_3}$ S-fold is indeed a 4d $\cN=3$ theory with moduli space labelled by $\CC^6/G_4$, the coordinates on $\CC^6$ being $v$. 
We may obtain further data about this tentative $\cN=3$ theory by considering the singular strata on its moduli space. These strata are in one-to-one correspondence with conjugacy classes of maximal cyclic subgroups of $G_4$ generated by a reflection. There is in fact only one such conjugacy class, of order 3 (see e.g. Table D.1 of \cite{lehrer2009unitary}), which we can take to be the fixed locus of $t$, namely 
\bea
\label{eq:d4sfixed}
t v = v \hspace{0.5 in} \Rightarrow \hspace{0.5 in} v_2 = \half (1+\sqrt{3})(1+i) v_1~.
\eea
This gives rise to the following simple Hasse diagram,
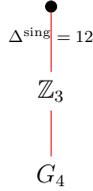
\[
\begin{tikzpicture}
\begin{scope}[scale=1.5,xshift=5.4cm,yshift=-0.7cm]
\node[bbc,scale=.5] (p0a) at (0,.75){};
\node[scale=.8] (p1) at (0,00) {$\ZZ_3$};
\node[scale=.8] (t1b) at (0,0.5) {{\scriptsize$\Delta^{\mathrm{sing}} = {12}$}};
\node[scale=.8] (p0b) at (0,-.75) {$G_{4}$};
\draw[red] (p0a) -- (p1);
\draw[red] (p1) -- (p0b);
\end{scope}\end{tikzpicture}
\]
which can also be obtained via the methods outlined in Appendix \ref{app:Hasse}. Recall that $\Delta^{\mathrm{sing}}$ appearing in the above diagram represents the order of the homogeneous polynomial whose vanishing defines the singular locus.\footnote{More explicitly, the invariant coordinates on the moduli space are of order 4 and 6, as discussed in Appendix \ref{app:CRGspecifics}. On the locus (\ref{eq:d4sfixed}) these take the form 
\bea
u_4 = -4(3 + 2 \sqrt{3}) v_1^4~, \hspace{0.5in} u_6 = 2(5 + 3 \sqrt{3}) (1+i) v_1^6~,
\eea 
and hence in terms of the invariant polynomials the singular locus is defined by an order-12 equation $u_4^3 - 12 \sqrt{3} \,i\,u_6^2 = 0$.}
 The $\ZZ_3$ in the diagram denotes a rank-1 theory with moduli space $\CC^3/\ZZ_3$. 
There are two potential candidates for this rank-1 theory---a non-trivial $\ZZ_3$ S-fold $\cS_{\varnothing,3}^{(1)}$, or a $\ZZ_3$ gauging of $U(1)$.

To decide between the two possibilities, it is useful to review some of the known constraints on the allowed structure for moduli spaces of 4d $\cN\geq 2 $ theories.  First of all, given the Coulomb branch dimensions of a theory (which are equal to the dimensions of the invariant polynomials of the relevant CCRG) it is possible to compute the central charge via the Shapere-Tachikawa formula \cite{Shapere:2008zf}, which for a rank-$r$ theory reads 
\bea
\label{eq:ShapTach}
2(2a-c) + {r \over 2} = \sum_{i=1}^r \Delta_i~. 
\eea
Using the fact that any $\cN=3$ theory has $a=c$ \cite{Aharony:2015oyb}, this allows us to solve for $c$.  

Given the central charge, one may then apply the formula of \cite{Martone:2020nsy,Martone:2021ixp}, which relates the central charge of the theory at the origin of the moduli space to the data of the rank-1 theories living on the codimension-1 singular strata,\footnote{We refer to a locus as ``codimension-1" if it is defined by a single equation in the coordinates $(v_1,v_2,\dots v_N)$ spanning $\CC^{3N}/\Gamma$, even though each $v_i$ is technically a coordinate on $\CC^3$ so that such a locus is really complex codimension-3.}
\bea
\label{eq:centralchargeform}
12 \,c = 2r + h_{\mathrm{ECB}} + \sum_{i \in \cI} \Delta_i^{\mathrm{sing}} b_i~.
\eea
Here $r$ is the rank of the theory at the origin of the moduli space, while $h_{\mathrm{ECB}}$ is the dimension of the extended Coulomb branch, which for $\cN=3$ theories is simply the full dimension of the moduli space, i.e. $ h_{\mathrm{ECB}}=r$. The sum is over the set $\cI$ of all codimension-1 singular strata. The quantity $\Delta_i^{\mathrm{sing}}$  is the degree of homogeneity of the polynomial specifying the stratum, and the quantities $b_i$ are built from the data of the rank-1 theories living on the strata via
\bea
b_i := {12 c_i - h_i - 2 \over \Delta_i}~
\eea
with $c_i$, $h_i$, $\Delta_i$ being the central charge, extended Coulomb branch dimension, and Coulomb branch scaling dimensions for the rank-1 theory. This data can be found in e.g. \cite{Martone:2021ixp}.

Let us now try to identify the particular rank-1 theories living on the codimension-1 stratum in the $G_4$ Hasse diagram. As mentioned above, based purely on geometric considerations this may be either a non-trivial $\ZZ_3$ S-fold theory or a $\ZZ_3$ gauging of a $U(1)$ theory. In the case of a $\ZZ_k$ S-fold, the quantities $b_i$ appearing above are known to take the following values \cite{Argyres:2015ffa,Argyres:2015gha,Argyres:2016xmc},
\bea
\label{eq:Sfoldbs}
\ZZ_k \hspace{0.1 in} \mathrm{S}\hbox{-}\mathrm{fold}: \qquad b =  { 6(k-1) \over k} ~. 
\eea
On the other hand, for a $\ZZ_k$ discrete gauging of $U(1)$, it is easy to show that 
\bea
\label{eq:U1bis0}
U(1) / \ZZ_k: \qquad b = 0 ~. 
\eea

If the codimension-1 stratum in the $G_4$ Hasse diagram corresponded to a $\ZZ_3$ gauging of a $U(1)$ theory, the central charge formula  (\ref{eq:centralchargeform}) would predict that $12\, c = 4$, which would mean that the theory we have constructed is just four free vector multiplets. We find it unlikely that the M-theory construction outlined here would give a free theory, so we do not consider this option further.\footnote{Ultimately this is only an assumption though. There do exist surprising cases in which M-theory on complicated Calabi-Yau singularities produce discrete quotients of free hypermultiplets \cite{Closset:2020scj,Closset:2020afy,Collinucci:2021ofd,DeMarco:2021try}, and we are not able to conclusively rule out a similar phenomenon here.}

We instead focus on the case in which the singular stratum hosts the non-trivial rank-1 S-fold $\cS_{\varnothing,3}^{(1)}$.  A simple check that this is consistent is that the central charge predicted by the central charge formula (\ref{eq:centralchargeform}) matches with the central charge predicted by the Shapere-Tachikawa formula. Indeed in the current case the central charge formula, together with (\ref{eq:Sfoldbs}), gives 
\bea
12\, c = 6 + 4 \times 12 = 54~,
\eea
whereas the Shapere-Tachikawa formula, together with the fact that the scaling dimensions on the moduli space $\CC^2/G_4$ are given by the degrees of $G_4$, i.e. $(\Delta_u, \Delta_v) = (4,6)$, gives
\bea
2c+ 1 = \Delta_u + \Delta_v \hspace{0.5 in} \Rightarrow \hspace{0.5 in} 12\, c = 54~,
\eea
which matches. 
%We thus expect that the $G_4$ theory we have identified is a new theory (and in particular is not a discrete gauging of a known theory), with the non-trivial S-fold $\cS_{\varnothing,3}^{(1)}$ living on its single codimension-1 stratum. 

We may gain further evidence for the presence of $\cS^{(1)}_{\varnothing, 3}$ on the singular stratum by analyzing the behavior of the (2,0) theory on the codimension-1 singular locus. To do so, we begin by inserting (\ref{eq:d4sfixed}) into $\Phi$ and using our freedom of Weyl transformations to write $\Phi = v_1 (1,\,1,\,1,\,i \sqrt{3})$. We now aim to understand the moduli space of the (2,0) theory at this location. To do so, we search for the set of $w \in \cW(\mathfrak{d}_4)$ such that $w \Phi = \Phi$. 
It can be shown by explicit implementation in GAP or Magma that this set contains six elements, and in fact forms the group $S_3 = \cW(\mathfrak{a}_2)$. This suggests that on this fixed locus the type-$\mathfrak{d}_4$ (2,0) theory actually becomes the type-$\mathfrak{a}_2$ (2,0) theory. S-folds of the latter are well-studied and are indeed expected to give the rank-1  $\cS_{\varnothing,3}^{(1)}$ theory, though this again is not conclusive.

%In fact, the same arguments will hold for all cases with a single rank-1 stratum. In all such cases it seems likely that the corresponding theories are genuinely new, and not just discrete gauging of known theories.

\subsubsection{Type-$\mathfrak{e}_6$ }
\label{sec:e6case}

\begin{figure}[tbp]
\begin{center}
\begin{tikzpicture}[decoration={markings,
mark=at position .5 with {\arrow{>}}}]
\begin{scope}[scale=1.5]
\node[bbc,scale=.5] (p0a) at (0,0) {};
\node[scale=.5] (p0b) at (0,-1.4) {};
\node[scale=.8] (t0b) at (0,-1.5){$G_5$};
\node[scale=.7] (p1) at (-.8,-.7) {$\ZZ_3$ };
\node[scale=.7] (p2) at (.8,-.7) { $\ZZ_3$ };
\node[scale=.7] (p3) at (0,-.6){};
\node[scale=.7] (p4) at (0,-.8){};
\node[scale=.8] (t2b) at (-.6,-.3) {{\scriptsize$\Delta^{\mathrm{sing}} = {12}$}};
\node[scale=.8] (t3b) at (.55,-.3)  {{\scriptsize$\Delta^{\mathrm{sing}} = {12}$}};
\draw[red] (p0a) -- (p1);
\draw[red] (p0a) -- (p2);
\draw[red] (p1) -- (p0b);
\draw[red] (p2) -- (p0b);
\end{scope}
\begin{scope}[scale=1.5,xshift=2.7cm,yshift=-0.7cm]
\node[bbc,scale=.5] (p0a) at (0,.75){};
\node[scale=.8] (p1) at (0,0) {$\ZZ_4$};
\node[scale=.8] (t1b) at (0,0.5) {{\scriptsize$\Delta^{\mathrm{sing}} = {24}$}};
\node[scale=.8] (p0b) at (0,-.75) {$G_8$};
\draw[red] (p0a) -- (p1);
\draw[red] (p1) -- (p0b);
\end{scope}
\begin{scope}[scale=1.5,xshift=5.4cm,yshift=-0.7cm]
\node[bbc,scale=.5] (p0a) at (0,1.655){};

\node[scale=.8] (p0) at (0,0.8) {$\ZZ_3$};
%\node[scale=.8] (t1c) at (0,0.5) {{\scriptsize$\Delta^{\mathrm{sing}} = {12}$}};

\node[scale=.8] (p1) at (0,0) {$G_4$};
\node[scale=.8] (t1b) at (0,1.3) {{\scriptsize$\Delta^{\mathrm{sing}} = {36}$}};
\node[scale=.8] (p0b) at (0,-.75) {$G_{25}$};
\draw[red] (p0) -- (p0a);
\draw[red] (p0) -- (p1);
\draw[red] (p1) -- (p0b);
\end{scope}
\end{tikzpicture}
\caption{Hasse diagrams for the moduli spaces of the $\cN=3$ theories obtained from the type-$\mathfrak{e}_6$ (2,0) theory. }
\label{fig:rank2e6}
\end{center}
\end{figure}
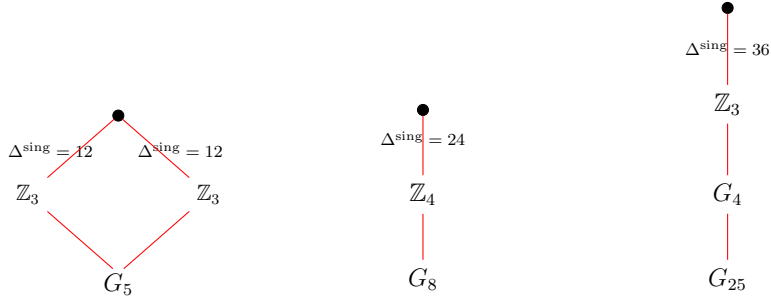

We next consider $\cN=3$ theories obtained by starting with the type-$\mathfrak{e}_6$ (2,0) theory and doing an appropriate S-folding. This allows us to obtain rank-2 theories with moduli space $\CC^6/G_5$ and $\CC^6/G_8$, as well as a rank-3 theory with moduli space $\CC^9/G_{25}$.\footnote{Note that $\cN=3$ theories labelled by $G_5$ and $G_8$ were also proposed in \cite{Cecotti:2021ouq}, though string theory constructions were not provided.} In particular, the operations implementing the S-foldings on the moduli space are
\bea
&G_5:& \hspace{0.5 in} \mathds{O}_5 = e^{2 \pi i/6} \mathds{1}~,
\no\\
&G_8:& \hspace{0.5 in} \mathds{O}_8 = e^{2 \pi i/4} \mathds{1}~,
\no\\
&G_{25}:& \hspace{0.5 in} \mathds{O}_{25} = e^{2\pi i/3}\, \mathds{1}~.
\eea

We may now proceed exactly as before. First demanding (\ref{eq:orbifoldconst}) with $ \mathds{O}= \mathds{O}_5$ and $ w \in \cW(\mathfrak{e}_6)$, we can write\footnote{For technical reasons we use an eight-dimensional matrix representation of $\cW(\mathfrak{e}_6)$ instead of the usual six-dimensional one.}
\bea
\Phi = \binom{v}{Mv} ~, \hspace{0.5 in} M ={ \tiny{{\scriptstyle 1\over \scriptstyle 12 + 6 \sqrt{3}} \left( \begin{matrix}  (1-i \sqrt{3})(6 + 3 \sqrt{3}) &  (1+i \sqrt{3})(6 + 3 \sqrt{3}) \\ -3 +(i-2)\sqrt{3} &{ (3-12 i)+(2-7i) \sqrt{3} } \\ {-(9+9i)-(5+5i) \sqrt{3}} & {(9-3i)+(5-i) \sqrt{3} } \\ (3-9i) + (1-5i) \sqrt{3} & -(1+i)(3 + \sqrt{3})\end{matrix} \right) }}~,
\eea
where $v=(v_1, v_2)$. Looking for the subset of the Weyl group $\cW(\mathfrak{e}_6)$ whose action on $\Phi$ can be recast as an action on the two-dimensional space spanned by $v$, we find 72 elements, which form a group generated by 
\bea
s = \left( \begin{matrix} \half (1- i \sqrt{3}) &\half(1+i\sqrt{3})  \\ 1 & 0 \end{matrix} \right) ~,\hspace{0.2in} t={{ \scriptstyle 1 \over \scriptstyle 12 + 6 \sqrt{3}}\left(  \begin{matrix} \scriptstyle -5-3 \sqrt{3} - i (11 + 7 \sqrt{3}) & \scriptstyle 8 + 5 \sqrt{3} - i \\ \scriptstyle 4 + 2 \sqrt{3} - i ( 2 + 2 \sqrt{3}) & \scriptstyle 11+ 6 \sqrt{3} + i (2 + \sqrt{3}) \end{matrix} \right) }~.\no\\
\eea
These satisfy 
\bea
s^3 = t^3 =1 ~, \hspace{0.5 in} stst = tsts~,
\eea
which are indeed the defining relations of $G_5$. 

We may now consider the singular loci in the moduli space $\CC^6/G_5$. Since $G_5$ has two distinct conjugacy classes of maximal cyclic subgroups generated by a reflection (both of order 3), we expect two distinct singular strata, % The theories living on them should be of Kodaira type $IV/IV^*$, since these are the only cases with order 3 monodromy. 
which can be taken to be the fixed loci of $s$ and $t$. We may now use the techniques in Appendix \ref{app:Hasse} to determine that the theories on both of these strata must have moduli spaces $\CC^3/\ZZ_3$, c.f. Figure \ref{fig:rank2e6}. The only candidates for the rank-1 theories living on these strata are a $\ZZ_3$ gauging of $U(1)$ or the non-trivial S-fold $\cS_{\varnothing,3}^{(1)}$. Let us now determine which combinations are allowed.

Assuming that the theory is not completely free, the only options are to have both strata host $\cS_{\varnothing,3}^{(1)}$, or for one stratum to host $\cS_{\varnothing,3}^{(1)}$ and the other to host a $\ZZ_3$ gauging of $U(1)$. In the former case, the central charge formula (\ref{eq:centralchargeform}) predicts 
\bea
 ( \cS_{\varnothing,3}^{(1)}, \cS_{\varnothing,3}^{(1)}):\qquad 12 c = 6 + 4\times 12 + 4\times 12 = 102
\eea
which is in perfect agreement with the result from the Shapere-Tachikawa formula. On the other hand, if one of the strata hosts a $\ZZ_3$ gauging of $U(1)$, we predict 
\bea
 ( \cS_{\varnothing,3}^{(1)}, U(1)/\ZZ_6):\qquad 12 c = 6 + 4\times 12  = 54~.
\eea
This central charge is identical to the one obtained for the $G_4$ theory in Section \ref{sec:G4}. It is thus possible that the current theory is simply a $\ZZ_3$ gauging of the $G_4$ theory obtained before. Indeed, $G_5$ as a group can be thought of as a $\ZZ_3$ extension of $G_4$, so this interpretation is plausible. 
%It is possible to determine which interpretation is the correct by analyzing the behavior of the (2,0) theory on the co-dimension one strata as done previously, but we will not perform this analysis here. 

The exact same steps can be repeated for the $\cN=3$ theories labeled by $G_8$ and $G_{25}$. Their central charges and scaling dimensions are given in Table \ref{tab:mainresults}, and their Hasse diagrams are shown in Figure \ref{fig:rank2e6}. The only slight modification to the above manipulations is that for $G_{25}$ the resulting theory is now rank-3. This in particular means that the moduli space can have codimension-2 singular loci hosting rank-2 theories. 

Since the $G_8$ and $G_{25}$ theories have only a single codimension-1 stratum, the theories on the strata must be non-trivial (assuming that the $G_8$ and $G_{25}$ theories themselves are not free). We may check that this is a consistent interpretation by verifying that the central charge formula (\ref{eq:centralchargeform}), which e.g. for $G_{25}$ gives
\bea
12 c = 9 + 4 \times 36= 153~,
\eea
 matches with the result from the Shapere-Tachikawa formula.

 It is furthermore natural to assume that the theory on the codimension-2 stratum for $G_{25}$, which has moduli space $\CC^6/G_4$, is the same as the $G_4$ theory obtained in Section \ref{sec:G4} via an S-folding of the type-$\mathfrak{d}_4$ (2,0) theory. This can be supported by considering the behavior of the (2,0) theory on the codimension-2 singular locus. Indeed, it can be shown that the subgroup of $\cW(\mathfrak{e}_6)$ which leaves the fixed locus invariant is none other than $\cW(\mathfrak{d}_4)$, meaning that the type-$\mathfrak{e}_6$ (2,0) theory becomes the type-$\mathfrak{d}_4$ (2,0) theory at these points. Arguments identical to those in Section \ref{sec:G4} can then be run--- asking for the subset of $\cW(\mathfrak{d}_4)$ which commutes with the action of $\mathds{O} = e^{2\pi i/3}\, \mathds{1}$ (realized as an element of  $\cW(\mathfrak{e}_6)$) gives precisely $G_4$.

%Note that in these cases there is only a single codimension-1 stratum, and hence (assuming that the theories are not completely free), the rank-1 theories on these strata must be non-trivial S-folds. This would suggest that the theories identified here are indeed new, and in particular are not discrete gaugings of known theories. 

\subsubsection{Type-$\mathfrak{e}_7$ }
\label{sec:G26}

\begin{figure}[tbp]
\begin{center}
\begin{tikzpicture}[decoration={markings,
mark=at position .5 with {\arrow{>}}}]

\begin{scope}[scale=1.5,xshift=4cm,yshift=-0.7cm]
\node[bbc,scale=.5] (p0) at (0,1.655){};

\node[scale=.8] (p1) at (-0.9,0.9) {$\ZZ_3$};
\node[scale=.8] (p3) at (0.9,0.9) {$\ZZ_2$};

\node[scale=.8] (p4) at (-0.9,0.15) {$G_4$};
\node[scale=.8] (p5) at (0.9,0.15) {$G(3,1,2)$};

\node[scale=.8]  at (-0.6,1.34) {{\scriptsize$\Delta^{\mathrm{sing}} = {36}$}};
\node[scale=.8] (t1b) at (0.6,1.34) {{\scriptsize$\Delta^{\mathrm{sing}} = {18}$}};
\node[scale=.8] (p6) at (0,-.75) {$G_{26}$};
\draw[red] (p0) -- (p1);
\draw[red] (p0) -- (p3);
\draw[red] (p1) -- (p4);
\draw[red] (p1) -- (p5);
\draw[red] (p3) -- (p5);
\draw[red] (p5) -- (p6);
\draw[red] (p4) -- (p6);
\end{scope}\end{tikzpicture}
\caption{Hasse diagram for the moduli space of the $\cN=3$ theory obtained from the type-$\mathfrak{e}_7$ (2,0) theory. }
\label{fig:E67Hasse}
\end{center}
\end{figure}
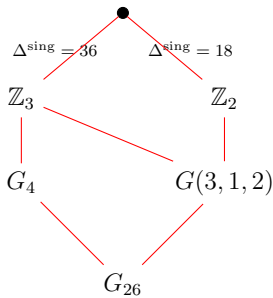

We may obtain a theory labelled by $G_{26}$ by beginning with the type-$\mathfrak{e}_7$ (2,0) theory and doing a $\ZZ_6$ S-folding, i.e. $\mathds{O} = e^{2 \pi i \over 6} \mathds{1}$. As in the case of $G_{25}$, the result is a rank-3 theory. The Hasse diagram of the theory is shown in Figure \ref{fig:E67Hasse}. We see that there are two codimension-1 singular loci, hosting rank-1 theories with moduli space $\CC^3/\ZZ_k$ for $k=2,3$. As was the case before, there are again seemingly two consistent choices for the theories living on each of these strata, namely a $\ZZ_k$ gauging of $U(1)$ or a non-trivial theory $\cS_{\varnothing, k}^{(1)}$. Let us begin by assuming that both of the rank-1 theories are non-trivial. Then the central charge formula (\ref{eq:centralchargeform}) gives 
\bea
12 c = 9 + 4 \times 36 + 3 \times 18 = 207~,
\eea
which matches with the results from the Shapere-Tachikawa formula. If on the other hand we take the theory on the $\ZZ_2$ stratum to be a $\ZZ_2$ gauging of $U(1)$, we instead find 
\bea
12 c = 9 + 4 \times 36 = 153~,
\eea
which we recognize as the central charge for the $G_{25}$ theory obtained in Section \ref{sec:e6case}. This suggests that the $G_{26}$ moduli space identified here could also be consistently interpreted as the moduli space of a $\ZZ_2$ gauging of the $G_{25}$ theory. Indeed, as a group $G_{26}$ is a $\ZZ_2$ extension of $G_{25}$. %Unfortunately will not be able to make any further comments regarding this ambiguity.  

On the other hand, if we keep the theory on the $\ZZ_2$ stratum as $\cS_{\varnothing, 2}^{(1)}$ and instead replace the theory on the $\ZZ_3$ stratum with a $\ZZ_3$ gauging of $U(1)$, then we have 
\bea
12 c = 9 + 3 \times 18 = 63~. 
\eea
The Shapere-Tachikawa formula then predicts $\sum_i \Delta_i = 12$. The list of rank-3 CCRGs satisfying this equality is limited, and in fact rules out all ECCRGs. As for classical CCRGs $G(m,p,3)$, we have $\sum_i \Delta_i = 3m(1 + {1\over p})$ and equating this to 12 allows us to solve for $m$ as a function of $p$. Further requiring that the CCRG have order which is twice the order of $G_{26}$ gives ${6 m^3\over p }= 2 \times 1296$.  Plugging in the expression for $m$ and solving for $p$, one finds that there is no integer solution. Thus this interpretation of the moduli space is in fact inconsistent. 

To summarize, the $G_{26}$ theory could be interpreted in two ways, namely as
\begin{enumerate}
\item A theory with moduli space $\CC^{9}/ G_{26}$ and non-trivial rank-1 theories on all of its codimension-1 strata.
\item A $\ZZ_2$ gauging of a theory with moduli space $\CC^{9}/ G_{25}$.
\end{enumerate}
It would be interesting to determine which is the correct interpretation. 

\subsubsection{Type-$\mathfrak{e}_8$ }
\label{sec:e8sec}

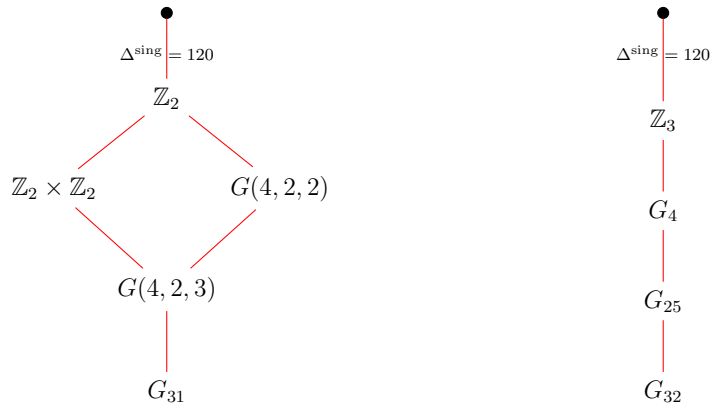
\begin{figure}[tbp]
\begin{center}
\begin{tikzpicture}[decoration={markings,
mark=at position .5 with {\arrow{>}}}]
\begin{scope}[scale=1.5,xshift=0cm,yshift=-0.7cm]
\node[bbc,scale=.5] (p0) at (0,1.655){};

\node[scale=.8] (p2) at (0,0.9) {$\ZZ_2$};

\node[scale=.8] (p5) at (-1,0.1) {$\ZZ_2 \times \ZZ_2$};
\node[scale=.8] (p6) at (1,0.1) {$G(4,2,2)$};

\node[scale=.8] (p7) at (0,-0.8) {$G(4,2,3)$};

\node[scale=.8]  at (0,1.3) {{\scriptsize$\Delta^{\mathrm{sing}} = {120}$}};
\node[scale=.8] (p8) at (0,-1.7) {$G_{31}$};
\draw[red] (p0) -- (p2);
\draw[red] (p2) -- (p5);
\draw[red] (p2) -- (p6);
\draw[red] (p5) -- (p7);
\draw[red] (p6) -- (p7);
\draw[red] (p7) -- (p8);

\end{scope}
\begin{scope}[scale=1.5,xshift=4.4cm,yshift=-0.7cm]
\node[bbc,scale=.5] (p0) at (0,1.655){};

\node[scale=.8] (p1) at (0,0.7) {$\ZZ_3$};

\node[scale=.8] (p2) at (0,-0.1) {$G_4$};

\node[scale=.8] (p3) at (0,-0.9) {$G_{25}$};

\node[scale=.8]  at (0,1.3) {{\scriptsize$\Delta^{\mathrm{sing}} = {120}$}};
\node[scale=.8] (p4) at (0,-1.7) {$G_{32}$};
\draw[red] (p0) -- (p1);
\draw[red] (p1) -- (p2);
\draw[red] (p2) -- (p3);
\draw[red] (p3) -- (p4);
\end{scope}

\end{tikzpicture}
\caption{Hasse diagrams for the moduli space of the $\cN=3$ theories obtained from the type-$\mathfrak{e}_8$ (2,0) theory.}
\label{fig:E8Hasse}
\end{center}
\end{figure}
Finally, the type-$\mathfrak{e}_8$ (2,0) theory can be used to obtain $\cN=3$ theories labelled by ECCRGs $G_{31}$ and $G_{32}$. To do so, we do respectively a $\ZZ_4$ or $\ZZ_6$ S-fold, i.e. $\mathds{O} = e^{2 \pi i \over 4} \mathds{1}$ or $e^{2 \pi i \over 6} \mathds{1}$. In each case, the result is a rank-4 theory, with the Hasse diagrams given in Figure \ref{fig:E8Hasse}. In each case there is only a single codimension-1 locus, and hence assuming that the theories are not completely free we conclude that the rank-1 theories on this loci must be non-trivial. The central charges may be computed via the central charge formula (\ref{eq:centralchargeform}) to get
\bea
&G_{31}:& \qquad 12 c = 12 + 3 \times 120 = 372~,
\no\\
&G_{32}:& \qquad 12 c = 12 + 4 \times 120 = 492~,
\eea
both of which agree with the results of the Shapere-Tachikawa formula. 

Interestingly, the $G_{31}$ moduli space contains a codimension-2 locus which hosts a rank-2 theory with moduli space $\CC^6/ G(4,2,2)$. It also contains a codimension-3 locus which hosts a rank-3 theory with moduli space $\CC^9/ G(4,2,3)$. Since the $G_{31}$ theory is not a discrete gauging, it is likewise expected that the $G(4,2,2)$ and $G(4,2,3)$ theories are not either. We thus expect the existence of non-trivial $\cN=3$ theories with moduli spaces labelled by $G(4,2,2)$ and $G(4,2,3)$. 
In Appendix \ref{sec:dtype(2,0) theory}, it will be shown that theories with moduli space labelled by $G(4,2,2)$ (resp. $G(4,2,3)$) can be obtained from  $\ZZ_4$ S-folding of the type-$\mathfrak{d}_8$ (resp. type-$\mathfrak{d}_{12}$) (2,0) theory, and it is natural to assume that those are the same theories as the ones identified here. This should be verifiable by analyzing the behavior of the (2,0) theory on the corresponding singular loci, though we do not carry out this analysis here.

\section{$\cN=3$ theories from outer automorphism twists}
\label{sec:nonsimplac}
We now turn to the case with outer automorphism twists. In this case the M-theory construction is complicated by a more subtle action of S-duality, as was foreshadowed in Section \ref{sec:generalization}. In this section we elaborate on this action, and use the results to identify new $\cN=3$ theories.

\subsection{S-duality for non-simply-laced groups}

S-duality is a statement about the equivalence of  a theory with gauge algebra $\mathfrak{g}$ at one coupling to a theory with gauge algebra  $\mathfrak{g}^\vee$  at another coupling.\footnote{Here $\mathfrak{g}^\vee$ is the Langlands dual of $\mathfrak{g}$.} For simply-laced algebras, $\mathfrak{g}^\vee \cong \mathfrak{g}$ and the duality maps a theory at coupling $\tau$ to a theory with the same algebra at coupling $S\tau$, where
\bea
S :\,\, \tau \mapsto - {1\over \tau}~.
\eea
 This together with $T: \tau \rightarrow \tau + 1$ gives rise to the full Montonen-Olive duality group $SL(2, \ZZ)$. 

For non-simply-laced algebras, the discussion is more subtle. Whereas before $\mathfrak{g}^\vee$ was isomorphic to $\mathfrak{g}$, for non-simply-laced cases we have 
\bea
\mathfrak{b}_N^\vee = \mathfrak{c}_N~, \hspace{0.2 in}\mathfrak{c}_N^\vee = \mathfrak{b}_N~,\hspace{0.2 in}\mathfrak{g}_2^\vee = \mathfrak{g}_2~, \hspace{0.2 in}\mathfrak{f}_4^\vee = \mathfrak{f}_4~. 
\eea
This in particular means that for  $\mathfrak{b}_N$ and $\mathfrak{c}_N$ there is no unambiguous way to say {how} $\tau$ transforms under the duality. Indeed, when $\mathfrak{g}^\vee \cong \mathfrak{g}$ it is possible to say how the duality acts on $\tau$ without distinguishing whether we are talking about the $\tau$ of a theory with algebra $\mathfrak{g}$ or the $\tau$ of a theory with algebra $\mathfrak{g}^\vee$, but this does not hold if the two algebras are distinct. In most of the remainder of this work we will restrict to  $\mathfrak{g} = \mathfrak{g}_2$ or $\mathfrak{f}_4$, where this issue does not arise. 

A further subtlety is that for theories with non-simply-laced gauge algebras, S-duality actually identifies a theory with algebra $\mathfrak{g}$ at coupling $\tau$ to a theory with algebra $\mathfrak{g}^\vee$ at coupling $\widetilde{S}\tau$, where $\widetilde{S}$ is defined as \cite{Vafa:1997mh,Argyres:2006qr}
\bea
\label{eq:Stildedef}
\widetilde S:\,\, \tau \rightarrow - {1\over n_\mathfrak{g} \tau}~, \hspace{0.5 in} n_\mathfrak{g} = \left\{ \begin{matrix} \,\,2 &\hspace{0.5in} \mathfrak{g}= \mathfrak{b}_N,\, \mathfrak{c}_N, \mathfrak{f}_4 \\ \,\,3 &  \mathfrak{g}=\mathfrak{g}_2  \end{matrix} \right. ~. 
\eea
As an element of $SL(2, \RR)$, this can be written as
\bea
\widetilde{S} = \left( \begin{matrix} 0 & 1/\sqrt{n_\mathfrak{g}} \\ - \sqrt{n_\mathfrak{g}} & 0 \end{matrix} \right)~.
\eea
What is the group that is generated by $T$ and $\widetilde{S}$? We may begin by noting that $\widetilde{S}T\widetilde{S}$ is equivalent to $ST^{n_\mathfrak{g}}S$. This together with $T$ generates the congruence subgroup $\Gamma_0(n_\mathfrak{g})\subset SL(2,\ZZ)$, and thus the full Montonen-Olive duality group in this case is an extension of $\Gamma_0(n_\mathfrak{g})$ by $\widetilde{S}$. This gives a so-called ``Hecke group" $\cH(n_\mathfrak{g})$ defined by the relations 
\bea
\label{eq:Heckegroup}
(\widetilde{S})^2 = (\widetilde{S} T)^{2n_\mathfrak{g}} = C~, \hspace{0.5in} C^2 =1 ~.
\eea
At special points on the boundary of the fundamental domain of the Hecke group, part of the group becomes an enhanced symmetry of the theory. In particular, for $\mathfrak{g}_2$ the fixed points and enhanced symmetry groups are
\bea
\label{eq:g2enhancedsymm}
\tau &=& {i \over \sqrt{3}}: \hspace{0.6in} \widetilde{\ZZ}_4 := \langle \widetilde{S} \rangle~, 
\no\\
 \tau& =& {i \pm \sqrt{3} \over 2 \sqrt{3}}: \qquad \widetilde{\ZZ}_{12} := \langle \widetilde{S} T\rangle \quad \mathrm{or} \quad  \ZZ_{6} := \langle (\widetilde{S} T)^2\rangle ~,
\eea
whereas for $\mathfrak{f}_4$ they are
\bea
\label{eq:f4enhancedsymm}
 \tau &=& {i \over \sqrt{2}}: \hspace{0.45in} \widetilde{\ZZ}_4 := \langle \widetilde{S} \rangle~,
 \no\\
  \tau &=& {i \pm 1\over 2}: \qquad \widetilde{\ZZ}_8 := \langle \widetilde{S} T\rangle \quad \mathrm{or} \quad  \ZZ_{4} := \langle (\widetilde{S} T)^2\rangle~.
\eea
%In particular, there is a coupling at which the $\widetilde{S}$ transformation---and hence the outer automorphism of $\mathfrak{f}_4$---becomes a symmetry of the theory. But at a generic point on the moduli space of the theory at this coupling, this enhanced symmetry is spontaneously broken by the action (\ref{eq:tildSgf}). 

As opposed to the simply-laced case, in the non-simply-laced case S-duality involves a non-trivial action not only on the couplings and charges, but also on the moduli space. 
If we denote the coordinates on the moduli space by $\Phi$ and the electric-magnetic charge vector by $(\mathbf{e}, \mathbf{m})$, we have \cite{Argyres:2006qr,Kapustin:2006pk}
\bea
\label{eq:SonPhiem}
&\widetilde{S}:& \qquad \Phi \mapsto \cR\, \Phi~,
\no\\
&\vphantom{.}& \hspace{0.3 in} (\mathbf{e}, \mathbf{m}) \mapsto (\cR\,\mathbf{e}, \cR\,\mathbf{m})\cdot\left( \begin{matrix} 0 & -1/\sqrt{n_\mathfrak{g}} \\  \sqrt{n_\mathfrak{g}} & 0 \end{matrix} \right)~,
\eea
where the matrix $\cR$ takes the following form 
\bea
\mathfrak{g}_2: \quad \cR = {1\over 2\sqrt{3}} \left(\begin{matrix} \sqrt{3} &1 \\ 1& -\sqrt{3} \end{matrix}  \right) ~,\hspace{0.5 in} \mathfrak{f}_4: \quad \cR = {1\over \sqrt{2}}\left(\begin{matrix} 1 & 1 & & \\ -1 & 1 & & \\ & & 1 & 1 \\ & & -1 & 1  \end{matrix} \right)~. 
\eea
Note that in both cases the square of $\cR$ is an element of the Weyl group, and thus the square of $\widetilde{S}$ acts trivially on the moduli space. We will also want the action of $\widetilde{S}$ on the invariant polynomials, which is \cite{Argyres:2006qr,Evtikhiev:2020yix}
\bea
\label{eq:tildSgf}
&\mathfrak{g}_2:& \qquad (u_2, u_6) \mapsto (u_2,-u_6)~,
\no\\
&\mathfrak{f}_4:& \qquad  (u_2,u_6,u_8, u_{12}) \mapsto (u_2,-u_6,u_8, -u_{12})~.
\eea
In both cases, these are equivalent to the action of the outer automorphism of the corresponding Dynkin diagram, c.f. Appendix \ref{app:CCRGs}.

\subsection{ECCRGs as subgroups of $\cW(\mathfrak{g}_2)$ and $\cW(\mathfrak{f}_4)$}
\label{sec:ECCRGsubgroupsnonsimp}

The $\mathsf{T}_k$ and $\mathsf{B}_k$ transformations defined in Section \ref{sec:Mtheory} have avatars as elements of Hecke groups in the non-simply-laced case, with $k$ now taking the values given in (\ref{eq:g2enhancedsymm}) and (\ref{eq:f4enhancedsymm}). As in Section \ref{sec:Mtheory}, we can obtain $\cN=3$ theories by quotienting by the diagonal combination of $\mathsf{T}_k \mathsf{B}_k$ and an R-symmetry transformation $\mathsf{R}_k$, the latter which may be chosen as before. Unlike in the simply-laced case, now both $\mathsf{T}_k \mathsf{B}_k$ and $\mathsf{R}_k$ have non-trivial action on the moduli space.

Our goal now will be to understand the moduli spaces of the $\cN=3$ theories that can be obtained from S-folds and outer automorphism twists of exceptional (2,0) theories. As before, we expect them to be labelled by ECCRGs, with the relevant ECCRGs being subgroups of Weyl groups. To identify these subgroups, we again turn to the invariant polynomials. The reader interested in only the results can refer to Table \ref{tab:subgroups} and skip to the next subsection.

\paragraph{$\mathfrak{g}_2$:} We begin by considering $\cW(\mathfrak{g}_2)$, which has invariant polynomials of degree 2 and 6. As per (\ref{eq:g2enhancedsymm}), we may may consider doing a $\widetilde{\ZZ}_4$, $\ZZ_6$, or $\widetilde{\ZZ}_{12}$ S-folding.  However, the latter seems incompatible with the action of $\mathsf{R}_k$ (which is defined for $k=3,4,6$) and hence we consider only the first two. Beginning with the $\widetilde{\ZZ}_4$ S-folding, we have 
\bea
& \mathsf{R}_4:& \qquad(u_2, u_6)\mapsto (-u_2, -u_6)~,
\no\\
&\mathsf{T}_4\mathsf{B}_4:&\qquad(u_2, u_6)\mapsto (u_2, -u_6)~,
\eea
 c.f. (\ref{eq:tildSgf}). The combined transformation $\mathsf{R}_4\mathsf{T}_4\mathsf{B}_4$ thus leaves a single degree-6 invariant, which is the invariant for $G(6,1,1) \cong \ZZ_6$.
 
 As for the $\ZZ_6$ S-folding, we have 
 \bea
  &\mathsf{R}_6:&\qquad (u_2, u_6)\mapsto (e^{2\pi i/3} u_2, u_6)~,
  \no\\
  &\mathsf{T}_6\mathsf{B}_6:&\qquad(u_2, u_6)\mapsto(u_2, u_6)~,
 \eea
 where we have noted that $\mathsf{T}_6\mathsf{B}_6$ contains \textit{two} actions of $\widetilde{S}$, c.f. (\ref{eq:g2enhancedsymm}), and thus acts trivially on the moduli space. Hence we again project out the degree-2 invariant, and are left with a degree-6 invariant. In summary then, the only subgroup of $\cW(\mathfrak{g}_2)$ which we identify is $\ZZ_6$. 

\paragraph{$\mathfrak{f}_4$:} We now consider $\cW(\mathfrak{f}_4)$, which has invariant polynomials of degree $2,6,8,$ and $12$. As per (\ref{eq:f4enhancedsymm}), we may do a $\widetilde{\ZZ}_4$, $\ZZ_4$ (with different generator), or $\widetilde{\ZZ}_{8}$ S-folding. We expect the latter to be inconsistent with the required $\mathsf{R}_k$, so we consider only the first two S-foldings.  Beginning with the $\widetilde{\ZZ}_4$ S-folding, we have 
\bea
&\mathsf{R}_4:&\qquad (u_2, u_6, u_8, u_{12}) \mapsto (-u_2, -u_6, u_8, u_{12})~, 
\no\\
&\mathsf{T}_4\mathsf{B}_4:& \qquad (u_2, u_6, u_8, u_{12})\mapsto (u_2, -u_6, u_8, -u_{12})~,
\eea
c.f. (\ref{eq:tildSgf}). Hence the combined transformation projects out $u_2$ and $u_{12}$, leaving a group with invariants of degree 6 and 8. This group is $G_{12}$. 

Next considering the $\ZZ_4$ S-fold, the generator now involves two copies of $\widetilde{S}$ and hence $\mathsf{T}_4\mathsf{B}_4$ act trivially on the moduli space. Since $\mathsf{R}_4$ acts in the same way as before, this action projects out the $u_2$ and $u_6$, leaving invariants of degree 8 and 12. These are the correct invariants for $G_8$. 

%Finally, considering the $\widetilde{\ZZ_8}$ S-folding, we have 
%\bea
%&\mathsf{R}_8:&\qquad (u_2, u_6, u_8, u_{12}) \rightarrow (i\, u_2, -i\, u_6, u_8, -u_{12})
%no\\
%&\mathsf{T}_8\mathsf{B}_8:& \qquad (u_2, u_6, u_8, u_{12})\mapsto (u_2, -u_6, u_8, -u_{12})~,
%\eea
%The combined action projects out $u_2$ and $u_6$, leaving invariants of degree 8 and 12. This is again $G_8$. 

\subsection{More exceptional $\cN=3$ theories}
\label{sec:ECCRGfromnonsimp}
Having identified the ECCRGs which can appear in our construction, we now discuss the $\cN=3$ theories realizing them. 

\subsubsection{Type-$\mathfrak{g}_2$}
Considering the type-$\mathfrak{d}_4$ (2,0) theory together with a triality twist and a $\widetilde{\ZZ}_4$ or $\ZZ_6$  S-folding gives a rank-1 theory with moduli space $\CC^3/\ZZ_6$. Since in this case the stratification of the CB is trivial, we have no way of checking whether these S-foldings both give rise to the same theory or to different theories with the same moduli space, nor even if the theories obtained in this way are truly new, rather than $\ZZ_6$ gaugings of free theories. Since we find the latter interpretation unlikely, we optimistically conjecture that these twisted compactifications of the type-$\mathfrak{d}_4$ (2,0) theory give rise to new interacting rank-1 $\cN=3$ theories which, by the Shapere-Tachikawa formula, have central charge $12 c = 33$. 

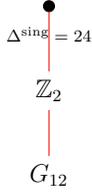
\begin{figure}[tbp]
\begin{center}
\begin{tikzpicture}[decoration={markings,
mark=at position .5 with {\arrow{>}}}]
\begin{scope}[scale=1.5,xshift=0cm,yshift=-0.7cm]
\node[bbc,scale=.5] (p0a) at (0,.75){};
\node[scale=.8] (p1) at (0,00) {$\ZZ_2$};
\node[scale=.8] (t1b) at (0,0.5) {{\scriptsize$\Delta^{\mathrm{sing}} = {24}$}};
\node[scale=.8] (p0b) at (0,-.75) {$G_{12}$};
\draw[red] (p0a) -- (p1);
\draw[red] (p1) -- (p0b);
\end{scope}\end{tikzpicture}
\caption{Hasse diagram for the moduli space of the $G_{12}$ theory obtained from outer automorphism twist of the type-$\mathfrak{e}_6$ (2,0) theory. }
\label{fig:F4Hasse}
\end{center}
\end{figure}

\subsubsection{Type-$\mathfrak{f}_4$}
By considering the type-$\mathfrak{e}_6$ (2,0) theory in the presence of outer automorphism twist and a $\widetilde{\ZZ}_4$ S-fold generated by $\widetilde{S}$, it is possible to get an $\cN=3$ theory with moduli space $\CC^6/G_{12}$. This may be confirmed by imposing (\ref{eq:orbifoldconst}) with
\bea
\mathds{O}= e^{2 \pi i/4} \times {1\over \sqrt{2}} \left(\begin{matrix} 1 & 0 & 0 & -1 \\ 0 & -1 & -1 & 0 \\ 0 & -1 & \,1 & 0 \\ -1 & 0 & 0 & -1 \end{matrix} \right)~,
\eea
where the first factor implements the $\ZZ_4$ S-folding $\mathsf{O}_4$ on the moduli space, and the second factor implements the outer automorphism coming from $\widetilde{S}$, c.f. Appendix \ref{app:CCRGs}. One then proceeds in the manner of Section \ref{sec:orientifold} to show that the moduli space of the theory restricted to the locus satisfying (\ref{eq:orbifoldconst}) is indeed $G_{12}$.

Having done so we may then identify the Hasse diagram, which is given in Figure \ref{fig:F4Hasse}. Because there is only a single codimension-1 stratum, the rank-1 theory living on this stratum must be a non-trivial S-fold, at least assuming that the $G_{12}$ theory is not completely free. This would suggest that the theory identified in this way is indeed new, and in particular is not a discrete gauging of a known theory. A simple check of this proposal is that taking the rank-1 theory to be the non-trivial theory $\cS_{\varnothing, 2}^{(1)}$ and computing the central charge via the central charge formula (\ref{eq:centralchargeform}),
\bea
12 c = 6 + 3 \times 24 = 78~,
\eea
gives a result in agreement with that from the Shapere-Tachikawa formula.

One could also consider starting from the type-$\mathfrak{e}_6$ (2,0) theory with outer automorphism twist and doing a $\ZZ_4$ S-fold generated by $(\widetilde{S}T)^2$. In this case one will obtain a theory with moduli space $\CC^6/G_8$. It is natural to expect that this theory is the same as the one obtained in Section \ref{sec:e6case} from the type-$\mathfrak{e}_6$ (2,0) theory without outer automorphism twist. It should be possible to verify this by analyzing the behavior of the (2,0) theory on the singular locus, but we will not do this here.

\section*{Acknowledgements}

We would like to thank Antonio Amariti, Stefan Hohenegger, and Yuji Tachikawa for useful correspondence, as well as Philip Argyres, Sergio Cecotti, Michele del Zotto, Robert Moscrop, Yuji Tachikawa, and Yunqin Zheng for comments on the draft. MM is supported in part by the NSF grant PHY-1915093, by the Simons Foundation grant 815892 and STFC grant ST/T000759/1. GZ is supported in part by the Simons Foundation grant 815892.

\newpage
%%%%%%%%%%%%%%%%%%%%%%%%%%%%%%%%%%%%
%%%%%%%%%%%%%%%%%%%%%%%%%%%%%%%%%%%%
\appendix 
%%%%%%%%%%%%%%%%%%%%%%%%%%%%%%%%%%%%
%%%%%%%%%%%%%%%%%%%%%%%%%%%%%%%%%%%%

\section{Complex reflection groups}
\label{app:CCRGs}

In this appendix we summarize some aspects of complex reflection groups. More information can be found in \cite{lehrer2009unitary} (see also the appendix in \cite{Tachikawa:2019dvq} for an account aimed at physicists). We also refer the interested reader to \cite{Aharony:2016kai,Caorsi:2018zsq,Bonetti:2018fqz,Cecotti:2015hca,Tachikawa:2019dvq,Evtikhiev:2020yix} for some previous appearances of complex reflections groups in physics. 

Complex reflection groups (CRGs) are groups generated by complex reflections, i.e. unimodular transformations of $\mathbb{C}^r$ that act non-trivially only on one combination of coordinates. As such, they define not only an abstract group, but also a specific representation of it. This representation is given by $r\times r$ unitary matrices determining the action of the group on the $r$ coordinates of $\mathbb{C}^r$, which importantly have only \textit{one} non-unit eigenvalue $e^{\frac{2\pi i}{k}}$. The dimension $r$ on which the group acts irreducibly (i.e. such that there are no fixed points except for the origin) is called the \textit{rank} of the complex reflection group.

A special case of CRGs is when we can replace $\mathbb{C}^r$ by $\mathbb{R}^r$ and complex reflections by real reflection, which are just multiplications of a coordinate by $-1$. Such cases are refereed to as \textit{real reflection groups} or \textit{Coxeter groups}. We can also restrict to the groups that preserve a lattice inside $\mathbb{C}^r$, which are referred to as \textit{crystallographic complex reflection groups}. Finally, we can combine the real and crystallographic cases to get crystallographic real reflection groups, which are precisely the Weyl groups of simple Lie groups.

We can introduce coordinates $z_i$ to describe the space $\mathbb{C}^r$. In general the coordinates will transform under the action of the CRG acting on $z_i$, but it might be possible to find some polynomial combination of the coordinates that remains invariant under all of the elements of the group. These are referred to as the \textit{invariant polynomials} of the group, and are usually specified by their \textit{degree}, i.e. the degree of the corresponding polynomial in the coordinates. For instance, real reflection groups can always be embedded inside $O(r)$, and as such the distance from the origin squared gives a polynomial invariant of degree two for every real reflection group. Invariant polynomials are useful when considering the quotient of the space $\mathbb{C}^r$ by the group in question, as they provide a natural parametrization of that space. 

Invariant polynomials exist for more general groups than just complex reflection groups, but for complex reflection groups they have a simple ring structure. Specifically, given two invariant polynomials of degrees $d_1$ and $d_2$, we can take their product and get a new invariant polynomial of degree $d_1+d_2$. In general, the ring of invariant polynomials is given by a collection of generators obeying several relations. However, for complex reflection groups this polynomial ring is freely generated, and in fact these are the only groups with a freely generated invariant polynomial ring. The ring of invariant polynomials can then be generated by the multiplication of a collection of $r$ polynomials of degree $d_i$. %The list of degrees, $d_i$, is refereed to simply as the degree or the dimension of the invariant polynomials. 

\subsection{Diagrammatic representation}

CRGs can be given diagrammatic representations \cite{Brou1998COMPLEXRG}, which in the context of real reflection groups are known as Coxeter diagrams. For groups that are the Weyl groups of Lie algebras, these are similar to the Dynkin diagrams of the associated Lie algebras, except that the direction of the arrow in the non-simply-laced case is irrelevant. The diagrams are given by nodes and lines connecting them, where to both the nodes and the lines are associated a number. The diagram gives an abstract description of the group in terms of its generators, given by the nodes, and its braiding relations, given by the lines. 

Specifically, a node labeled by a number $p$ is associated with a generator of order $p$. It is customary to suppress this number when it is equal to $2$. Generators $s$ and $t$ connected by a line with number $n\geq 3$ obey a braiding relation of the form
\be \nonumber
\underbrace{s t s t ...}_{n} \,=\, \underbrace{t s t s ...}_{n} ~.
\ee  
When $n=2$, i.e. when the generators commute, no line is drawn. In some places in the literature, a different notation is used wherein the number labeling the line is replaced by multiple separate lines. In this notation, for braiding relations of order $3$ one uses a single line, while braiding relations of order $2n$ are denoted by $n$ separate line (see below for examples).

In the particular case of Weyl groups, the generators are given by reflections along the simple roots, which in turn are associated with nodes in the Dynkin diagram. As such we can indeed associate a generator of the Weyl group with each node in the Dynkin diagram. As each one is a reflection, they are all of order two. Additionally, their commutation relation is dictated by the angle between the corresponding roots, which is represented by the lines connecting the different nodes. In particular, reflections along orthogonal roots should commute. For real reflection groups, all the generators are order two, but complex reflection groups allow for generators with higher order. 

Finally, note that all the relations so far involved at most two generators. However, there are some cases where we shall also need to introduce relations involving three generators. These are usually represented by a circle connecting the three generators. Specifically, the notation:
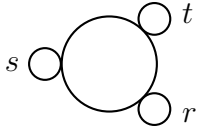
\[\begin{tikzpicture}[decoration={markings,
mark=at position .5 with {\arrow{>}}}]
\begin{scope}[scale=1.5]
\draw[thick] (0,0) circle (12pt);
\draw[thick] (0.4,0.4) circle (4pt);
\node[right] at (0.55,0.45) {$t$};
\draw[thick] (0.4,-0.4) circle (4pt);
\node[right] at (0.55,-0.45) {$r$};
\draw[thick] (-0.57,0) circle (4pt);
\node[left] at (-0.7,0) {$s$};
\end{scope}
\end{tikzpicture}\]
implies that the three generators $s$, $r$, and $t$ obey the relation $s t r = t r s = r s t$. We can also consider circles with multiplicity $n$, in which case the corresponding relation is instead
\be \nonumber
\underbrace{s t r s t r ...}_{2n}\, = \,\underbrace{t r s t r s ...}_{2n} \,= \,\underbrace{r s t r s t ...}_{2n} ~.
\ee 

\subsection{Specific groups}
\label{app:CRGspecifics}
We now discuss some properties of the specific groups that appear in this paper.

\subsubsection*{The Weyl group of $\mathfrak{d}_4$}

The Weyl group of $\mathfrak{d}_4$ is a semi-direct product of the group $(\mathbb{Z}_2)^3$ with the symmetric group $S_4$. We denote it by $\cW(\mathfrak{d}_4)$. Its Coxeter diagram is
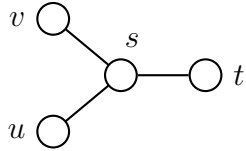
\[\begin{tikzpicture}[decoration={markings,
mark=at position .5 with {\arrow{>}}}]
\begin{scope}[scale=1.5]
\draw[thick] (0.15,0)--(0.6,0);
\draw[thick] (0,0)--(-0.6,-0.5);
\draw[thick] (0,0)--(-0.6,+0.5);
\filldraw[thick,fill=white] (0,0) circle (4pt);
\filldraw[thick,fill=white] (0.75,0) circle (4pt);
\filldraw[thick,fill=white] (-0.6,-0.5) circle (4pt);
\filldraw[thick,fill=white] (-0.6,+0.5) circle (4pt);
\node[right] at (0.9,0) {$t$};
\node[above] at (0.1,0.15) {$s$};
\node[left] at (-0.75,-0.5) {$u$};
\node[left] at (-0.75,0.5) {$v$};
\end{scope}
\end{tikzpicture}\]
All of its generators are of order $2$ (since it is a real reflection group), and they all either commute or obey a third order braiding relation,
\be \nonumber
[t,u]=[t,v]=[u,v]=0~,\hspace{0.3in} tst=sts~, \hspace{0.3in}usu=sus~, \hspace{0.3in}vsv=svs~.
\ee
If we write the coordinates on the space $\mathbb{C}^4$ as $z_i$ for $i=1,\dots,4$, this structure can be represented by permutations of $z_i$ and $z_{i+1}$ for $i=1,2,3$, and the combination of a reflection and permutation given by $z_3 \rightarrow - z_4$, $z_4 \rightarrow - z_3$. In terms of matrices these are given by
\be \nonumber
t = \begin{pmatrix}
  0 & 1 & 0 & 0\\
  1 & 0 & 0 & 0\\
  0 & 0 & 1 & 0\\
	0 & 0 & 0 & 1
\end{pmatrix}~,   \hspace{0.2 in}
s = \begin{pmatrix}
  1 & 0 & 0 & 0\\
  0 & 0 & 1 & 0\\
  0 & 1 & 0 & 0\\
	0 & 0 & 0 & 1
\end{pmatrix}~, \hspace{0.2 in}
u = \begin{pmatrix}
  1 & 0 & 0 & 0\\
  0 & 1 & 0 & 0\\
  0 & 0 & 0 & 1\\
	0 & 0 & 1 & 0
\end{pmatrix}~, \hspace{0.2 in}
v = \begin{pmatrix}
  1 & 0 & 0 & 0\\
  0 & 1 & 0 & 0\\
  0 & 0 & 0 & -1\\
  0 & 0 & -1 & 0
\end{pmatrix}~,
\ee 
which indeed obey the braiding relations determined from the Coxeter diagram. Overall, the full group is given by all permutations of $z_1,z_2,z_3,$ and $z_4$, together with all possible reflections of an even number of the coordinates. As an abstract group it is generated by the three basic generators of $S_4$, together with the reflection $z_3 \rightarrow - z_3$, $z_4 \rightarrow - z_4$. However, the latter element acts as a reflection on two coordinates and so cannot be used as a generator of a reflection group.

A special property of $\cW(\mathfrak{d}_4)$ is that it has a non-abelian group of diagram automorphisms, given by the group $S_3$ associated with the permutations of the three nodes $t, u$, and $v$. In the Lie group, these correspond to the triality outer automorphism permuting its three eight-dimensional representations. We can span the group by two generators, one exchanging the two spinor representations (which can be generalized to higher rank) and one implementing the triality. The element exchanging the two spinor representations can be chosen to act on the basic reflections as $u \leftrightarrow v$ with $s$ and $t$ invariant. This can be implemented by the matrix:  
\be \nonumber
\begin{pmatrix}
  1 & 0 & 0 & 0\\
  0 & 1 & 0 & 0\\
  0 & 0 & 1 & 0\\
  0 & 0 & 0 & -1
\end{pmatrix}~.
\ee 
On the other hand, the triality transformation acts by $t \rightarrow u \rightarrow v \rightarrow t$ with $s$ invariant. This can be implemented by the matrix:  
\be \nonumber
\frac{1}{2}\begin{pmatrix}
  1 & 1 & 1 & -1\\
  1 & 1 & -1 & 1\\
  1 & -1 & 1 & 1\\
  1 & -1 & -1 & -1
\end{pmatrix}~.
\ee 

Th group $\cW(\mathfrak{d}_4)$ has four independent invariant polynomials. A straightforward exercise shows that
\bea \nonumber
& & u_2 = z^2_1 + z^2_2 + z^2_3 + z^2_4~ , \\ \nonumber & & u_4 = \frac{\sqrt{3}}{24} (z^4_1 + z^4_2 + z^4_3 + z^4_4 - 2 z^2_1 z^2_2 - 2 z^2_1 z^2_3 - 2 z^2_1 z^2_4 - 2 z^2_2 z^2_3 - 2 z^2_2 z^2_4 - 2 z^2_3 z^2_4 ) ~, \\ \nonumber & & \widetilde{u}_4 = z_1 z_2 z_3 z_4 ~, \\ \nonumber & & u_6 = z^6_1 + z^6_2 + z^6_3 + z^6_4 + 15 z^2_1 z^2_2 z^2_3 + 15 z^2_1 z^2_3 z^2_4 + 15 z^2_1 z^2_2 z^2_4 + 15 z^2_2 z^2_3 z^2_4 ~
\eea
are indeed invariant. 
Note that under the $S_3$ diagram automorphism, we have $u_2$ and $u_6$ invariant while $u_4$ and $\widetilde{u}_4$ transform as the two dimensional irreducible representation of $S_3$, given by the matrices:
\be \nonumber
\begin{pmatrix}
  1 & 0\\
  0 & -1
\end{pmatrix}~,\hspace{0.5 in}
\frac{1}{2}\begin{pmatrix}
  -1 & -\sqrt{3} \\
  \sqrt{3} & -1
\end{pmatrix}
\ee 
corresponding to the elements of order $2$ and $3$ respectively.

\subsubsection*{The dihedral group}

We next consider the dihedral group $I_n$, which is the symmetry group of the regular $n$-gon. As an abstract group it is given by the semi-direct product of $\mathbb{Z}_n$ and $\mathbb{Z}_2$. For generic $n$ it is not crystallographic, and hence is not a Weyl group. The only exceptions are when $n=3,4,$ and $6$, in which case it is the Weyl group of $\mathfrak{a}_2$, $\mathfrak{c}_2$ and $\mathfrak{g}_2$ respectively. In the current paper we shall be mostly interested in the case of $n=6$, corresponding to the Weyl group of $\mathfrak{g}_2$, but will keep the discussion here more general for completeness. The Coxeter diagram is
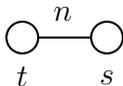
\[\begin{tikzpicture}[decoration={markings,
mark=at position .5 with {\arrow{>}}}]
\begin{scope}[scale=1.5]
\draw[thick] (0,0)--(0.6,0);
\filldraw[thick,fill=white] (0,0) circle (4pt);
\filldraw[thick,fill=white] (0.75,0) circle (4pt);
\node[above] at (0.36,0.05) {$n$};
\node[below] at (0.75,-0.2) {$s$};
\node[below] at (0,-0.16) {$t$};
\end{scope}
\end{tikzpicture}\]
Note that this group is still a real reflection group, so it can be spanned by elements of order $2$. It is convenient to choose these elements to be two reflections along axes related by a $\frac{2\pi}{n}$ rotation. These obey the braiding relation $\underbrace{s t s ...}_{n} = \underbrace{t s t ...}_{n}\,$, and are given in terms of matrices by
\be \nonumber
t=\begin{pmatrix}
  1 & 0\\
  0 & -1
\end{pmatrix}~,\hspace{0.5 in}
s=\begin{pmatrix}
  \cos \frac{2\pi}{n} & \,\,\,\,\sin \frac{2\pi}{n} \\
  \sin \frac{2\pi}{n} & -\cos \frac{2\pi}{n}
\end{pmatrix}~,
\ee
where the matrices are taken to act on the two coordinates $z_1,z_2$ of $\mathbb{C}^2$. 
Overall, the full group is given by an order $n$ rotation of the $z_i$ coordinates, a reflection $z_2\rightarrow -z_2$, and the reflections related to it by the rotation.

The dihedral group possesses a diagram automorphism of order $2$, given by reflecting the diagram. The action on the generators is $s \leftrightarrow t$. This can be implemented by the matrix
\be \nonumber
R=\begin{pmatrix}
  \cos\frac{\pi}{n} & \,\,\,\,\sin\frac{\pi}{n} \\
  \sin\frac{\pi}{n} & -\cos\frac{\pi}{n}
\end{pmatrix}~.
\ee
For $n=3$ this is related to the charge conjugation outer automorphism of the $\mathfrak{a}_2$ Lie algebra, although it should be noted that from the point of view of the Weyl group this can also be generated by an inner automorphism. For $n=4$ or $6$, this gives an automorphism of the Weyl group of $\mathfrak{c}_2$ and $\mathfrak{g}_2$ that is not related to an outer automorphism of the Lie algebra. This is because the Lie algebras are sensitive to the length of the roots, while the reflections they generate are not. 

There are two independent invariant polynomials of these groups, which are of order $2$ and $n$. Choosing the basis of $\mathbb{C}^2$ as $z_1, z_2$, it is straightforward to see that these are given by:
\be \nonumber
u_2 = z^2_1 + z^2_2 ~, \hspace{0.5 in}u_n = \prod^{n-1}_{j=0} \left(z_1 \cos\frac{2\pi j}{n} +z_2\sin\frac{2\pi j}{n} \right) ~.
\ee
The first follows since the transformations are orthogonal, while the latter is the product of the $n$ vertices of the invariant $n$-gon. Under the diagram automorphism we have that $u_2$ is invariant while $\widetilde{u}_n \rightarrow -\widetilde{u}_n$, where $\widetilde{u}_n$ is equal to $u_n$ for $n$ odd, but for $n$ even is a linear combination of $u_6$ and $(u_2)^{\frac{n}{2}}$.

\subsubsection*{The Weyl group of $\mathfrak{f}_4$}

We next consider the Weyl group $\cW(\mathfrak{f}_4)$ of the exceptional Lie algebra $\mathfrak{f}_4$. It is a semi-direct product of the Weyl group of $\mathfrak{d}_4$ and the symmetric group $S_3$, where the latter acts on the former via the outer automorphism of $\mathfrak{d}_4$. Its Coxeter diagram is
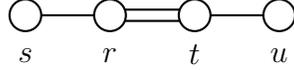
\[\begin{tikzpicture}[decoration={markings,
mark=at position .5 with {\arrow{>}}}]
\begin{scope}[scale=1.5]
\draw[thick] (-0.1,-0.05)--(0.7,-0.05);
\draw[thick] (-0.1,0.05)--(0.7,0.05);
\draw[thick] (-0.75,0)--(0,0);
\draw[thick] (0.75,0)--(1.5,0);
\filldraw[thick,fill=white] (-0.75,0) circle (4pt);
\filldraw[thick,fill=white] (0,0) circle (4pt);
\filldraw[thick,fill=white] (0.75,0) circle (4pt);
\filldraw[thick,fill=white] (1.5,0) circle (4pt);
\node[below] at (-0.75,-0.2) {$s$};
\node[below] at (0.75,-0.16) {$t$};
\node[below] at (0,-0.2) {$r$};
\node[below] at (1.5,-0.2) {$u$};
\end{scope}
\end{tikzpicture}\]

Because it is a Weyl group, all the generators are of order $2$. The four generators obey the braiding relations $s r s = r s r$, $t u t = u t u$, and $r t r t = t r t r$, with all other elements commuting. Taking the coordinates on $\mathbb{C}^{4}$ to be $z_i$, we may realize the group via transformations of $z_i$ by
\be \nonumber
s = \begin{pmatrix}
  0 & 1 & 0 & 0\\
  1 & 0 & 0 & 0\\
  0 & 0 & 1 & 0\\
	0 & 0 & 0 & 1
\end{pmatrix}~,   \hspace{0.2in}
r = \begin{pmatrix}
  1 & 0 & 0 & 0\\
  0 & 0 & 1 & 0\\
  0 & 1 & 0 & 0\\
	0 & 0 & 0 & 1
\end{pmatrix}~,   \hspace{0.2in}
t = \begin{pmatrix}
  1 & 0 & 0 & 0\\
  0 & 1 & 0 & 0\\
  0 & 0 & -1 & 0\\
	0 & 0 & 0 & 1
\end{pmatrix}~,   \hspace{0.2in}
u = \frac{1}{2}\begin{pmatrix}
  1 & -1 & -1 & 1\\
  -1 & 1 & -1 & 1\\
  -1 & -1 & 1 & 1\\
  1 & 1 & 1 & 1
\end{pmatrix}~.
\ee 
%That these matrices generate the Weyl group of $F_4$ can be understood from our discussion of the Weyl group of $SO(8)$. 
Note that the matrices $t$ and $u$ here generate the $S_3$ outer automorphisms of $\mathfrak{d}_4$, while $s$ and $r$ generate an $S_3$ subgroup of $\cW(\mathfrak{d}_4)$. The additional generators required to get the full Weyl group of $\mathfrak{d}_4$ are generated by the action of $t$ and $u$ on $s$. 

There is an order-$2$ diagram automorphism for $\cW(\mathfrak{f}_4)$ given by reflecting the diagram. The action on the Weyl group generators  is $s \leftrightarrow u$, $r \leftrightarrow t$. This can be implemented by the matrix

\be
\frac{1}{\sqrt{2}}\begin{pmatrix}
  1 & 0 & 0 & -1\\
  0 & -1 & -1 & 0\\
  0 & -1 & 1 & 0\\
  -1 & 0 & 0 & -1
\end{pmatrix}~.
\ee 
This gives an automorphism of the Weyl group of $\mathfrak{f}_4$ that is not related to an outer automorphism of the Lie algebra. This is because the Lie algebras are sensitive to the length of the roots while the reflections they generate are not. 

The group has four independent invariant polynomials of orders $2$, $6$, $8$, and $12$. These can be understood as the two invariant polynomial of $\cW(\mathfrak{d}_4)$ of order $2$ and $6$, which are invariant under the $S_3$ outer automorphism group, plus the two independent invariants of $S_3$ of orders $2$ and $3$, but made from the two order $4$ invariants of $\cW(\mathfrak{d}_4)$. The explicit expressions can be written down using the expressions found above for the invariant polynomials of $\cW(\mathfrak{d}_4)$ and the dihedral group $I_3$. The diagram automorphism acts on the  invariant polynomials as $(u_2,\,u_6,\, u_8,\,u_{12}) \rightarrow(u_2,\, -u_6,\, u_8,\,- u_{12})$.

\subsubsection*{The group $G(k,p,N)$}

We next consider the complex reflection groups $G(k,p,N)$, with $p$ a divisor of $k$. This group is of rank $N$. If we take the coordinates on $\mathbb{C}^N$ to be $z_i$ with $i=1,\dots,N$, then we can take its action to be given by permutations of the $z_i$ coordinates, together with a transformation $\mathrm{diag}(e^{\frac{2\pi i n_1}{k}},\,e^{\frac{2\pi i n_2 }{k}},\, \dots,\,e^{\frac{2\pi i n_N }{k}})$ with $\sum_{i=1}^N n_i = 0\,$ mod $p$. For $k=1,2$ these are real reflection groups, corresponding to the Weyl group of $\mathfrak{a}_{N-1}$ for $(k,p)=(1,1)$, $\mathfrak{c}_{N}$ for $(k,p)=(2,1)$, and $\mathfrak{d}_{N}$ for $(k,p)=(2,2)$. It is also a real reflection group for $k=p$, $N=2$, in which case it becomes the dihedral group studied above. It is crystallographic when $k=1,2,3,4$, and $6$.

The diagrammatic representation of  $G(k,p,N)$ depends on $p$, and we refer the reader to \cite{Brou1998COMPLEXRG} for the various cases. It has $N$ invariant polynomials of degrees $k$, $2k$, $3k$, ... , $(N-1)k$, and $\frac{k N}{p}$.    

\subsubsection*{The group $G_4$}

We next consider the exceptional complex reflection group $G_4$. As an abstract group it is equal to the binary tetrahedral group, i.e. the lift to $SU(2)$ of the subgroup of $SO(3)$ corresponding to the rotation group of the tetrahedron (as an abstract group, the latter is the alternating group of four elements, $A_4$). Subgroups of $SU(2)$ can be associated with Lie groups via the McKay correspondence, and the one associated with the binary tetrahedral group is $E_6$. This group is crystallographic. 

Its diagramatic representation is
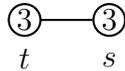
\[\begin{tikzpicture}[decoration={markings,
mark=at position .5 with {\arrow{>}}}]
\begin{scope}[scale=1.5]
\draw[thick] (0,0)--(0.6,0);
\filldraw[thick,fill=white] (0,0) circle (4pt);
\filldraw[thick,fill=white] (0.75,0) circle (4pt);
\node at (0,0) {$3$};
\node at (0.75,0) {$3$};
\node[below] at (0.75,-0.2) {$s$};
\node[below] at (0,-0.16) {$t$};
\end{scope}
\end{tikzpicture}\]
and as such it can be spanned by two elements of order $3$ obeying the braiding relation $s t s = t s t$.
As a complex reflection group, it can be represented in terms of matrices by
\be \nonumber
s=\frac{e^{\frac{\pi i}{3}}}{2}\begin{pmatrix}
  1+i & 1+i \\
  -1+i & 1-i
\end{pmatrix}~,\hspace{0.5 in}
t=\frac{e^{\frac{\pi i}{3}}}{2}\begin{pmatrix}
  1+i & -1-i \\
  1-i & 1-i
\end{pmatrix}~,
\ee
which indeed obey $s^3 = t^3 = 1$ and $s t s = t s t$.  

There is an order-$2$ diagram automorphism given by reflecting the diagram. The action on the basic reflections is $s \leftrightarrow t$, and it can be implemented by the matrix
\be\no
\begin{pmatrix}
  1 & \,\,\,\,0 \\
  0 & -1
\end{pmatrix}~.
\ee
The group has two independent invariant polynomials of orders $4$ and $6$. These can be understood as the product of the four vertices and six lines making the tetrahedron. Using $z_1, z_2$ as coordinates on $\mathbb{C}^2$, the invariant polynomials can be taken to be
\bea \nonumber
 u_4 = z^4_1 + z^4_2 + 2 \sqrt{3} i z^2_1 z^2_2 ~,\hspace{0.5 in}u_6 = z_1 z_2 (z^4_1 - z^4_2) ~.
\eea
The diagram automorphism acts on the invariants as $(u_4,\,u_6) \rightarrow (u_4,\, -u_6)$.

\subsubsection*{The group $G_5$}

We next consider the exceptional complex reflection group $G_5$. It can be thought of as $G_4$ extended by the element $e^{\frac{\pi i}{3}} \mathds{1}_{2\times 2}$. This group is crystallographic. Its diagramatic representation is
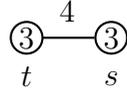
\[\begin{tikzpicture}[decoration={markings,
mark=at position .5 with {\arrow{>}}}]
\begin{scope}[scale=1.5]
\draw[thick] (0,0)--(0.6,0);
\filldraw[thick,fill=white] (0,0) circle (4pt);
\filldraw[thick,fill=white] (0.75,0) circle (4pt);
\node at (0,0) {$3$};
\node at (0.75,0) {$3$};
\node[above] at (0.36,0.05) {$4$};
\node[below] at (0.75,-0.2) {$s$};
\node[below] at (0,-0.16) {$t$};
\end{scope}
\end{tikzpicture}\]
As such, it can be generated by two elements of order $3$ obeying the braiding relation $s t s t = t s t s$. These can be represented in terms of matrices by
\be \nonumber
s=\frac{e^{\frac{\pi i}{3}}}{2}\begin{pmatrix}
  1-i & -1-i \\
  1-i & 1+i
\end{pmatrix}~,\hspace{0.5 in}
t=\frac{e^{\frac{\pi i}{3}}}{2}\begin{pmatrix}
  1+i & -1-i \\
  1-i & 1-i
\end{pmatrix}~.
\ee

There is an order-$2$ diagram automorphism given by reflecting the diagram. The action on the basic reflections is $s \leftrightarrow t$, and it can be implemented by the matrix
\be\no
\frac{1}{\sqrt{2}}\begin{pmatrix}
  0 & 1-i \\
  1+i & 0
\end{pmatrix}~.
\ee

The group has two independent invariant polynomials of orders $6$ and $12$, where the latter is the the order $4$ invariant of $G_4$, raised to the third power so as to be invariant under $e^{\frac{\pi i}{3}} \mathds{1}_{2\times 2}$. If we use $z_1, z_2$ as coordinates on $\mathbb{C}^2$, the invariants can be chosen to be
\bea \nonumber
u_6 = z_1 z_2 (z^4_1 - z^4_2) ~, \hspace{0.5 in} u_{12} = (z^4_1 + z^4_2 + 2 \sqrt{3} \,i \,z^2_1 z^2_2)^3~.
\eea
The diagram automorphism acts on the invariants as $(u_6,\, \widetilde{u}_{12}) \rightarrow (u_6,\, -\widetilde{u}_{12})$, where $\widetilde{u}_{12}$ is some appropriate combination of $u_{12}$ and $u^2_6$. 

\subsubsection*{The group $G_8$}

We next consider the exceptional complex reflection group $G_8$. This is the first of the complex reflection groups based on the binary octahedral group, which is the lift to $SU(2)$ of the subgroup of $SO(3)$ corresponding to the rotational group of the cube (which in turn is equal to $S_4$ as an abstract group). Note however that $G_8$ is not \textit{equal} to the binary octahedral group, having twice as many elements. This group is crystallographic. It has the diagramatic representation
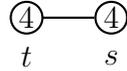
\[\begin{tikzpicture}[decoration={markings,
mark=at position .5 with {\arrow{>}}}]
\begin{scope}[scale=1.5]
\draw[thick] (0,0)--(0.6,0);
\filldraw[thick,fill=white] (0,0) circle (4pt);
\filldraw[thick,fill=white] (0.75,0) circle (4pt);
\node at (0,0) {$4$};
\node at (0.75,0) {$4$};
\node[below] at (0.75,-0.2) {$s$};
\node[below] at (0,-0.16) {$t$};
\end{scope}
\end{tikzpicture}\]
As such, it can be spanned by two elements of order $4$ obeying the braiding relation $s t s = t s t$. These can be represented in terms of matrices by
\be \nonumber
s=\frac{1}{2}\begin{pmatrix}
  1-i & 1-i \\
  -1+i & 1-i
\end{pmatrix}~,\hspace{0.5 in}
t=\begin{pmatrix}
  1 & 0 \\
  0 & -i
\end{pmatrix}~.
\ee
These indeed obey $s^4 = t^4 = 1$ and $s t s = t s t$.   

There is an order-$2$ diagram automorphism given by reflecting the diagram. The action on the basic reflections is $s \leftrightarrow t$, and it can be implemented by the matrix

\be\no
\frac{1}{\sqrt{2}}\begin{pmatrix}
  1 & -i \\
  i & -1
\end{pmatrix}~.
\ee

The group has two independent invariant polynomials of degrees $8$ and $12$. These can be understood as the product of the eight vertices and twelve lines making the cube. Using $z_1, z_2$ as coordinates on $\mathbb{C}^2$, they can be chosen to be
\bea \nonumber
u_8 = z^8_1 + z^8_2 + 14 \,z^4_1 z^4_2 ~, \hspace{0.5 in} u_{12} = z^{12}_1 + z^{12}_2 - 33\, z^4_1 z^4_2 (z^4_1 + z^4_2)~.
\eea
The diagram automorphism acts on the polynomial invariants as $(u_8,\,u_{12}) \rightarrow (u_8,\,-u_{12})$.

\subsubsection*{The group $G_{12}$}

We next consider the exceptional complex reflection group $G_{12}$. This group is also based on the binary octahedral group, having the same order, though the two are not isomorphic. In fact, it is isomorphic to $GL(2,3)$, the group of $2$-dimensional square invertible matrices over a field of order $3$. The binary octahedral group and $GL(2,3)$ are different extensions of $S_4$ by $\mathbb{Z}_2$. This group is crystallographic. It has the diagramatic representation
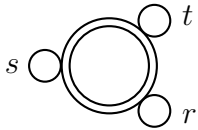
\[\begin{tikzpicture}[decoration={markings,
mark=at position .5 with {\arrow{>}}}]
\begin{scope}[scale=1.5]
\draw[thick] (0,0) circle (12pt);
\draw[thick] (0,0) circle (10pt);
\draw[thick] (0.4,0.4) circle (4pt);
\node[right] at (0.55,0.45) {$t$};
\draw[thick] (0.4,-0.4) circle (4pt);
\node[right] at (0.55,-0.45) {$r$};
\draw[thick] (-0.57,0) circle (4pt);
\node[left] at (-0.7,0) {$s$};
\end{scope}
\end{tikzpicture}\]
Despite being rank-$2$, it requires three reflections to span, all of which have order $2$. They obey the braiding relation $s t r s = t r s t = r s t r$. These can be represented in terms of matrices by
\be \nonumber
s=\frac{1}{\sqrt{2}}\begin{pmatrix}
  -1 & i \\
  -i & 1
\end{pmatrix}~,
\hspace{0.3 in}
t=\frac{1}{\sqrt{2}}\begin{pmatrix}
  -1 & -1 \\
  -1 & 1
\end{pmatrix}~,
\hspace{0.3 in}
r=\frac{1}{\sqrt{2}}\begin{pmatrix}
  0 & -1+i \\
  -1-i & 0
\end{pmatrix}~.
\ee
If we multiply these matrices by $i$, then they generate the binary octahedral group, though this changes the order of the elements and as such the two groups are not isomorphic.   

The diagram has an order-$3$ automorphism given by rotation by $120^\circ$. The action on the basic reflections is $s \rightarrow t \rightarrow r \rightarrow s$, and it can be implemented by the matrix:
\be\no
\frac{e^{\frac{\pi i}{12}}}{\sqrt{2}}\begin{pmatrix}
  i & i \\
  -1 & 1
\end{pmatrix}~.
\ee

The group has two independent invariant polynomials of orders $6$ and $8$. These can be understood as the product of the six faces and eight vertices making the cube. Using $z_1, z_2$ as coordinates on $\mathbb{C}^2$, these can be chosen to be
\bea \nonumber
u_6 = z_1 z_2 (z^4_1 - z^4_2)~,\hspace{0.5 in} u_8 = z^8_1 + z^8_2 + 14 z^4_1 z^4_2~.
\eea
The diagram automorphism acts on the polynomial invariants as $(u_6,\, u_8) \rightarrow (u_6,\, e^{\frac{2\pi i}{3}} u_8)$.

\subsubsection*{The group $G_{25}$}

We next consider the exceptional complex reflection group $G_{25}$. This group is crystallographic. Its diagramatic representation is
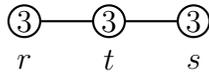
\[\begin{tikzpicture}[decoration={markings,
mark=at position .5 with {\arrow{>}}}]
\begin{scope}[scale=1.5]
\draw[thick] (0,0)--(-0.6,0);
\draw[thick] (0,0)--(0.6,0);
\filldraw[thick,fill=white] (-0.75,0) circle (4pt);
\filldraw[thick,fill=white] (0,0) circle (4pt);
\filldraw[thick,fill=white] (0.75,0) circle (4pt);
\node at (-0.75,0) {$3$};
\node at (0,0) {$3$};
\node at (0.75,0) {$3$};
\node[below] at (-0.75,-0.2) {$r$};
\node[below] at (0.75,-0.2) {$s$};
\node[below] at (0,-0.16) {$t$};
\end{scope}
\end{tikzpicture}\]
As such, it can be spanned by three elements of order $3$ obeying the braiding relations $s t s = t s t$, $r t r = t r t$, and $s r = r s$. These can be represented through matrices by
\be \nonumber
s=\begin{pmatrix}
  1 & 0 & 0 \\
  0 & 1 & 0 \\
	0 & 0 & e^{\frac{2\pi i}{3}}
\end{pmatrix}~,\hspace{0.3 in}
t=\frac{-i}{\sqrt{3}}\begin{pmatrix}
  e^{\frac{2\pi i}{3}} & e^{\frac{4\pi i}{3}} & e^{\frac{4\pi i}{3}} \\
  e^{\frac{4\pi i}{3}} & e^{\frac{2\pi i}{3}} & e^{\frac{4\pi i}{3}} \\
	e^{\frac{4\pi i}{3}} & e^{\frac{4\pi i}{3}} & e^{\frac{2\pi i}{3}}
\end{pmatrix}~,\hspace{0.3 in}
r=\begin{pmatrix}
  1 & 0 & 0 \\
  0 & e^{\frac{2\pi i}{3}} & 0 \\
	0 & 0 & 1
\end{pmatrix}~.
\ee

There is an order-$2$ diagram automorphism given by reflecting the diagram. The action on the basic reflections is $s \leftrightarrow r$, with $t$ invariant. It can be implemented by the matrix
\be\no
\begin{pmatrix}
  1 & 0 & 0 \\
  0 & 0 & 1 \\
	0 & 1 & 0
\end{pmatrix}~.
\ee

The group has three independent invariant polynomials of orders $6$, $9$, and $12$. These can be chosen to be
\bea \nonumber
& & u_6 = z^6_1 + z^6_2 + z^6_3 - 10 \,(z^3_1 z^3_2 + z^3_1 z^3_3 + z^3_2 z^3_3) ~, \\ \nonumber & & u_9 = (z^3_1 - z^3_2)(z^3_1 - z^3_3)(z^3_2 - z^3_3) ~, \\ \nonumber & & u_{12} = z^{12}_1 + z^{12}_2 + z^{12}_3 - 110\, (z^9_1 z^3_2 + z^9_1 z^3_3 + z^9_2 z^3_1 + z^9_2 z^3_3 + z^9_3 z^3_1 + z^9_3 z^3_2) \\ \nonumber & &\hspace{0.5 in} + 462\, (z^6_1 z^6_2 + z^6_1 z^6_3 + z^6_2 z^6_3) ~.
\eea
The diagram automorphism acts on the polynomial invariants as $(u_6,\, u_9,\, u_{12}) \rightarrow (u_6,\, -u_9,\, u_{12}) $.

\subsubsection*{The group $G_{26}$}

We next consider the exceptional complex reflection group $G_{26}$. It is equal to the semi-direct product of $G_{25}$ with $\mathbb{Z}_2$, where the latter acts on the former by its diagram automorphism. This group is crystallographic. Its diagramatic representation is
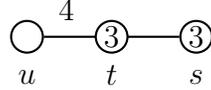
\[\begin{tikzpicture}[decoration={markings,
mark=at position .5 with {\arrow{>}}}]
\begin{scope}[scale=1.5]
\draw[thick] (0,0)--(-0.6,0);
\draw[thick] (0,0)--(0.6,0);
\filldraw[thick,fill=white] (-0.75,0) circle (4pt);
\filldraw[thick,fill=white] (0,0) circle (4pt);
\filldraw[thick,fill=white] (0.75,0) circle (4pt);
\node at (0,0) {$3$};
\node at (0.75,0) {$3$};
\node[above] at (-0.4,0.05) {$4$};
\node[below] at (-0.75,-0.2) {$u$};
\node[below] at (0.75,-0.2) {$s$};
\node[below] at (0,-0.16) {$t$};
\end{scope}
\end{tikzpicture}\]
As such, it can be spanned by three elements, two of order $3$ and the other of order $2$, obeying the braiding relations $s t s = t s t$, $u s u s = s u s u$,  and $t u = u t$. It can be represented in terms of matrices by
\be \nonumber
s=\begin{pmatrix}
  1 & 0 & 0 \\
  0 & 1 & 0 \\
	0 & 0 & e^{\frac{2\pi i}{3}}
\end{pmatrix}~,\hspace{0.3 in}
t=\frac{-i}{\sqrt{3}}\begin{pmatrix}
  e^{\frac{2\pi i}{3}} & e^{\frac{4\pi i}{3}} & e^{\frac{4\pi i}{3}} \\
  e^{\frac{4\pi i}{3}} & e^{\frac{2\pi i}{3}} & e^{\frac{4\pi i}{3}} \\
	e^{\frac{4\pi i}{3}} & e^{\frac{4\pi i}{3}} & e^{\frac{2\pi i}{3}}
\end{pmatrix}~,\hspace{0.3 in}
u=\begin{pmatrix}
  1 & 0 & 0 \\
  0 & 0 & 1 \\
	0 & 1 & 0
\end{pmatrix}~.
\ee
This follows in a rather straightforward manner from our discussion on the group $G_{25}$. The element $u$ generates the $\mathbb{Z}_2$ automorphism.%and through its action on $s$, with the other elements also generate $G_{25}$.   

The group has three independent invariant polynomials of orders $6$, $12$, and $18$, where the latter is just the square of the order $9$ invariant of $G_{25}$.

\subsubsection*{The group $G_{31}$}

We next consider the exceptional complex reflection group $G_{31}$. This group is crystallographic. Its diagramatic representation is
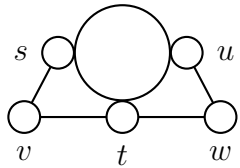
\[\begin{tikzpicture}[decoration={markings,
mark=at position .5 with {\arrow{>}}}]
\begin{scope}[scale=1.5]
\draw[thick] (0.57,0)--(0.87,-0.57)--(0,-0.57);
\draw[thick] (-0.57,0)--(-0.87,-0.57)--(0,-0.57);
\draw[thick] (0,0) circle (12pt);
\filldraw[thick,fill=white] (0.57,0) circle (4pt);
\filldraw[thick,fill=white] (-0.57,0) circle (4pt);
\filldraw[thick,fill=white] (-0.87,-0.57) circle (4pt);
\filldraw[thick,fill=white] (0.87,-0.57) circle (4pt);
\filldraw[thick,fill=white] (0,-0.57) circle (4pt);
\node[left] at (-0.74,0) {$s$};
\node[right] at (0.74,0) {$u$};
\node[below] at (-0.87,-0.73) {$v$};
\node[below] at (0,-0.73) {$t$};
\node[below] at (0.87,-0.73) {$w$};
\end{scope}
\end{tikzpicture}\]
Despite being of rank $4$, it requires five reflections to span, all of which have order $2$. They obey the braiding relations $s v s = v s v$, $t v t = v t v$, $w u w = u w u$, $t w t = w t w$, and $s u t = t s u = u t s$. This can be represented in terms of matrices by

\bea \nonumber
& & s=\begin{pmatrix}
  \frac{1}{2} & -\frac{1}{2} & \frac{1}{\sqrt{2}} & 0 \\
  -\frac{1}{2} & \frac{1}{2} & \frac{1}{\sqrt{2}} & 0 \\
	\frac{1}{\sqrt{2}} & \frac{1}{\sqrt{2}} & 0 & 0 \\
	0 & 0 & 0 & 1
\end{pmatrix}~,\hspace{0.3 in}
v=\begin{pmatrix}
  0 & 1 & 0 & 0 \\
  1 & 0 & 0 & 0 \\
	0 & 0 & 1 & 0 \\
	0 & 0 & 0 & 1
\end{pmatrix}~,\hspace{0.3 in}
t=\begin{pmatrix}
\frac{1}{2} & \frac{-i}{2} & \frac{1+i}{2\sqrt{2}} & \frac{-1+i}{2\sqrt{2}} \\
  \frac{i}{2} & \frac{1}{2} & \frac{1-i}{2\sqrt{2}} & \frac{1+i}{2\sqrt{2}} \\
	\frac{1-i}{2\sqrt{2}} & \frac{1+i}{2\sqrt{2}} & \frac{1}{2} & \frac{-i}{2} \\
	\frac{-1-i}{2\sqrt{2}} & \frac{1-i}{2\sqrt{2}} & \frac{i}{2} & \frac{1}{2}
\end{pmatrix}~, \\ \nonumber
& &\hspace{1 in} w=\begin{pmatrix}
  1 & 0 & 0 & 0 \\
  0 & 1 & 0 & 0 \\
	0 & 0 & 0 & 1 \\
	0 & 0 & 1 & 0
\end{pmatrix},\hspace{0.3 in}
u=\begin{pmatrix}
  1 & 0 & 0 & 0 \\
  0 & 0 & \frac{1}{\sqrt{2}} & \frac{1}{\sqrt{2}} \\
	0 & \frac{1}{\sqrt{2}} & \frac{1}{2} & -\frac{1}{2} \\
	0 & \frac{1}{\sqrt{2}} & -\frac{1}{2} & \frac{1}{2}
\end{pmatrix}.
\eea

The group has an order-$2$ diagram automorphism given by reflecting the diagram. However, due to the circular relation $t$ is not mapped to itself, but rather we have that $s \leftrightarrow u$, $w \leftrightarrow v$, and $t \leftrightarrow t^T$. Note that $t^T = t (u s)^2$ so this action indeed maps group elements to themselves. It can be implemented by the matrix

\be
\begin{pmatrix}
  0 & 0 & 0 & 1 \\
  0 & 0 & 1 & 0 \\
  0 & 1 & 0 & 0 \\
  1 & 0 & 0 & 0
\end{pmatrix}~.
\ee
The subgroup of $G_{31}$ that commutes with this action appears to be generated by $v w$ and $t u s$, which in turn generate the group $G_9$.

The group has four independent invariant polynomials of orders $8$, $12$, $20$, and $24$. The results for the subgroup commmuting with the diagram automorphism suggests that the latter acts on the polynomial invariants as $(u_{8},\,u_{12},\,u_{20},\,u_{24})\rightarrow(u_{8},\,-u_{12},\,-u_{20},\,u_{24})$.

\subsubsection*{The group $G_{32}$}

We next consider the exceptional complex reflection group $G_{32}$. This group is crystallographic, and has the following diagrammatic representation
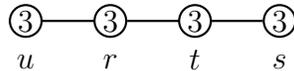
\[\begin{tikzpicture}[decoration={markings,
mark=at position .5 with {\arrow{>}}}]
\begin{scope}[scale=1.5]
\draw[thick] (-1.5,0)--(-0.6,0);
\draw[thick] (0,0)--(-0.6,0);
\draw[thick] (0,0)--(0.6,0);
\filldraw[thick,fill=white] (-1.5,0) circle (4pt);
\filldraw[thick,fill=white] (-0.75,0) circle (4pt);
\filldraw[thick,fill=white] (0,0) circle (4pt);
\filldraw[thick,fill=white] (0.75,0) circle (4pt);
\node at (-1.5,0) {$3$};
\node at (-0.75,0) {$3$};
\node at (0,0) {$3$};
\node at (0.75,0) {$3$};
\node[below] at (-1.5,-0.2) {$u$};
\node[below] at (-0.75,-0.2) {$r$};
\node[below] at (0.75,-0.2) {$s$};
\node[below] at (0,-0.16) {$t$};
\end{scope}
\end{tikzpicture}\]
As such, it can be spanned by four elements of order $3$ obeying the braiding relations $s t s = t s t$, $r t r = t r t$, and $u r u = r u r$, with the rest commuting. This can be represented through matrices by
\bea \nonumber
& & u=\begin{pmatrix}
  1 & 0 & 0 & 0\\
  0 & 1 & 0 & 0\\
	0 & 0 & e^{\frac{2\pi i}{3}} & 0\\
	0 & 0 & 0 & 1
\end{pmatrix}~,\hspace{0.3 in}
r=\frac{-i}{\sqrt{3}}\begin{pmatrix}
  e^{\frac{2\pi i}{3}} & e^{\frac{4\pi i}{3}} & e^{\frac{4\pi i}{3}} & 0 \\
  e^{\frac{4\pi i}{3}} & e^{\frac{2\pi i}{3}} & e^{\frac{4\pi i}{3}} & 0 \\
	e^{\frac{4\pi i}{3}} & e^{\frac{4\pi i}{3}} & e^{\frac{2\pi i}{3}} & 0 \\
	0 & 0 & 0 & \sqrt{3} i
\end{pmatrix}~,\hspace{0.3 in}
t=\begin{pmatrix}
  1 & 0 & 0 & 0 \\
  0 & e^{\frac{2\pi i}{3}} & 0 & 0 \\
	0 & 0 & 1 & 0 \\
	0 & 0 & 0 & 1
\end{pmatrix}~, \\ \nonumber
& & \hspace{1.75in}s=\frac{-i}{\sqrt{3}}\begin{pmatrix}
  e^{\frac{2\pi i}{3}} & e^{\frac{\pi i}{3}} & 0 & e^{\frac{\pi i}{3}} \\
  e^{\frac{\pi i}{3}} & e^{\frac{2\pi i}{3}} & 0 & e^{\frac{4\pi i}{3}} \\
	0 & 0 & \sqrt{3} i & 0 \\
	e^{\frac{\pi i}{3}} & e^{\frac{4\pi i}{3}} & 0 & e^{\frac{2\pi i}{3}}
\end{pmatrix}~.
\eea

There is an order-$2$ diagram automorphism given by reflecting the diagram. The action on the basic reflections is $s \leftrightarrow u$, $t \leftrightarrow r$. It can be implemented by the matrix

\be
\frac{1}{\sqrt{3}}\begin{pmatrix}
  0 & 1 & -1 & -1 \\
  1 & 1 & 1 & 0 \\
	-1 & 1 & 0 & 1 \\
	-1 & 0 & 1 & -1
\end{pmatrix}~.
\ee
The subgroup of $G_{32}$ that commutes with this action is generated by $u s$ and $r t r$, which in turn generate the group $G_{10}$. 

The group has four independent invariant polynomials of orders $12$, $18$, $24$, and $30$. The results for the subgroup commmuting with the diagram automorphism suggests that the latter acts on the polynomial invariants as $(u_{12},\,u_{18},\,u_{24},\,u_{30})\rightarrow(u_{12},\,-u_{18},\,u_{24},\,-u_{30})$.

\section{Hasse diagrams}
\label{app:Hasse}
In this appendix we describe a simple method for constructing Hasse diagrams of moduli spaces of the form $\CC^{3r}/\Gamma$ with $\Gamma$ a CCRG. We illustrate these methods by means of some simple examples. 

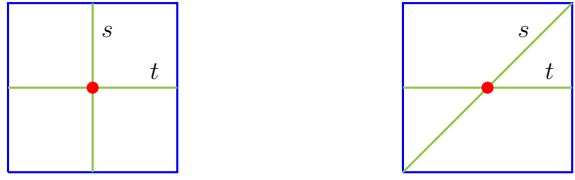
\begin{figure}[tbp]
\begin{center}
\begin{tikzpicture}[decoration={markings,
mark=at position .5 with {\arrow{>}}}]
\begin{scope}[scale=1.5]
\draw[blue,thick] (-1.5,0) -- (0,0)--(0,1.5)--(-1.5,1.5)--(-1.5,0);
\draw[LimeGreen,thick] (-1.5,0.75)--(0,0.75);
\node[right,scale=.8] at  (-0.75,1.25){$s$};
\draw[LimeGreen,thick] (-0.75,0)--(-0.75,1.5);
\node[above,scale=.8] at  (-0.2,0.75){$t$};
\node[bbc,scale=.5,red] at (-0.75,0.75){};
\end{scope}
\begin{scope}[scale=1.5,xshift=3.5cm]
\draw[blue,thick] (-1.5,0) -- (0,0)--(0,1.5)--(-1.5,1.5)--(-1.5,0);
\draw[LimeGreen,thick] (-1.5,0.75)--(0,0.75);
\node[left,scale=.8] at  (-0.3,1.25){$s$};
\draw[LimeGreen,thick] (-1.5,0)--(0,1.5);
\node[above,scale=.8] at  (-0.2,0.75){$t$};
\node[bbc,scale=.5,red] at (-0.75,0.75){};
\end{scope}
\end{tikzpicture}
\caption{Schematic pictures of the  spaces $\CC^6/\ZZ_2 \times \ZZ_2$ and $\CC^6/G(3,1,2)$. The green lines represent the fixed loci of the generators $s,t$ that they are labelled by. The red dot is the intersection of the two fixed loci.}
\label{fig:rank2modspaces}
\end{center}
\end{figure}

Begin by considering a rank-2 theory with moduli space $\CC^6/ \ZZ_2 \times \ZZ_2$, e.g. $\cN=4$ $SU(2)\times SU(2)$ SYM. We may denote the generators of the two $\ZZ_2$ factors by $s$ and $t$. We will work with a coordinate system $(v_1,v_2)$ on $\CC^6$ such that the fixed loci of the two generators are 
\bea
s:\quad v_1=0 ~, \hspace{0.5 in } t: \quad v_2=0~.
\eea
We refer to these loci as ``codimension-1," since they are defined by a single equation in $(v_1,v_2)$, even though each $v_i$ is technically a coordinate on $\CC^3$ so that the loci are really complex codimension-3. 
The space $\CC^6/ \ZZ_2 \times \ZZ_2$ is shown schematically on the left-hand side of Figure \ref{fig:rank2modspaces}, with $s$ and $t$ acting as reflections along the horizontal and vertical axes, respectively. Let us focus on the fixed locus of $t$, corresponding to the horizontal green line in the figure. This is a codimension-1 locus in the full moduli space, and hence the theory which lives on it has a one-dimensional moduli space (by which we really mean one copy of $\CC^3$), corresponding to motion in the transverse, i.e. $v_2$, direction. Locally this moduli space is $\CC^3$, but globally there will be identifications descending from the $\ZZ_2 \times \ZZ_2$ quotient of the full space. To determine the quotient group, we ask for the subgroup of $\ZZ_2 \times \ZZ_2$ which fixes a generic point on the singular locus. Since a generic point on this locus is given by a vector $(v_1, 0)$, the elements of $\ZZ_2\times \ZZ_2$ which leave this invariant are clearly only $1$ and $t$ itself. Thus the moduli space on the fixed locus is of the form $\CC^3/\ZZ_2$, where the $\ZZ_2$ in question is the one generated by $t$. Analogous comments hold for $s$.

This data is enough for us to construct the following Hasse diagram: 
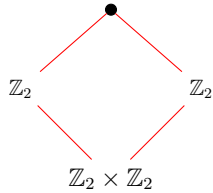
\[
\begin{tikzpicture}
\begin{scope}[scale=1.5]
\node[bbc,scale=.5] (p0a) at (0,0) {};
\node[scale=.8] (p0b) at (0,-1.5){$\ZZ_2 \times \ZZ_2$};
\node[scale=.7] (p1) at (-.8,-.7) {$\ZZ_2$ };
\node[scale=.7] (p2) at (.8,-.7) { $\ZZ_2$ };
\draw[red] (p0a) -- (p1);
\draw[red] (p0a) -- (p2);
\draw[red] (p1) -- (p0b);
\draw[red] (p2) -- (p0b);
\end{scope}
\end{tikzpicture}
 \]
In this diagram, the point at the top represents the generic point in the geometry $\CC^6/\ZZ_2 \times \ZZ_2$, which has trivial moduli space. As we have seen above, the geometry contains two singular loci, which host theories with moduli spaces $\CC^3/ \ZZ_2$. These are represented by the two entries in the middle row. Finally, on the intersection point of these two singular loci lives a rank-2 theory with moduli space being the full geometry $\CC^6/\ZZ_2 \times \ZZ_2$. This is the bottom row.

We next consider a rank-2 theory with moduli space $\CC^6/ G(3,1,2)$. Such a theory can be realized by considering a pair of $\mathrm{D}3$-branes probing a $\ZZ_3$ S-fold. Note that 
\bea
G(3,1,2) = \langle s,t \,\, | \,\, s^2 = t^3 = 1 \,, \, s t s t = t s t s \rangle~,
\eea
with an explicit matrix representation being given by 
\bea
\renewcommand\arraystretch{0.7}
s = \left( \begin{matrix}0 & 1 \\ 1 & 0 \end{matrix} \right) ~, \hspace{0.5 in}t = \left( \begin{matrix}1 & 0 \\ 0 & e^{2 \pi i /3} \end{matrix} \right) ~.
\eea
The fixed loci of each of these are 
\bea
s: \quad v_2 = v_1 ~, \hspace{0.5 in} t: \quad v_2 = 0~,
\eea
giving rise to the right-hand side of Figure \ref{fig:rank2modspaces}. 

Let us again begin with the fixed locus of $t$. As before, the theory on this locus is rank-1, and has moduli space $\CC^3/\Gamma$ where $\Gamma$ is the subgroup of $G(3,1,2)$ such that a generic point on the t-fixed locus is left invariant. Since such generic point is of the form $(v_1,0)$, the relevant subgroup is that containing matrices of the form $\left( \begin{smallmatrix} 1 & 0 \\ \# & \# \end{smallmatrix} \right)$. As for the $s$ fixed locus, the generic point is now of the form $(v_1,v_1)$. To analyze the moduli space here we may begin by doing a change of basis via $p = \left( \begin{smallmatrix} 1 & 1 \\ 1 & -1\end{smallmatrix} \right)$ which maps the $s$ fixed locus onto the $t$ fixed locus. Then one again looks for matrices fixing the latter. 

To summarize, the moduli spaces of the fixed loci are of the form $\CC^3/\Gamma$, where $\Gamma$ is
\bea
&t:&\quad  \Gamma = \left\langle \left( \begin{matrix} 1 & 0 \\ \# & \# \end{matrix} \right)  \right\rangle = \ZZ_3~,
\no\\
&s:&\quad \Gamma = \left\langle p^{-1} \left( \begin{matrix} 1 & 0 \\ \# & \# \end{matrix} \right) p \,\,\Big | \,\, p =  \left( \begin{matrix} 1 & 1 \\ 1 & -1\end{matrix} \right) \right\rangle = \ZZ_2~,
\eea
with the final equalities obtained by explicitly scanning through elements of  $G(3,1,2)$ using GAP. 

Finally we can consider the intersection of the two fixed loci. This point is simply the origin, and it hosts a rank-2 theory which is none other than the theory that we started with. Another way of saying this is that the relevant $\Gamma$ is the subset of $G(3,1,2)$ which leaves the origin fixed, which is of course $G(3,1,2)$ itself. 

This data is then enough for us to construct the following Hasse diagram: 
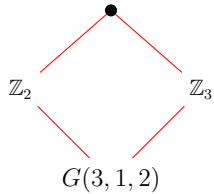
\[
\begin{tikzpicture}
\begin{scope}[scale=1.5]
\node[bbc,scale=.5] (p0a) at (0,0) {};
\node[scale=.8] (p0b) at (0,-1.5){$G(3,1,2)$};
\node[scale=.7] (p1) at (-.8,-.7) {$\ZZ_2$ };
\node[scale=.7] (p2) at (.8,-.7) { $\ZZ_3$ };
\draw[red] (p0a) -- (p1);
\draw[red] (p0a) -- (p2);
\draw[red] (p1) -- (p0b);
\draw[red] (p2) -- (p0b);
\end{scope}
\end{tikzpicture}
 \]

\begin{figure}[tbp]
\begin{center}
\begin{tikzpicture}[decoration={markings,
mark=at position .5 with {\arrow{>}}}]
\begin{scope}[scale=1.5]

\draw[red,thick]  (-1.65,1.35)--(0.35,1.35);

\draw[blue,thick] (-2,0) -- (0,0)--(0,2)--(-2,2)--(-2,0);
\draw[blue,dashed] (-1.3,0.7) -- (0.7,0.7);
\draw[blue,dashed] (-1.3,2.7)--(-1.3,0.7);
\draw[blue,thick] (0.7,2.7)--(0.7,0.7);
\draw[blue,thick] (0.7,2.7)--(-1.3,2.7);
\draw[blue,dashed] (-2,0) -- (-1.3,0.7);
\draw[blue,thick] (0.7,0.7) -- (0,0);
\draw[blue,thick] (0.7,2.7) -- (0,2);
\draw[blue,thick] (-2,2) -- (-1.3,2.7);

\draw[LimeGreen] (-2,1)--(0,1);
\draw[LimeGreen] (-2,1) -- (-1.3,1.7);
\draw[LimeGreen] (0,1) --(0.7,1.7);
\draw[LimeGreen] (-1.3,1.7)--(0.7,1.7);

\draw[LimeGreen,thick] (-2,0)--(0,0);
\draw[LimeGreen,thick] (0.7,2.7)--(-1.3,2.7);
\draw[LimeGreen]  (-2,0)--(-1.3,2.7);
\draw[LimeGreen]  (0,0)--(0.7,2.7);

\node[above,scale=.8] at  (-1.6,0.96){$s$};
\node[above,scale=.8] at  (0.2,2.4){$u$};
\node[above,scale=.8] at  (-0.65,1.35){$s \cap u$};

\end{scope}
\end{tikzpicture}
\caption{Schematic picture of the  space $\CC^9/G_{26}$. The green planes represent the fixed loci of the generators $s$ and $u$. Each of these is codimension-1, and hosts a rank-1 theory. The red line is the intersection of the two fixed loci, which hosts a rank-2 theory. The origin hosts a rank-3 theory.}
\label{fig:rank3modspace}
\end{center}
\end{figure}
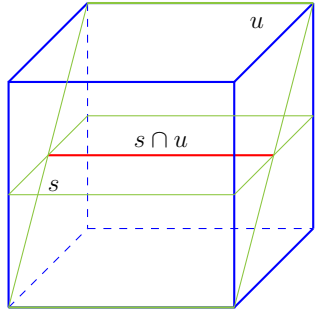

As a final example, we consider a rank-3 theory with moduli space $\CC^9/G_{26}$. The group $G_{26}$ is described abstractly by the relations 
\bea
G_{26} = \langle s, t, u \,\, \big | \,\, s^3 = t^3 = u^2 = 1\,, \, sts=tst \, , \, susu=usus \,, \,tu=ut \,   \rangle~, 
\eea
which can be given an explicit matrix realization by 
\bea
\renewcommand\arraystretch{0.7}
s= \left( \begin{matrix} 1 & 0 & 0 \\ 0 & 1 & 0 \\ 0 & 0 & e^{2 \pi i \over 3} \end{matrix} \right) ~, \hspace{0.3 in}t= - {i \over \sqrt{3}} \left( \begin{matrix} e^{2 \pi i \over 3} & e^{4 \pi i \over 3} & e^{4 \pi i \over 3} \\ e^{4 \pi i \over 3} & e^{2 \pi i \over 3} & e^{4 \pi i \over 3} \\ e^{4 \pi i \over 3} & e^{4 \pi i \over 3} & e^{2 \pi i \over 3}\end{matrix} \right) ~, \hspace{0.3 in}u= \left( \begin{matrix} 1 & 0 & 0 \\ 0 & 0 & 1 \\ 0 & 1 & 0 \end{matrix} \right) ~.
\eea
The fixed loci are now located at 
\bea
s: \,\,\, v_3 =0 ~, \hspace{0.4 in} t: \,\,\,  v_3 = - (v_1 + v_2)~, \hspace{0.4 in} u: \,\,\,  v_3 = v_2~. 
\eea
The $s$ and $u$ fixed loci, together with their intersection, are shown in Figure \ref{fig:rank3modspace}. 

Each of these loci is codimension-1 and hosts a rank-1 theory. The moduli spaces of these theories are given by $\CC^3/\Gamma$, with $\Gamma$ a subgroup of $G_{26}$ identified using the steps above. Explicitly, we have 
\bea
&s:&\quad  \Gamma = \left\langle \left( \begin{matrix} 1 & 0 & 0  \\ 0 & 1 & 0 \\ \# & \# & \# \end{matrix} \right)  \right\rangle = \ZZ_3~,
\no\\
&t:&\quad \Gamma = \left\langle p^{-1} \left( \begin{matrix} 1 & 0 & 0  \\ 0 & 1 & 0 \\ \# & \# & \# \end{matrix} \right) p \,\,\Big | \,\, p =  \left( \begin{matrix} 1 & 0 & 0 \\ 0 & 1 & 0 \\ 1 & 1 & 1\end{matrix} \right) \right\rangle = \ZZ_3~,
\no\\
&u:&\quad \Gamma = \left\langle p^{-1} \left( \begin{matrix} 1 & 0 & 0  \\ 0 & 1 & 0 \\ \# & \# & \# \end{matrix} \right) p \,\,\Big | \,\, p =  \left( \begin{matrix} 1 & 0 & 0 \\ 0& 1 & 1 \\ 0 & 1 & -1\end{matrix} \right) \right\rangle = \ZZ_2~. 
\eea
Here we have used GAP to identify the relevant subgroups.

We next consider the intersection of two such codimension-1 loci. The theory on this intersection is rank-2, and has moduli space $\CC^6/\Gamma$, where $\Gamma$ is again a subgroup of $G_{26}$ which leaves invariant a generic point on the intersection. By entirely analogous steps as above, we have 
\bea
&s \cap u:&\quad \Gamma = \left\langle \left( \begin{matrix} 1 & 0 & 0  \\ \# & \#  & \#  \\ \# & \# & \# \end{matrix} \right) \right\rangle =G(3,1,2)~,
\no\\
&s \cap t:&\quad \Gamma = \left\langle p^{-1} \left( \begin{matrix} 1 & 0 & 0  \\ \# & \#  & \#  \\ \# & \# & \# \end{matrix} \right) p \,\,\Big | \,\, p =  \left( \begin{matrix} 1 & -1 & 0 \\ 1 & 1 & 0 \\ 0 & 0 & 1\end{matrix} \right) \right\rangle =G_4~,
\no\\
&t \cap u:&\quad \Gamma = \left\langle p^{-1} \left( \begin{matrix} 1 & 0 & 0  \\ \# & \#  & \#  \\ \# & \# & \# \end{matrix} \right) p \,\,\Big | \,\, p =  \left( \begin{matrix} 1 & 0 & 0 \\ 1 & 1 & 1 \\ 0 & 1 & -1\end{matrix} \right) \right\rangle =G(3,1,2)~. 
\eea
Again we have used GAP to identify these subgroups. Finally, the intersection of all three fixed loci is simply the origin, which has the full geometry as its moduli space. 

In the current case, before constructing the Hasse diagram we must ask if any of the fixed loci discussed above are identical. Indeed, it can be shown that $s$ and $t$ are actually in the same conjugacy class in $G_{26}$, and hence that the corresponding fixed loci are actually identical in the orbifold geometry. This can also be seen by evaluating the invariant polynomials of $G_{26}$ given in Appendix \ref{app:CCRGs} on the two fixed loci and checking that they are the same. Thus we need only consider one such strata, though this stratum can still have self-intersection giving rise to a rank-2 theory, as seen by $s \cap t$ above. In conclusion, the Hasse diagram we find is the one given in Figure \ref{fig:E67Hasse}.

%%%%%%%%%%%%%%%%%%%%%%%%%%%%%%%%%
\section{S-folds of type-$\mathfrak{d}$ (2,0) theories}
%%%%%%%%%%%%%%%%%%%%%%%%%%%%%%%%%
\label{sec:dtype(2,0) theory}
In this appendix we use the techniques developed in the main text to understand the moduli spaces of $\cN=3$ theories obtained via S-folding of type-$\mathfrak{d}_N$ (2,0) theories. The $\cN=3$ theories which can be obtained from these theories have moduli spaces $\CC^{3N}/\Gamma$ with $\Gamma$ a subgroup of the Weyl group $\cW(\mathfrak{d}_N)$. Concretely, we will find that these subgroups are all either $G(k,1,N)$ or $G(k,2,N)$ with $k=4,6$, and hence that no ECCRGs make an appearance. As usual, we will show this by studying the action of S-foldings on the invariant polynomials of $\cW(\mathfrak{d}_N)$. 

To begin, note that the invariant polynomials of $\cW(\mathfrak{d}_N)$ have degrees 
\bea
\mathfrak{d}_N: \quad 2,\, 4,\, 6,\, 8 ,\, \dots,\, 2N-2,\, N~.
\eea
 We shall focus here on the $\ZZ_3, \ZZ_4$, and $\ZZ_6$ S-foldings. Note that there is also a $\widetilde{\ZZ}_2$ action sending the invariant polynomial of degree $N$ to minus itself, which corresponds to the diagram automorphism of $\mathfrak{d}_N$. For $N$ odd, it is equivalent to the usual $\ZZ_2$ orientifolding, that is the $k=2$ S-fold. However, for $N$ even, the $\ZZ_2$ S-fold acts trivially and $\widetilde{\ZZ}_2$ is a distinct action. As a result, for $N$ odd the $\ZZ_3$ S-folding can be thought of as a combination of this $\widetilde{\ZZ}_2$ action with the $\ZZ_6$ S-folding, while for $N$ even the $\ZZ_3$ and $\ZZ_6$ S-folding are equivalent. Let us now see how each of these operations acts on the invariant polynomials.

\paragraph{$N$ even: }
First consider the case of $N=2n$ being even. Starting with the $\ZZ_4$ S-folding, this operation leaves the following invariant polynomials 
\bea
\mathfrak{d}_{2n}/\ZZ_4: \quad 4,\, 8,\, 12 ,\, \dots,\, 4n-4,\,\,\,\left\{ \begin{matrix} \varnothing && n\,\,\, \mathrm{odd} \\ 2n && n\,\,\, \mathrm{even} \end{matrix}\right. ~. 
\eea
We recognize these as the invariant polynomials of the following CCRGs, 
\bea
n\,\,\, \mathrm{odd}: \quad G(4,1,n-1) ~,\hspace{0.5 in}n\,\,\, \mathrm{even}: \quad G(4,2,n)~.
\eea
We may also consider the $\ZZ_4$ S-folding together with the action of $\widetilde{\ZZ}_2$, which gives 
\bea
\mathfrak{d}_{2n}/\widetilde{\ZZ}_2 \times \ZZ_4 : \quad 4,\, 8,\, 12,\, \dots,\, 4n-4,\,\,\,\left\{ \begin{matrix} 2n && n\,\,\, \mathrm{odd} \\  \varnothing&& n\,\,\, \mathrm{even} \end{matrix}\right. ~.
\eea
This gives the same results as before, but with $n$ even and odd interchanged.

We next consider the $\ZZ_6$ S-fold, which leaves the following invariant polynomials
\bea
\mathfrak{d}_{2n}/\ZZ_6: \quad 6,\, 12,\, 18,\, \, \dots,\,\,\,\left\{ \begin{matrix} 4n-6,\, 2n && n\in 3 \ZZ \\ 4n-4 && n \in 3 \ZZ+1 \\ 4n-2 && n \in 3 \ZZ+ 2\end{matrix}\right. ~. 
\eea
These we recognize as the invariant polynomials of the following CCRGs, 
\bea
&n= &3\ell: \hspace{0.6 in} G(6,2,2\ell) ~,
\no\\
&n= &3\ell+1: \qquad G(6,1,2\ell)~,
\no\\
&n = &3 \ell+2: \qquad G(6,1,2\ell+1)~ .
\eea
The $\ZZ_3 $ S-folding is equivalent to the $\ZZ_6$ S-folding in this case.

\paragraph{$N$ odd: } We next consider the case of $N=2n+1$ being odd. In this case doing both the $\ZZ_4$ or $\widetilde{\ZZ}_2 \times \ZZ_4 $ S-folding gives the same results, namely 
\bea
\mathfrak{d}_{2n+1}/\ZZ_4\,\,\, \&\,\,\, \mathfrak{d}_{2n+1}/\widetilde{\ZZ}_2 \times\ZZ_4 : \quad 4, \, 8, \, 12, \, \dots, 4n
\eea
and hence for any $n$ this gives the invariants of $G(4,1,n)$. 

On the other hand, now there is a distinction between the $\ZZ_3$ and $\ZZ_6$ S-folds. In particular, the $\ZZ_3$ S-fold leaves
\bea
\mathfrak{d}_{2n+1}/\ZZ_3 :\quad 6,\, 12,\, 18,\, \, \dots,\,\,\,\left\{ \begin{matrix} 4n && n\in 3 \ZZ \\ 4n-4,\, 2n+1 && n \in 3 \ZZ+1 \\ 4n-2 && n \in 3 \ZZ+ 2\end{matrix}\right. ~. 
\eea
These we recognize as the invariant polynomials of the following CCRGs, 
\bea
&n= &3\ell: \hspace{0.6 in} G(6,1,2\ell) ~,
\no\\
&n= &3\ell+1: \qquad G(6,2,2\ell+1)~,
\no\\
&n = &3 \ell+2: \qquad G(6,1,2\ell+1)~ .
\eea
Further including $\widetilde{\ZZ}_2$  to get the $\ZZ_6$ S-fold, we see that we project out the degree $2n+1$ invariant in the case of $n \in 3 \ZZ+1$, and hence obtain the group $G(6,1,2\ell)$ instead of $G(6,2,2\ell+1)$. 

\begin{table}[tp]
\begin{center}
\begin{tabular}{c|c|c|c}
$ \Gamma$ & $(2,0)$ construction & $12 c$ & $\{ \Delta_i\}$
\\\hline
$G(4,1,n)$ & $\mathfrak{d}_{2n+1}/\ZZ_4$ & $3n(4n+3)$ & $\{4,\,8,\,12,\,\dots,\,4n \}$
\\
& $\mathfrak{d}_{2n+2}/\ZZ_4$,\, $n$ even &\dittotikz  &\dittotikz
\\
& $\mathfrak{d}_{2n+2}/\widetilde{\ZZ}_2 \times \ZZ_4$,\, $n$ odd & \dittotikz& \dittotikz
\\
$G(4,2,n)$ & $\mathfrak{d}_{2n}/\ZZ_4$,\, $n$ even & $3n(4n-1)$ & $\{4,\,8,\,12,\,\dots,\,4n-4,\, 2n \}$
\\
& $\mathfrak{d}_{2n}/\widetilde{\ZZ}_2 \times \ZZ_4$,\, $n$ odd &\dittotikz & \dittotikz
\\
$G(6,1,n)$ & $\mathfrak{d}_{3n+1}/\ZZ_6$ & $3n(6n+5)$& $\{6,\,12,\,18,\,\dots,\, 6n\}$
\\
& $\mathfrak{d}_{3n+2}/\ZZ_6$ & \dittotikz& \dittotikz
\\
& $\mathfrak{d}_{3n+3}/\ZZ_6$,\, $n$ even& \dittotikz& \dittotikz
\\
$G(6,2,n)$ & $\mathfrak{d}_{3n}/\ZZ_6$,\, $n$ even & $3n(6n-1)$& $\{6,\,12,\,18,\,\dots,\, 6n-6, \, 3n\}$ 
\\
& $\mathfrak{d}_{3n}/\ZZ_3$,\, $n$ odd & \dittotikz& \dittotikz
\end{tabular}
\end{center}
\caption{4d $\cN=3$ theories obtainable from type-$\mathfrak{d}_N$ (2,0) theories. }
\label{tab:typedresults}
\end{table}%

\paragraph{Summary: }

 Starting from the type-$\mathfrak{d}_N$ (2,0) theory, we see that we can obtain $\cN=3$ theories labelled by CCRGs $G(4,1,n)$, $G(4,2,n)$, $G(6,1,n)$, and $G(6,2,n)$ by doing an appropriate S-folding---the list of S-foldings is given in Table \ref{tab:typedresults}. This table also gives the Coulomb branch dimensions of the corresponding theories, which are equal to the dimensions of the invariant polynomials of the CCRGs. From these one can compute the central charges via the Shapere-Tachikawa formula (\ref{eq:ShapTach}), together with the fact that any $\cN=3$ theory has $a=c$.

\paragraph{Moduli spaces: }
To further understand the tentative $\cN=3$ theories of type $G(4,1,n)$, $G(4,2,n)$, $G(6,1,n)$, and $G(6,2,n)$, it is useful to consider the singularity structure of their moduli spaces. In particular, let us focus on the codimension-1 singular loci. These loci can be obtained by considering the first layer in the Hasse diagram, which in turn can be constructed using the techniques of Appendix \ref{app:Hasse}. Using these techniques, we find:
\[
G(m,1,n): \hspace{0.1 in}\begin{tikzpicture}[baseline=-30]
\begin{scope}[scale=1.5]
\node[bbc,scale=.5] (p0a) at (0,0) {};
\node[scale=.7] (p1) at (-.8,-.7) {$\ZZ_2$ };
\node[scale=.7] (p2) at (.8,-.7) { $\ZZ_m$ };

\node[scale=.7] at (-.8,-.3) {$\Delta^{\mathrm{sing}} = m n (n-1)$ };
\node[scale=.7]  at (.8,-.3)  {$\Delta^{\mathrm{sing}} = m n $ };
\draw[red] (p0a) -- (p1);
\draw[red] (p0a) -- (p2);
\draw[red] (p1)--(-1,-1);
\draw[red] (p1)--(-0.6,-1);
\draw[red] (p2)--(1,-1);
\node[scale=.7] at (-.8,-1.2) {$\dots$ };
\node[scale=.7] at (1,-1.2) {$\dots$ };
\end{scope}
\end{tikzpicture}
\hspace{0.5 in}
G(2m,2,n): \hspace{0.1 in}
\begin{tikzpicture}[baseline=-30]
\begin{scope}[scale=1.5]
\node[bbc,scale=.5] (p0a) at (0,0) {};
\node[scale=.7] (p1) at (-.8,-.7) {$\ZZ_2$ };
\node[scale=.7] (p2) at (.8,-.7) { $\ZZ_m$ };

\node[scale=.7] at (-.8,-.3) {$\Delta^{\mathrm{sing}} = 2m n (n-1)$ };
\node[scale=.7]  at (.8,-.3)  {$\Delta^{\mathrm{sing}} = m n $ };
\draw[red] (p0a) -- (p1);
\draw[red] (p0a) -- (p2);
\draw[red] (p1)--(-1,-1);
\draw[red] (p1)--(-0.6,-1);
\draw[red] (p2)--(1,-1);
\node[scale=.7] at (-.8,-1.2) {$\dots$ };
\node[scale=.7] at (1,-1.2) {$\dots$ };
\end{scope}
\end{tikzpicture}
\]
The $\ZZ_k$ in the diagrams above represent rank-1 theories with moduli space $\CC^3 / \ZZ_k$. As we will discuss momentarily, these could be interpreted as either non-trivial $\ZZ_k$ S-fold theories, or as $\ZZ_\ell$ gaugings of $\ZZ_p$ S-folds, with $k=\ell\, p$ (which in the limiting case $\ell=1$ becomes a $\ZZ_k$ gauging of $U(1)$ gauge theory). One way to resolve this ambiguity is to analyze the behavior of the (2,0) theory on the complex codimension-1 loci. This analysis was performed for some ECCRG theories in the main text, but here we will only discuss the consistency at the geometric level.

To do so, we apply the formula in (\ref{eq:centralchargeform}). As we have just mentioned, the rank-1 theories living on the codimension-1 strata can be either non-trivial $\ZZ_k$ S-fold theories or some discretely gauged theories---here for simplicity we will consider only the case in which $\ell=1$, i.e. a $\ZZ_k$ gauging of a $U(1)$ theory. The values of $b_i$ in these cases were given in (\ref{eq:Sfoldbs}) and (\ref{eq:U1bis0}). 

Let us begin with the case of $G(k,1,n)$ for $k=4,6$ and assume that both $\ZZ_2$ and $\ZZ_m$ correspond to non-trivial rank-1 S-folds.\footnote{There is currently no known non-trivial rank-1 $\cN=3$ theory with moduli space $\CC^3/\ZZ_6$. For the purpose of this discussion though we will keep an open mind.} Then the formula (\ref{eq:centralchargeform}) would tell us that 
\bea
12 c\, =\, 3 n+ 3 m n (n-1) + 6 n (m-1) \,= \,3n (nm+m-1)~.
\eea
For $m=4,6$, we see that this matches precisely with the results in Table \ref{tab:typedresults}, which were obtained by an independent computation involving only the Shapere-Tachikawa formula. This is a rather non-trivial consistency check.

However, this alone cannot rule out the possibility of one of the rank-1 strata being a discrete gauging. Indeed, let us consider replacing the non-trivial S-fold theory on the $\ZZ_m$ stratum with a $\ZZ_m$ gauging of $U(1)$. Plugging (\ref{eq:U1bis0}) into (\ref{eq:centralchargeform}) then tells us that 
\bea
\label{eq:Zmdiscc}
12 c = 3 n (mn-m+1)~,
\eea
which is not equivalent to the value in Table \ref{tab:typedresults}. This of course is to be expected: instead of thinking of the theory as one with moduli space $\CC^{3n}/ G(m,1,n)$, we are now trying to think of it as a $\ZZ_m$ gauging of another theory with moduli space $\CC^{3n}/ G(\widetilde m,\widetilde p,n)$. This other theory will generically have different central charge, which is preserved under discrete gauging. This candidate ``other" theory can be identified as follows. First, we note that for $G(\widetilde m,\widetilde p,n)$ we have 
\bea
\sum \Delta_i \,\,=\,\, \widetilde m + 2 \widetilde m + \dots + (n-1) \widetilde m + {\widetilde m n \over \widetilde p}  \,\,= \,\,\half \widetilde m n (n-1) + {\widetilde m n \over \widetilde p} ~.
\eea
This can be inserted into the Shapere-Tachikawa formula to find the corresponding value of $12c$, which must be equivalent to (\ref{eq:Zmdiscc}). We must also require that the order of $\CC^{3n}/ G(\widetilde m,\widetilde p,n)$, namely $|G(\widetilde m,\widetilde p,n)| = {n! \widetilde m^n \over \widetilde p}$, be $1 \over n$ times the order of $G(m,1,n)$. These two equations are solved by $\widetilde m = \widetilde p = m$, which tells us that our proposed theory could also be consistently interpreted as a $\ZZ_m$ gauging of a theory with moduli space $\CC^{3n}/ G( m, m,n)$. Indeed, this is the interpretation put forward in \cite{Aharony:2016kai}. 

We may next try to keep the  $\ZZ_m$ stratum as a non-trivial S-fold, but replace the $\ZZ_2$ stratum with a discrete gauging. Following a similar analysis as above, one finds that this is in fact inconsistent, except in the case of $n=2$, in which case it is possible to interpret the theory as a $\ZZ_2$ gauging of $G(m,2,2)$. Finally, we could imagine replacing \textit{both} the $\ZZ_m$ and $\ZZ_2$ strata with discrete gaugings. Then the central charge formula (\ref{eq:centralchargeform}) would give simply $12c = 3 n$, i.e. $c = {n\over 4}$, which is the central charge of a theory of $n$ free vector multiplets. Thus in this case the theory is completely free. As done in the other cases, we shall mostly ignore this possibility as we find it unlikely for the M-theory construction described in the main text to give rise to a completely free theory. 

To summarize, we have seen that for $G(m,1,n)$ at generic $n$ there are two possibilities: 
\begin{enumerate}
\item A theory with moduli space $\CC^{3n}/ G(m,1,n)$ and non-trivial rank-1 S-folds on all of its codimension-1 strata.
\item A $\ZZ_m$ discrete gauging of a theory with moduli space $\CC^{3n}/ G(m,m,n)$.
\end{enumerate}
The latter interpretation is the one promoted in \cite{Aharony:2016kai}. The same analysis may be repeated for $G(2m,2,n)$, for which there are again two possibilities:
\begin{enumerate}
\item A theory with moduli space $\CC^{3n}/ G(2m,2,n)$ and non-trivial rank-1 S-folds on all of its codimension-1 strata.
\item A $\ZZ_m$ discrete gauging of a theory with moduli space $\CC^{3n}/ G(2m,2m,n)$.
\end{enumerate}
Thus at this stage it remains unclear if the $\cN=3$ theories obtained from the type-$\mathfrak{d}_N$ (2,0) theory are in fact new theories, or just discrete gaugings of the known $G(4,4,n)$ and $G(6,6,n)$ theories. Let us however note that studying the exceptional $\cN=3$ theories leads one to suspect that the $G(4,2,2)$ and $G(4,2,3)$ cases encountered here correspond to genuinely new theories, and not just discrete gaugings of known theories, c.f. Section \ref{sec:e8sec}. 

\bibliographystyle{JHEP}
\bibliography{bib.bib}

\end{document}